\documentclass[nolinenumbers]{aastex631}

\usepackage[utf8]{inputenc}
\usepackage{amsmath}
\usepackage{hyperref}
\usepackage{amssymb}
\usepackage{graphicx}
\usepackage{bm}
\usepackage{float}
\DeclareMathOperator{\arcosh}{arcosh}

\DeclareFontFamily{U}{wncy}{}
\DeclareFontShape{U}{wncy}{m}{n}{<->wncyr10}{}
\DeclareSymbolFont{mcy}{U}{wncy}{m}{n}
\DeclareMathSymbol{\Sh}{\mathord}{mcy}{"58}

\hypersetup{colorlinks=true, linkcolor=blue,
    filecolor=blue,      
    urlcolor=blue,citecolor=blue}
    
\begin{document}

\title{Integral field spectroscopy with the solar gravitational lens}

\author{Alexander Madurowicz}
\affiliation{Kavli Institute for Particle Astrophysics and Cosmology, Stanford University}
\author{Bruce Macintosh}
\affiliation{Kavli Institute for Particle Astrophysics and Cosmology, Stanford University}



\begin{abstract}
    The prospect of combining integral field spectroscopy with the solar gravitational lens (SGL) to spectrally and spatially resolve the surfaces and atmospheres of extrasolar planets is investigated. The properties of hyperbolic orbits visiting the focal region of the SGL are calculated analytically, demonstrating trade offs between departure velocity and time of arrival, as well as gravity assist maneuvers and heliocentric angular velocity. Numerical integration of the solar barycentric motion demonstrates that navigational acceleration of $\textrm{d}v \lesssim 80 \frac{\textrm{m}}{\textrm{s}} + 6.7 \frac{\textrm{m}}{\textrm{s}} \frac{t}{\textrm{year}}$ is needed to obtain and maintain alignment. Obtaining target ephemerides of sufficient precision is an open problem. The optical properties of an oblate gravitational lens are reviewed, including calculations of the magnification and the point-spread function that forms inside a telescope. Image formation for extended, incoherent sources is discussed when the projected image is smaller than, approximately equal to, and larger than the critical caustic. Sources of contamination which limit observational SNR are considered in detail, including the sun, the solar corona, the host star, and potential background objects. A noise mitigation strategy of spectrally and spatially separating the light using integral field spectroscopy is emphasized. A pseudoinverse-based image reconstruction scheme demonstrates that direct reconstruction of an Earth-like source from \textit{single} measurements of the Einstein ring is possible when the critical caustic and observed SNR are sufficiently large. In this arrangement, a mission would not require multiple telescopes or navigational symmetry breaking, enabling continuous monitoring of the atmospheric composition and dynamics on other planets.
\end{abstract}


\section{Introduction}
In the wake of modern discoveries of exoplanets around Sun-like stars \citep{Mayor1995},  advancements in exoplanet detection have revealed statistics on the vast populations of planets in our galactic neighborhood \citep{Akeson2013}.  Continued detection and characterization of these worlds has led others to question the possibility that other lifeforms inhabit these planets, and how we might infer their existence from spectroscopic indicators \citep{seager2013}. Perhaps the most direct path forward to the detection of biosignatures is the development of exoplanet direct imaging and spectroscopy \citep{pueyo2018}, which is the focus of two future flagship NASA mission proposals HabEx \citep{habex} and LUVOIR \citep{luvoir}, as well as the Astro2020 recommendation to combine these efforts into a single 6.5 m class observatory. For these telescope designs, the planets can be resolved spectrally, but spatially appear as a single point, and no current technology has the capability to truly resolve details on the planet's surface or in its atmosphere. One exception is inferring longitudinal cloud variations from temporal and rotational variability in extremely bright, nearby, substellar-mass brown dwarfs \citep{Crossfield2014}, but this technique cannot be easily extended to faint targets or latitudinal variations. However, an extremely ambitious idea to leverage the high magnification and angular resolution of the solar gravitational lens (SGL) to truly resolve a exoplanet has recently been conceptualized \citep{turyshev2020}.

This idea is not entirely new, as it is adapted from similar concepts to use the gravitational lens of the sun as a link for interstellar communication \citep{ehsleman1979}. However, recent theoretical advancements in the scattering of electromagnetic waves off of arbitrary oblate potentials \citep{TT2021a} has enabled the extremely accurate theoretical modeling of the capabilities of the solar gravitational lens. The presence of high order multipole moments in the gravitational potential result in the formation of a critical caustic in the image plane \citep{Loutsenko2018}, which modifies the effective magnification of the lens as well as altering the point-spread function formed in the image plane of a telescope observing at the SGL. This effect is critical for reconstruction of extended sources, as we demonstrate in Section 5. The asymmetric point spread function formed for points inside the critical caustic enables direct disentanglement of the source intensity distribution from single measurements of the Einstein ring's azimuthal profile. In this configuration the reconstruction does not require multiple laterally offset observations of the Einstein ring, which were required using a point-mass or monopole-only model of the gravitational potential \citep{madurowicz2020}.

The paper is organized into three major sections, 2, 3, and 4. Section 2 considers the orbital mechanics required to send a craft to the SGL as well as maintaining alignment with a target planet. Section 3 broadly covers optical effects related to the gravitational lensing in the presence of an oblate potential. Section 4 discusses the sources of noise which will limit real observations as well as potential mitigation strategies. The paper concludes with Section 5 which demonstrates a simple model of source reconstruction from single measurements of the Einstein ring.

\section{Orbital Considerations}

\subsection{Analytic hyperbolic trajectories}
There are multiple, equivalent mathematical formulations for unbound hyperbolic trajectories. The following formulas are typically found in textbooks on astrodynamics \citep{schaub2003} \citep{vallado2001}. The recognizable Cartesian parameterization 
\begin{equation}
    \frac{x^2}{a^2} - \frac{y^2}{b^2} = 1
\end{equation}
describes the shape of the hyperbola with two numbers $(a,b)$ named the semi-major and semi-minor axes. The graph has two asymptotes at $y = \pm\frac{b}{a} x$ and two branches opening along the x-axis which approach the asymptotes as the coordinates approach infinity.

A different formulation uses polar coordinates $(r,\theta)$ 
\begin{equation}
r = \frac{l}{1-e \cos(\theta)}    
\end{equation}
where the shape is described by two new numbers, the semi-latus rectum $l = -\frac{b^2}{a}$ and eccentricity $e = \sqrt{1 + (\frac{b}{a})^2}$. An important note is that the polar coordinates have their origin as the focus of the right branch, and so to match the Cartesian grid, the $(x,y)$ coordinates can be recovered with the relations $x = r \cos(\theta) - ae$ and $y = r \sin(\theta)$.

\begin{figure}
    \centering
    \includegraphics{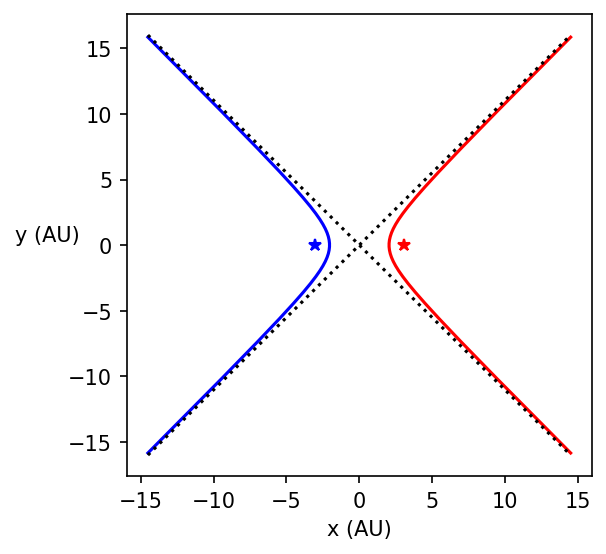}
    \caption{Hyperbola with $r_p = 1$ AU and $v_p = 47$ km/s, corresponding to an escape trajectory from an Earth-like starting position. The shape parameters in other bases are, $a \approx -2.04$  AU, $b \approx 2.25$ AU, $l \approx 2.49$ AU, $e \approx 1.49$.}
    \label{fig:hyperbola}
\end{figure}

However, for the purposes of this investigation, we wish to parameterize the hyperbola with two different numbers, the radius at periastron $r_p$ and the velocity at periastron $v_p$, so that one may think of the hyperbola shape phase space as a set of initial conditions for a rocket launching from an inner solar system before reaching the solar gravitational lens. Everywhere along the hyperbola, the velocity associated with the trajectory is governed by the vis-viva equation
\begin{equation}
    v^2 = \mu\Big(\frac{2}{r} - \frac{1}{a}\Big)
\end{equation}
where $\mu = GM_\odot$ is the standard gravitational parameter of the Sun. So simply plugging in $r_p, v_p$, and solving for $a$ allows determination of the semi-major axis
\begin{equation}
    a = \frac{\mu r_p}{2\mu - v_p^2 r_p}.
\end{equation}
Additionally, the specific angular momentum $h = r_p v_p = \sqrt{\mu l}$ of the orbit is conserved and can be used to find the semi-minor axis
\begin{equation}
    b = \frac{r_p^{3/2} v_p}{\sqrt{v_p^2r_p - 2\mu}}.
\end{equation}
However, this equation will return an imaginary number when $2\mu > v_p^2r_p$, in this case, the chosen $r_p, v_p$ do not correspond to a hyperbola. An example plot of a basic hyperbola is shown in Figure (\ref{fig:hyperbola}).

Using the $(r_p, v_p)$ basis for hyperbola shapes, we aim to answer a few basic questions about the properties of these orbits to inform a design of a craft which intends to visit the solar gravitational lens. The questions are: ``How long will it take to reach a heliocentric distance of $r_{\textrm{SGL}} \approx 600$ AU?" ``What is the angular velocity of the craft with respect to the Sun at that point in time?" and ``How much acceleration is needed to cancel the relative lateral motion to enter a radial hyperbolic trajectory?"

\begin{figure}
    \centering
    \includegraphics[width=\textwidth]{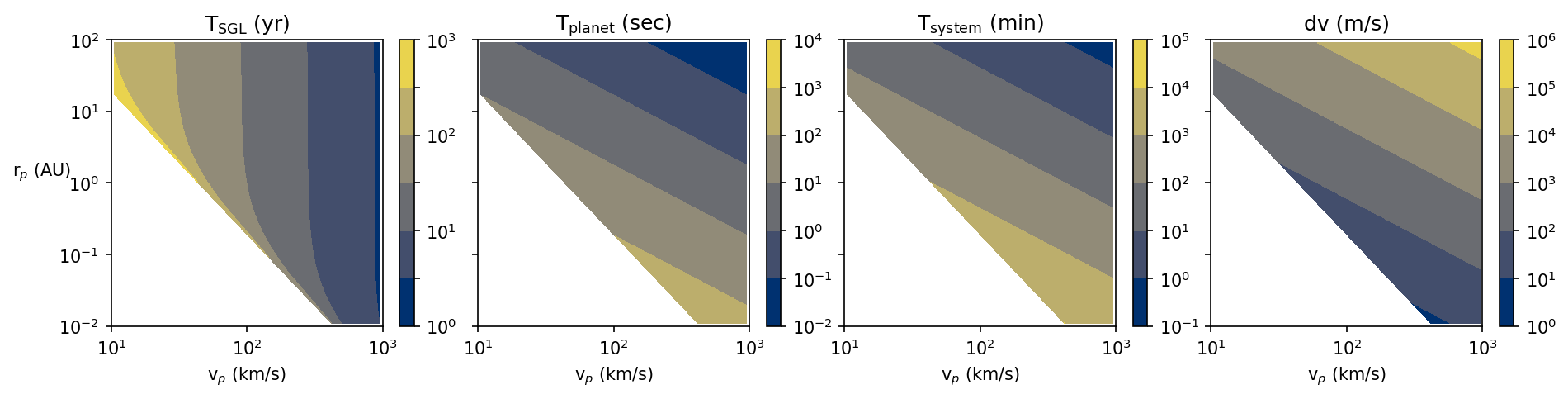}
    \caption{Results of various analytic calculations plotted as contours over the hyperbola shape phase space ($r_p$, $v_p$). (a) time from periastron passage to heliocentric distance $r_\textrm{SGL} = 600$ AU, (b) planet-crossing timescale in seconds, (c) system-crossing timescale in minutes, and (d) delta-v to enter radial orbit. The white region in the bottom-left corner corresponds to bound elliptical trajectories which occur when $2\mu > v_p^2r_p$. }
    \label{fig:tsgl}
\end{figure}

The first question can be answered using the relationship between mean anomaly $M$, eccentric (or hyperbolic) anomaly $E$, and time $t$. Specifically,
\begin{eqnarray}
E &=& \arcosh \frac{\cos \theta + e}{1 + e \cos \theta} \\
M &=& e\sinh E - E \\
t &=& \sqrt{\frac{-a^3}{\mu}}M.
\end{eqnarray}
Then it is just matter of finding $\theta$ corresponding to $r_\textrm{SGL}$ using
\begin{equation}
    \theta_{\textrm{SGL}} = \arccos{\frac{l/r_{\textrm{SGL}} - 1}{e}}
\end{equation}
to compute the time a craft would need to reach such an extreme heliocentric distance. Contours are plotted for a variety of cases in Figure (\ref{fig:tsgl} a).

The second question can be analytically derived using the conservation of specific angular momentum. Since $h = r \times (r \frac{\textrm{d}\theta}{\textrm{d}t})$,
\begin{equation}
    \frac{\textrm{d}\theta}{\textrm{d}t} = \frac{h}{r_\textrm{SGL}^2}.
\end{equation}
which is equivalent to Kepler's second law. We convert these angular velocities into planet-crossing and system-crossing timescales $T_{\textrm{planet}} = \theta_\textrm{planet}/\frac{\textrm{d}\theta}{\textrm{d}t}$ and $T_{\textrm{system}}=\theta_\textrm{system}/\frac{\textrm{d}\theta}{\textrm{d}t}$ using $\theta_\textrm{planet} = 2R_\textrm{Earth} / d_s$, and $\theta_\textrm{system} = 2a_\textrm{sys} / d_s$ where $d_s = 40.54$ ly is the system distance, $a_\textrm{sys} = 0.062$ AU is the target planet semi-major axis and $R_{\textrm{Earth}}$ is the Earth's radii. These estimates correspond to the TRAPPIST-1 system \citep{Gillon2017} and are plotted in Figure \ref{fig:tsgl} (b) and (c).

The last question can be tackled by means of computing the flight path angle $\phi_\textrm{fpa}$, which gives the angle between the velocity vector and the direction perpendicular to the radial vector. The flight path angle is computed with
\begin{equation}
    \phi_\textrm{fpa} = \arctan{\frac{e\sin\theta_\textrm{SGL}}{1+e\cos\theta_\textrm{SGL}}},
\end{equation} and the magnitude of the velocity vector $v_\textrm{SGL} = \sqrt{\mu(2/r_\textrm{SGL} - 1/a)}$ is computed with the vis-viva equation. Thus the delta-v required to eliminate the transverse motion and enter a radial hyperbolic orbit is \begin{equation}
    \textrm{d}v = v_\textrm{SGL} \sin(\frac{\pi}{2} - \phi_\textrm{fpa}).
\end{equation}
Values of which are plotted in Figure \ref{fig:tsgl} (d).

Examining the plots in Figure (\ref{fig:tsgl}), a few important trade offs in the orbit parameter space become apparent. Orbits with large velocity at periastron correspondingly have the shortest time to reach great heliocentric distance, while the radius at periastron has a small effect, with a slight preference for starting further out of the solar system. However, these wide-starting orbits have much greater angular velocity at $r_\textrm{SGL}$, and correspondingly smaller planet-crossing and system-crossing timescales. The orbits with the smallest angular velocity (and thus largest crossing timescales, as well as smallest delta-v to enter the radial trajectory) correspond to orbits with small radius at periastron and large velocity at periastron.

To make a concrete example, suppose a craft is launched in a similar manner to the New Horizons space craft which was launched in 2006 to investigate Pluto \citep{Guo2008}.  Its trajectory departed Earth with a velocity estimate of 16.2 km/s, which was the highest recorded departure speed at the time, which corresponds to a heliocentric velocity of ~46 km/s. Yet New Horizons didn't fly by Pluto at ~33 AU until nearly a decade later \citep{pluto2015}. If a similar rocket would be used to launch a mission to the solar gravitational lens, our calculations indicate it would take $\sim$130 years to reach a distance of 600 AU, only to cross the image of the target planet in $\sim$40 seconds, or the entire system in just $\sim$1000 minutes. However, the craft could correct the lateral motion with a delta-v of only $\sim$80 m/s to enter a pure radial trajectory and extend its potential observation time further. Additionally, New Horizons used a Jupiter fly-by maneuver to obtain a velocity boost of $\sim$4 km/s, which can shorten the long wait but also causes the craft to have a wider heliocentric trajectory with a larger angular velocity, which would cost additional delta-v in the end.

\subsection{Numerical methods for planet perturbations}

Regardless of how elegant the analytic solutions are, they are only a two-body approximation to the dynamics of a craft exiting the solar system. The graviational dynamics of all of the bodies in the solar system are known to cause motion of the Sun around the barycenter \citep{barycenter2018}, due to the gravitational interactions of all of the massive bodies in orbit. The N-body problem has been known since Newton, and is essentially intractable except for special cases \citep{Musielak_2014} or statistical ensembles \citep{Stone2019}. However, numerical techniques can be quite effective to solve the equations of motion, and so we use REBOUND \citep{rebound} to investigate these effects. REBOUND has implemented a high-order adaptive time step integrator called IAS15 \citep{reboundias15} which enables machine-precision accuracy of complex N body orbits at extremely cheap computational cost, and can be used to simulate and compute the motion of the sun around the solar system barycenter extremely accurately into the future.

To evaluate the magnitude of this effect, we initialize a REBOUND simulation with $N_\textrm{particles} = 10$ particles, one for the Sun, each of the eight planets, and Pluto, using the built-in tool to load barycentric orbit elements from the JPL HORIZONS online ephemeris database \citep{jplhorizons}. Then REBOUND integrates the equations of motion for the N-body system of particles forward in time for one hundred years, using $N_\textrm{timesteps} = 10^4$ samples. This results in a sampling rate of two per week, enough to resolve effects from Mercury's short 88 day orbit. The results of the simulation is a data cube of size $(N_\textrm{particles}, N_\textrm{timesteps}, 6)$  which contains the six x-, y-, and z- positions and velocities of each of the particles in the integration at every time step. 

To evaluate the effect of the barycentric motion of the sun on the pointing of the solar gravitational lens, we investigate the worst-case scenario. Since the majority of the motion of the Sun occurs in the invariable plane \citep{invariable_plane}, defined by the point of the barycenter and perpendicular to the total angular momentum, a target planet which is pointed directly along this angular momentum vector will suffer from the largest perturbations of the position of the Sun along the line of sight. Any other target pointing will be somewhat better from this perspective as the motion of the sun is projected into the plane of observation. In our coordinate system, this plane is defined by the $x$ and $y$ coordinates, and so the telescope craft visiting the solar gravitational lens has a position of $(0,0,r_\textrm{SGL})$ once it has reached the observation location and maneuvered to enter a purely radial orbit trajectory. This next part of the analysis ignores the additional $z$ axis motion for simplicity. Although it is a small effect, this could be corrected by adding an additional term for a time-dependent $z$ coordinate, $z = r_\textrm{SGL} + (t/T_\textrm{SGL})\sqrt{\mu(2/z - 1/a)}$, which would require solving the vis-viva equation for $v(z(t))$.

\begin{figure}
    \centering
    \includegraphics[width=\textwidth]{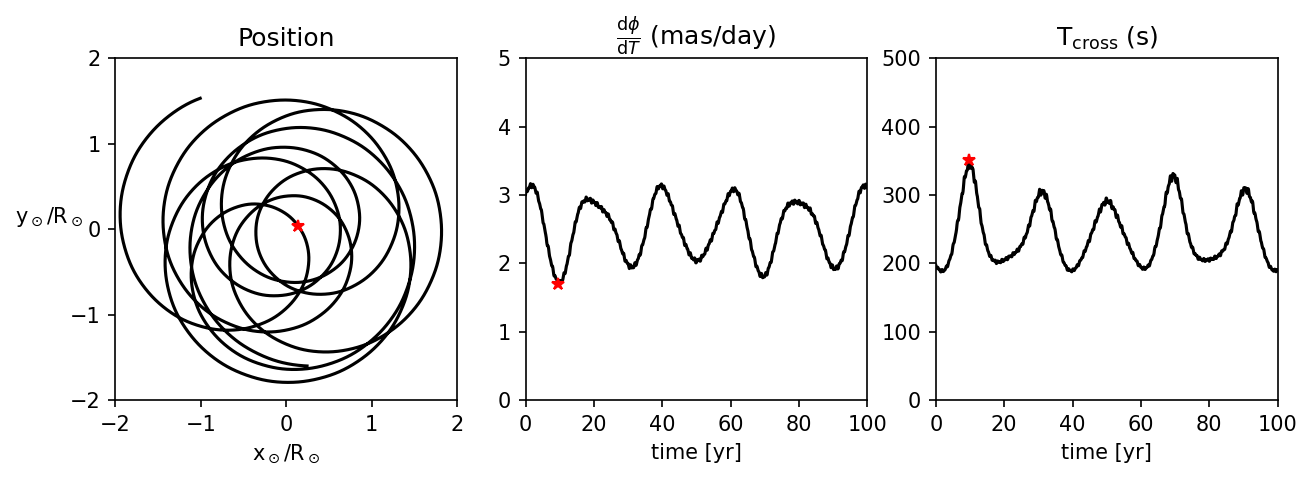}
    \caption{(a) $x-$ and $y-$ position of the Sun relative to the Solar System barycenter. (b) Angular velocity of the pointing vector for stationary craft at the SGL over time. (c) Planet-crossing timescale corresponding to the angular velocity of the pointing vector. The longest timescale in (c) is highlighted with a red star, as well as its corresponding locations in the companion plots.}
    \label{fig:barycenter}
\end{figure}

In Figure (\ref{fig:barycenter} a), the $(x,y)$ position of the Sun relative to the barycenter is plotted in units of solar radii, showing the reflex motion due to the perturbations from the planets causing a displacement of around $\sim$$1$-$2 R_\odot$. These values are used to estimate the angular velocity of the ``SGL pointing" vector defined as starting at the stationary craft at $(0,0,r_\textrm{SGL})$ and directed at the position of the Sun $(x_\odot, y_\odot, 0)$. The discretely-sampled position vector is the output of the REBOUND simulation. First, we compute the discrete $x$ and $y$ second-order position differences d$x = (x_{t+1} - x_{t-1})/2$ and d$y = (y_{t+1} - y_{t-1})/2$, using \texttt{np.gradient} which handles the edge cases using a first order finite difference, as well as the discrete time difference d$t = 100$ years$/10^4$ samples. The error on these estimates of the derivative is of order $\mathcal{O} (\textrm{d}t^2)$. Finally, the angular velocity of the pointing vector is computed with
\begin{equation}
    \frac{\textrm{d}\phi}{\textrm{d}t} = \frac{\sqrt{\textrm{d}x^2 + \textrm{d}y^2}}{r_\textrm{SGL}\textrm{d}t}
\end{equation}
which can be similarly converted into a planet-crossing timescale as before, using $T_{\textrm{cross}} = \theta_\textrm{planet}/\frac{\textrm{d}\phi}{\textrm{d}t}$. Both the angular velocity and planet-crossing timescale are plotted over time in Figure (\ref{fig:barycenter} b) and (\ref{fig:barycenter} c). From these estimates one can see the average angular velocity is around $2-3$ milliarcsecond per day, and that the corresponding planet-crossing timescales are between $200-300$ seconds. So to design a craft which obtains longer exposure times at ultra precise alignments, additional navigation acceleration will be needed to counteract this effect.

\begin{figure}
    \centering
    \includegraphics[width=\textwidth]{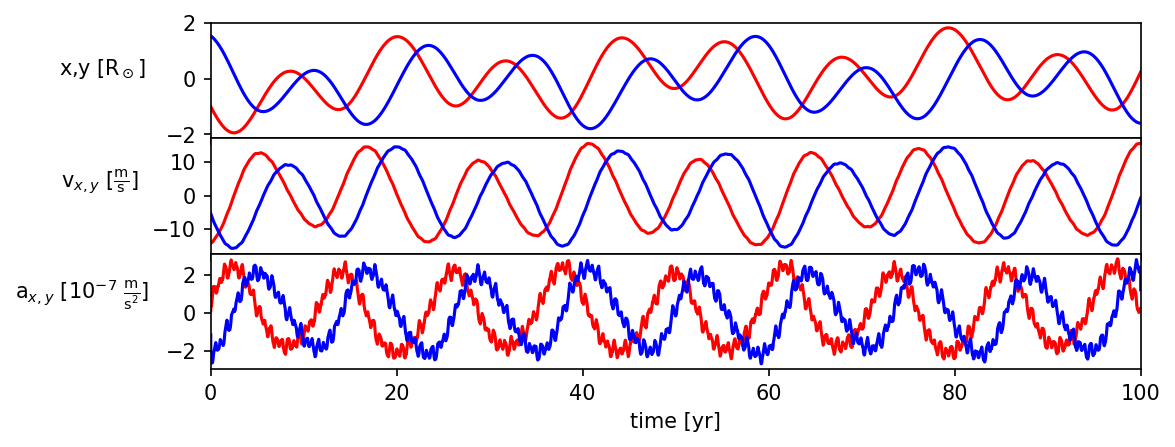}
    \caption{(a) Lateral position required for stationkeeping the solar gravitational lens for a fixed location on the sky. (b) Craft velocity required to maintain position computed as the discrete derivative of the position. (c) Corresponding acceleration.}
    \label{fig:derivatives}
\end{figure}

In order to compute the amount of acceleration necessary to counteract the barycentric motion of the sun, we first compute the location the craft must stationkeep at to keep the SGL aimed at a fixed point on the sky. If $d_s$ is the distance between the Sun and source or target planet, and $d_l = r_\textrm{SGL}$ is the distance between the craft and the sun or lens, a simple similar triangle argument shows the position $(x,y,z=d_l)$ required to stationkeep is enhanced by a small factor
\begin{eqnarray}
    x &=& x_\odot (1 + \frac{d_l}{d_s}) \\
    y &=& y_\odot (1 + \frac{d_l}{d_s})
\end{eqnarray}
relative to the position of the Sun due to the additional distance. This enhancement is a small number which depends on the distance to the target being considered. For d$_s$ = 40.54 ly and d$_l$ = 600 AU, $\frac{d_l}{d_s} \approx 2\times10^{-4}$. This position is shown as a function of time in Figure (\ref{fig:derivatives} a) in units of solar radii. Subsequently, the velocity $v_{x,y} = \frac{\textrm{d}x,y}{\textrm{d}t}$ and acceleration $a_{x,y} = \frac{\textrm{d}v_{x,y}}{\textrm{d}t}$ required to maintain this position are computed numerically using the discrete second order difference using \texttt{np.gradient}, and are plotted in Figures (\ref{fig:derivatives} b) and (\ref{fig:derivatives} c). 

Examining the figure, it is clear that a small, continuous acceleration of $\sim 2 \times 10^{-7}$ m/s$^2$ in an oscillatory fashion is necessary to counteract the lateral motion of the Sun around the barycenter to maintain a fixed pointing of the SGL. Considering the possibilities of rocket designs, it seems that ion thrusters may be ideally suited to this task. While the specifics depend on a number of design choices, such a beam divergence, mass of the ion, and the electrical current and voltage used to accelerate the ions \citep{goebel2008}, a reasonable estimate of thrust from such an ion thruster is around $\sim 100$ mN, which could give accelerations of $10^{-4}$ m/s$^2$ for a $1000$ kg craft over very long durations, due to the high specific impulse of the ions, which is more than two orders of magnitude needed.

\begin{figure}
    \centering
    \includegraphics[width=\textwidth]{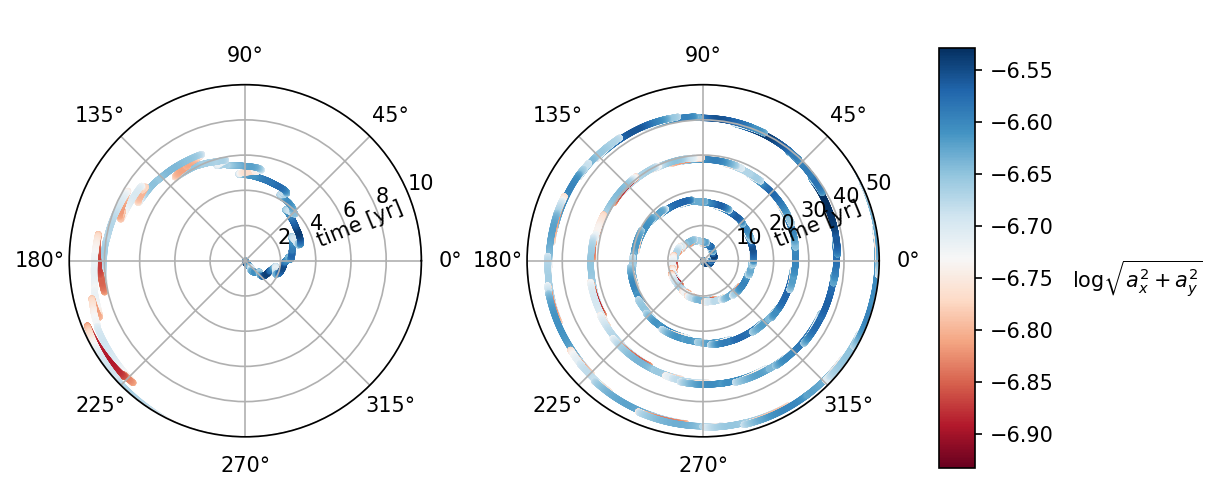}
    \caption{Visualization of the fixed pointing acceleration vector.}
    \label{fig:acceleration}
\end{figure}

An alternative visualization of this acceleration trajectory is visualized in Figure (\ref{fig:acceleration}). By converting to polar coordinates, the direction of the acceleration vector in the $xy$ plane is plotted as a function of time, which is now the radial coordinate. The log of the magnitude of the acceleration vector is represented by the color of the curve. One can see that acceleration curve spirals outward, as the vector rotates around with an approximate $10$ year period dominated by the effect of Jupiter. However, the perturbations of the other planets contribute as well, with consistent changes in the acceleration vector's direction and corresponding changes in magnitude.

To continue our analysis, we estimate the total delta-v needed to maintain this acceleration over time as the integral d$v_\textrm{total}(t) = \int_0^t |a_\textrm{total}(t')| \textrm{d}t'$ of the magnitude of the acceleration vector $|a_\textrm{total}(t)| = \sqrt{a_x(t)^2 + a_y(t)^2}$. This total delta-v is a linear function of the length of the time you would like to the maintain the alignment for, with a slope corresponding to the average value $\langle | a_\textrm{total} | \rangle = \frac{1}{t} \int_0^t | a_\textrm{total}(t') | \textrm{d}t'$. As a result, the approximate cost scaling to maintain alignment is
\begin{equation}
    \textrm{d}v_\textrm{total}(t) = \langle | a_\textrm{total} | \rangle t \approx 6.7 \frac{\textrm{m}}{\textrm{s} \cdot \textrm{year}} t,
\end{equation}
which demonstrates that the stationkeeping cost to counteract the solar barycentric motion is sufficiently small as to be reasonably possible. 

However, this entire analysis has so far ignored the motion of the target across the sky, composed of the relative proper motion of the system and the orbital motion of the target around its host star. A simple estimate of the acceleration that would be needed to counteract the orbital motion of a hypothetical planet with a face-on, circular orbit of semi-major axis $a$ and period $T$ is

\begin{equation}
    a_p = \frac{d_l}{d_s}\frac{4\pi^2 a}{T^2},
\end{equation}

which can dominate over or be comparable to the effect of the solar barycentric motion depending on the assumed planet properties. If $a = 1$ AU, $T = 1$ yr, $d_l = 600$ AU, and $d_s = 10$ pc, then $a_p \sim 6 \times 10^{-6}$ m/s$^2$ $\sim 60$ m/s/year, but if instead $d_s = 100$ pc, then $a_p \sim 0.2 \times 10^{-6}$ m/s$^2$ $\sim 6$ m/s/year. A more complete three dimensional analysis such as \citep{TT2021b} is necessary to evaluate the combined effect, as the orientation of the orbits is important to determine if this acceleration exacerbates or counteracts the solar barycentric motion. Ultimately, precise determination of the target ephemeris will be necessary to locate the exact position needed to send the craft, and predicting this location in advance may be rather difficult.

\subsection{Constraints on Target Ephemerides}
Different exoplanet detection techniques are capable of constraining different planetary orbital elements. Transit Photometry is able to determine the planet's period and to some extent its orbital inclination, as it must be nearly edge-on in order for a transit to occur \citep{Deeg2018}. Doppler spectroscopy or the radial velocity method measures the period and velocity semiamplitude, which can be used infer the orbit's eccentricity, mass-inclination product $m \sin i$ and, if the star's mass is known through another technique, the orbital semi-major axis \citep{Eggenberger2010}. If the planet is amenable to both transit and radial velocity, the combination of both techniques can further constrain the orbit. The Rossiter-Mclaughlin effect can be used to measure the relative azimuthal angle on the sky between the planet's orbit and the stellar spin axis \citep{Ohta_2005} \citep{Masuda2018}, using asymmetric distortions in the line profiles of the stellar spectrum as the planet crosses and sequentially obscures different subsets of the stellar disk. However, absolute determination of the planet's orbital position angle on the sky would still require determination of the orientation of the stellar spin axis, which may be possible using spectro-interferometry to measure the astrometric motion associated with stellar rotation \citep{Kraus_2020}, with  spectro-astrometric analysis of high resolution long-slit spectra \citep{Lesage2014}, or by combining projected spectroscopic rotation rates with photometric rotation periods \citep{Bryan_2020}. It may also be possible to make these measurements during the early part of the mission to the SGL, by visiting the image of the star first to aid in navigation before moving to the image of the target planet. \cite{TT2021b} discusses a strategy which uses astrometric observations of the host star using the SGL to help constrain the planet's position.

However, direct imaging of exoplanets can  constrain planetary orbital elements by measuring directly measuring the planet's astrometric position on the sky. Typical astrometric accuracies obtainable with current instruments on single telescopes are between $3$ and $22$ mas \citep{Konopacky_2016}, while the most precise measurements currently obtained could be around $\sim100$ $\mu$as \citep{gravity2019} using multiple telescopes combined into an interferometer. These uncertainties in measured precision are typically relative measurements, comparing the planet's position to the host star, and an absolute astrometric calibration would require precise determination of the stellar absolute astrometry which could be measured using GAIA \citep{gaia2021, gaia2018}. The uncertainty in angular position $\delta \theta$ on the sky directly translates into a spatial, lateral uncertainty $\delta x = \textrm{d}_l \delta \theta$ of the image of the target in the SGL image plane. The large enhancement due to the great heliocentric distance means that accurate determination of the angular position on the sky of the target is critical to locating the position of the image of the target in the SGL focal plane. An angular uncertainty of $\delta \theta \sim 20$ mas corresponds to an uncertainty in location of $\delta x \sim 8800$ km, while $\delta \theta \sim 100$ $\mu$as corresponds to $\delta x \sim 44$ km. Since the image of the target planet is rescaled from its original size $R_\textrm{image} = \frac{\textrm{d}_l}{\textrm{d}_s} R_\textrm{Earth}$ by the ``plate scale" of the SGL geometry, an Earth-sized planet at a distance of $40$ ly has an image only $\sim3$ km in diameter. With current levels of uncertainty in relative astrometry it could be missed even with navigation relative to the image of host star. A further caveat is those measurements are typically for Jupiter-sized planets, since no Earth-like planets have been directly imaged, although mission concepts like HabEx and LUVOIR have been proposed to potentially achieve this \citep{habex} \citep{luvoir}.

Additionally, the planet and star are continually in motion, due to the relative orbital motion of the system, and the proper motion and acceleration of the entire system within the galaxy. Typically stars have galactic velocities around $\sim 100$ km/s and accelerations around $\sim 10^{-10}$ m/s$^2$ \citep{silverwood_easther_2019}, which means determining the location needed to visit the image of either the star or planet would need to be predicted in advance, to plan an appropriate trajectory. Given that realistic timescales between launch and arrival at the SGL are order $\sim 100$ years, and that typical proper motions are of order $\sim .1$ arcsec/year, the location of the image of a target star now and later would be displaced by millions of kilometers, with uncertainties in current proper motion further worsening the issue. Additionally, predicting the future orbital motion of the relative motion of the planet and star is subject to additional uncertainty. 

\begin{figure}
    \centering
    \includegraphics[width=\textwidth]{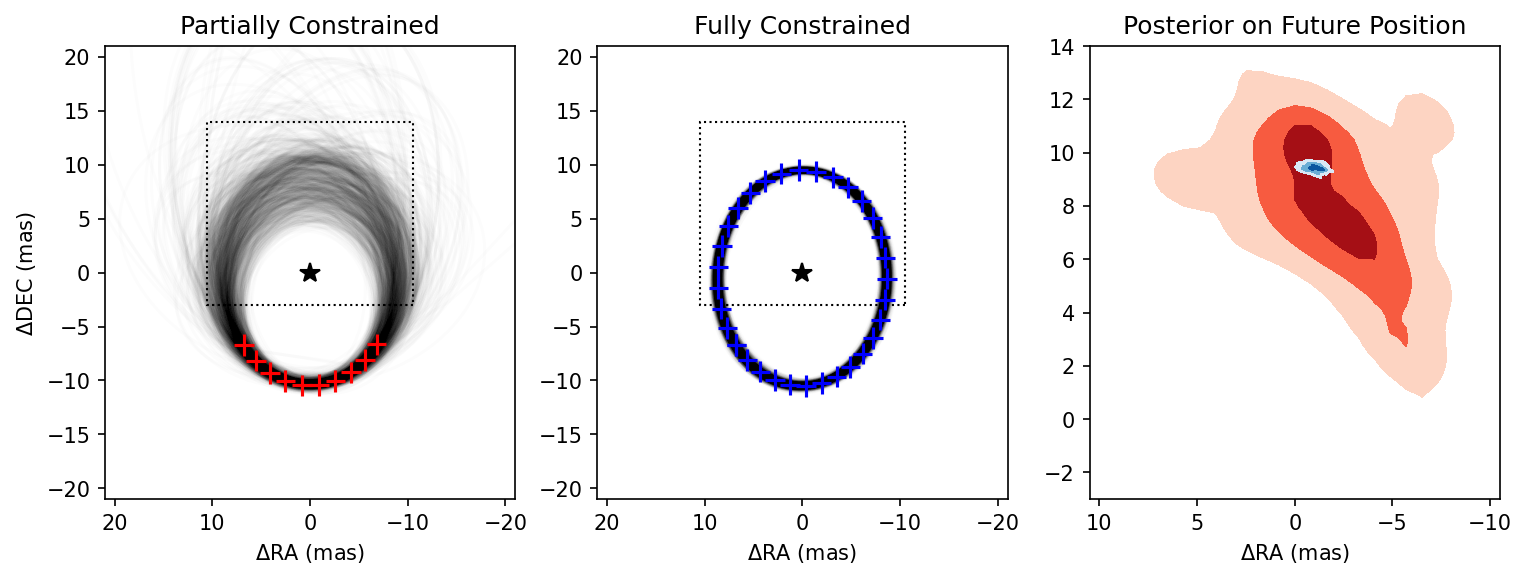}
    \caption{Predicting the future location (c) of a hypothetical target planet with partially (a) and fully constrained (b) orbits using simulated astrometric measurements and orbitize! \citep{Blunt_2020}. The first two plots show $10^3$ random samples from the orbit posterior distribution and the dotted box corresponds to the edges of the plot for the predicted future location posterior. Contours in (c) correspond to the 20$^\textrm{th}$, 50$^\textrm{th}$, and 80$^\textrm{th}$ percentiles.}
    \label{fig:orbit}
\end{figure}

Software such as $\texttt{orbitize!}$ \citep{Blunt_2020} can be used to fit astrometric observations with three dimensional Keplerian orbits, allowing estimates of position into the future, although the accuracy of prediction depends strongly on the orbital coverage of the data and its astrometric accuracy. To demonstrate this, Figure (\ref{fig:orbit}) was created. First, a series of simulated astrometric data are generated which correspond to a ground truth Keplerian orbit with a semi-major axis of 1 AU, an eccentricity of 0.05, an inclination of 30$^\circ$, for a system $100$ pc away. Either 10 or 33 samples were generated over 0.3 or 1.0 years with an observational uncertainty around $\sim 1$ mas which correspond to an orbital coverage of $33\%$ and $100\%$ for the partially and fully constrained cases. 

The simulated data points are fit with the affine-invariant Markov Chain Monte Carlo (MCMC) algorithm developed by \citep{ForemanMackey2013} which samples the posterior distributions of all orbital parameters. After a burn-in time of $10^4$ steps and a total number of $10^6$ samples, the posterior distributions have mostly converged, although with less precision on the true orbital elements for the partially constrained orbit. For the partially constrained orbit, attempting to predict the location at a time corresponding to unconstrained and unmeasured locations in the orbit causes the posterior distribution of predicted locations to become large. Even with observational uncertainties of order $\sim$ $1$ mas, the future location posterior has uncertainty an order of magnitude larger around $\sim$ $10$ mas. However, a fully constrained orbit predicts the future location with accuracy comparable to the observational uncertainty on the data points themselves, which is a necessary precursor for navigation to the precise location of the planet's image at the SGL. This demonstrates the importance of complete astrometric orbital coverage on constraining the planetary orbital elements and predicting its future location. 

\section{Optical Considerations}

\subsection{Gravitational field oblateness}

The simplest model for the gravitational potential of the Sun is that of a point mass, $\Phi = \frac{-GM}{r}$. This model has a very simple formula \citep{Narayan1996}
\begin{equation}
    \boldsymbol{\theta_g} = \frac{2R_S}{b}\hat{e}_b
\end{equation}
for the deflection of a photon as a function of impact parameter $b$. Here $R_S = \frac{2GM_\odot}{c^2}$ is the Sun's Schwarzschild radius and $\hat{e}_b$ is the unit vector in the direction of the impact parameter. Since a photon can travel from the source to the observer around the point mass and receive the same deflection for a path at any azimuthal angle, it is easy to see how gravitational lensing around a point source forms an Einstein Ring symmetrically around the point mass.

However, sophisticated helioseismological analyses \citep{Mecheri2009} require the gravitational field to deviate from spherical symmetry. The rotation of the Sun creates an oblate gravitational potential which is described by a zonal spherical harmonic expansion
\begin{equation}
    \Phi = \frac{-G M}{r} \Big[ 1 + \sum_{n=2}^{\infty} I_n \Big( \frac{R_\odot}{r}\Big)^n P_n(\cos\beta_S) \Big],
\end{equation}
where $R_\odot$ is the radius of the Sun, the $P_n$ are the Legendre Polynomials \citep{abramowitz1968}, $\beta_S$ is the solar co-latitude, and $I_n$ are the dimensionless multipole moments. Due to the North-South symmetry along the rotation axis, all terms with odd $n$ are zero. The largest even $n$ terms are $I_2 \approx 2 \times 10^{-7}$, $I_4  \approx -4 \times 10^{-9}$, $I_6 \approx -3 \times 10^{-10}$, $I_8 \approx 1 \times 10^{-11}$ \citep{Roxbrugh2001}.

The effect of this deviation from spherical symmetry on the resulting optical properties of the solar gravitational lens was first investigated by Loutsenko \citep{Loutsenko2018}. They demonstrated that the quadrupole moment $I_2$ results in the formation of an astroid-shaped caustic in the image plane with a diameter
\begin{equation}\label{eq:astroid}
    D_\textrm{astroid} = \frac{4 I_2 R_\odot^2}{\sqrt{2 R_S r}} \sin^2 \beta_S.
\end{equation}
 Turyshev and Toth \citep{TT2021a} extended their work and computed the effect of any high order multipole $I_n$ on the resulting optical properties of the gravitational lens. In general, the caustic curves are given by higher order hypocycloids \citep{TT2021c}, and the general case of photon deflection can be computed with 
\begin{equation}
    \boldsymbol{\theta}_g = \frac{2 R_S}{b} \Big[ \hat{e}_b - \sum_{n=2}^{\infty}I_n \Big(\frac{R_\odot}{b}\Big)^n \sin^n(\beta_S) \Big( \hat{e}_b \cos[n(\phi_\xi - \phi_S)] - \hat{e}_{\phi_{\xi}} \sin[n(\phi_\xi - \phi_S)]\Big)\Big],
\end{equation}
where $\phi_\xi$ and $\phi_S$ are the azimuthal orientations of the impact parameter and the solar rotation axis. When $I_n = 0$, this formula reduces to the well known example of a point mass, but includes new terms for the deflection along the $\hat{e}_{\phi_\xi}$ direction, which is perpendicular to the impact parameter, see Figure (2) in \citep{TT2021a}. This additional deflection gives the solar gravitational lens entirely new behavior which is not symmetric with respect to azimuthal rotation, and explains geometrically the formation of the Einstein Cross for $n = 2$.

Furthermore, they go on to solve Maxwell's equations by treating the new gravitational potential as a perturbation of the Newtonian metric, and using the angular eikonal method which takes inspiration from Mie scattering theory. In this formalism, the magnification of the lens $\mu$, defined as the ratio of the Poynting vectors for a plane wave at the source and resulting image, is \citep{TT2021d}
\begin{equation}
    \mu (\boldsymbol{x}) = 2 \pi k R_S |E(\boldsymbol{x})|^2,
\end{equation}
where $k = \frac{2\pi}{\lambda}$ is the wavevector and the quantity $E(\boldsymbol{x})$ is given by the diffraction integral \citep{TT2021d}
\begin{equation}\label{eq:E}
    E(\boldsymbol{x}) = \frac{1}{2 \pi} \int_0^{2\pi} \textrm{d}\phi_{\xi} \textrm{exp}\Big[ -i k \Big( \sqrt{\frac{2 R_S}{r}} \rho \cos(\phi_\xi - \phi) + 2 R_S \sum_{n=2}^{\infty} \frac{I_n}{n} \Big(\frac{R_\odot}{\sqrt{2 R_S r}}\Big)^n \sin^n(\beta_S) \cos[n(\phi_\xi - \phi_S)] \Big)\Big],
\end{equation}
where $\boldsymbol{x} = (\rho,\phi)$ are the polar coordinates in the image plane. If all of the multipole moments $I_n = 0$, this diffraction integral can be analytically computed and the result is that of the monopole term only, $E(\boldsymbol{x}) = J_0(k\sqrt{2R_S/r}\rho)$, where $J_0$ is the Bessel Function of the First Kind \citep{abramowitz1968}. However, it is not analytically tractable for non-zero $I_n$, as would be the case for a realistic Sun, and so numerical methods must be relied upon to evaluate Equation (\ref{eq:E}).

In this paper, we investigate Equation (\ref{eq:E}) with a discrete approximation to the integral over the azimuthal angles $\phi_\xi$. By replacing $\int_0^{2\pi}\textrm{d}\phi_\xi\textrm{exp}[...]$ with a discrete sum indexed by $j$
\begin{equation}\label{eq:approx}
    E(\boldsymbol{x}) \approx \frac{1}{2 \pi} \sum_{j=1}^{N_p} \Delta_{\phi_{\xi}} \textrm{exp}\Big[ -i k \Big( \sqrt{\frac{2 R_S}{r}} \rho \cos(\phi_{\xi,j} - \phi) + 2 R_S \sum_{n=2}^{\infty} \frac{I_n}{n} \Big(\frac{R_\odot}{\sqrt{2 R_S r}}\Big)^n \sin^n(\beta_S) \cos[n(\phi_{\xi,j} - \phi_S)] \Big)\Big],
\end{equation}
it is possible to evaluate the integral numerically. Here $\phi_{\xi,j} \in [0, 2\pi]$ is a discrete approximation to the continuous $\phi_\xi$ and $\Delta_{\phi_{\xi}} = \frac{2\pi}{N_p}$, where $N_p$ is the number of points to use for the azimuthal sampling. $N_p$ must be chosen carefully to be sufficiently large so that the approximation is valid. For ``easy" cases where $k$, $\beta_S$, and $\rho$ are relatively small, such as $\lambda = 4$ $\mu$m, $\beta_S = 10^\circ$, and $\rho = 0$ m, we find that $N_p \gtrsim 10^4$ is sufficient. ``Hard" cases where $k$, $\beta_S$, and $\rho$ are relatively large such as $\lambda = 0.5$ $\mu$m, $\beta_S = 90^\circ$, and $\rho = 10^3$ m can require significantly finer azimuthal sampling, up to $N_p \gtrsim 10^5$ to ensure the discrete approximation to the integral doesn't break down.

\subsection{The point-spread function and the effects of telescope diameter, wavelength, and solar co-latitude}

\begin{figure}
    \centering
    \includegraphics[width=\textwidth]{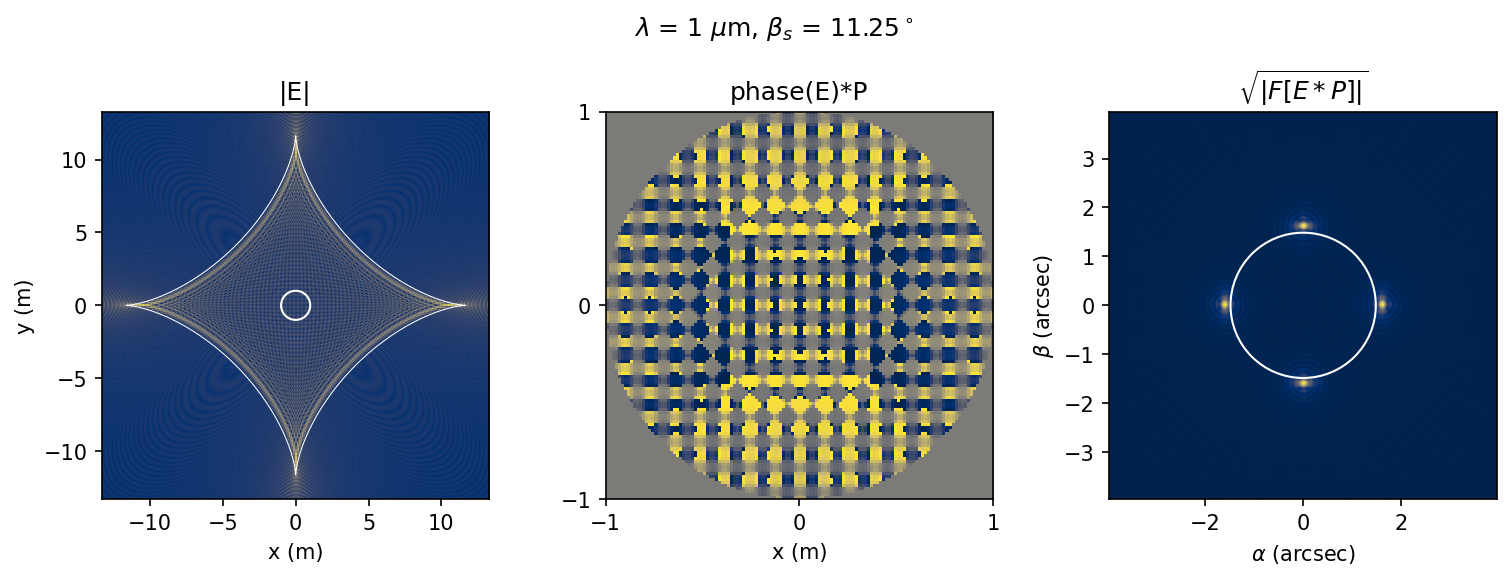}
    \includegraphics[width=\textwidth]{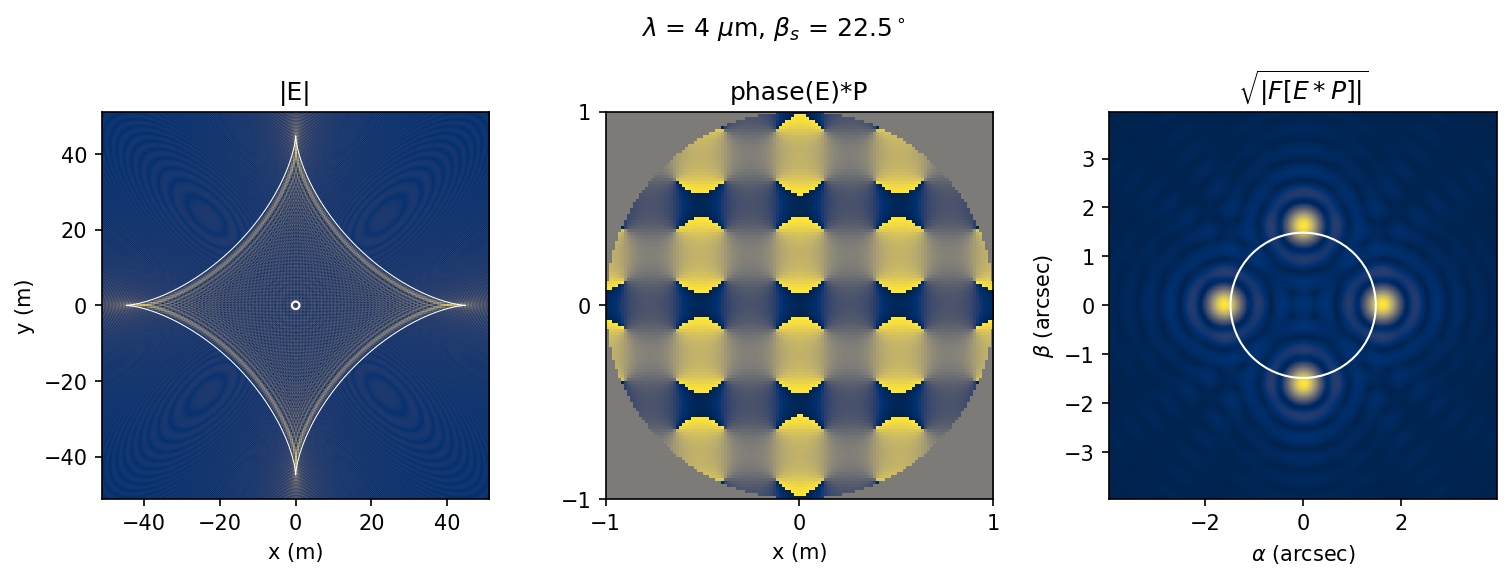}
    \includegraphics[width=\textwidth]{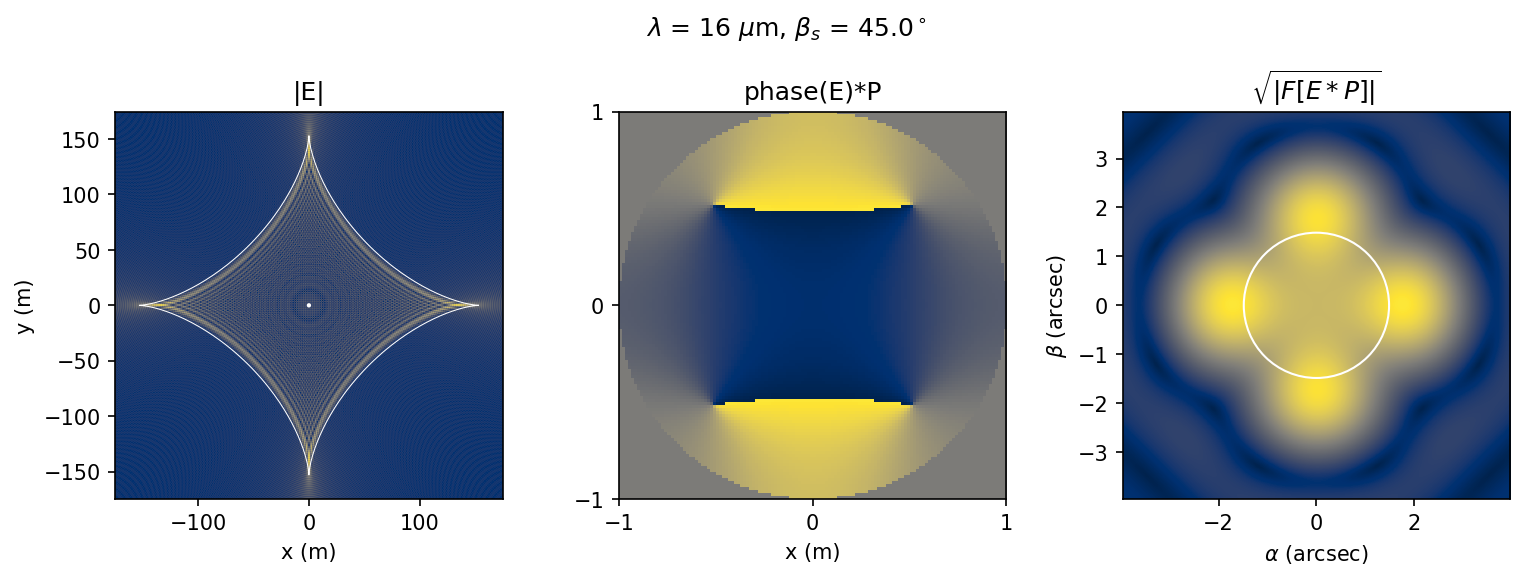}
    \caption{Three simulated cases showing from left to right, the magnitude of the electromagnetic field $|E|$ computed numerically with the approximation in Equation (\ref{eq:approx}), the phase of the electromagnetic field inside the pupil of the telescope, centered on the optical axis, and the resulting point spread function in the image plane of the telescope. In the left figures, the telescope pupil is plotted with a white circle and the geometric astroid curve governed by Equation (\ref{eq:astroid}) is plotted with a $5\%$ enhancement so as to not cover up the intensity image. In the rightmost images the angular size of the Solar disk is shown with a white circle.}
    \label{fig:SGL}
\end{figure}

The approximation in Equation (\ref{eq:approx}), depends on many parameters, the location in the image plane $\boldsymbol{x} = (\rho,\phi)$, the wavelength $k = 2\pi/\lambda $ of light, the multiple moments $I_n$, the mass $M_\odot$ and radius $R_\odot$ of the Sun, and the orientation of the solar rotation axis $(\beta_S, \phi_S)$. Furthermore,  once one has computed the result of Equation (\ref{eq:approx}), which gives the complex electromagnetic field in the SGL image plane rescaled by the magnification prefactors, it is rather straightforward to find the point spread function that would result in the image plane of a telescope using the Fraunhofer approximation and the techniques of Fourier Optics \citep{Hecht2002} \citep{goodman1996}. If $P(\boldsymbol{x}) = \rho < D_\textrm{tel}/2$ is the binary pupil transmission function of a circular aperture telescope, then the point-spread function in its image plane is
\begin{equation}\label{eq:PSF}
    \textrm{PSF}(\alpha,\beta) = \big| \mathcal{F}^{-1}(\alpha,\beta)[P(\boldsymbol{x})E(\boldsymbol{x})]\big|^2,
\end{equation}
where $(\alpha,\beta)$ are the angular coordinates in the conjugate plane, and $\mathcal{F}^{-1}[...]$ is the inverse Fourier Transform. The inverse Fourier Transform, which has a positive sign in the exponent $e^{+ik...}$ of the exponential basis, is used instead of the Forward Transform to remain consistent with observed orientation of asymmetric point-spread functions in high contrast imaging from the ground, see Section 2.4 in \citep{madurowicz2019}. To showcase the effect of these parameters on the resulting images formed, Figure (\ref{fig:SGL}) was created. The figure contains three separate cases, with different wavelengths $\lambda = (1, 4, 16) \mu$m and solar co-latitude $\beta_S = (11.25, 22.5, 45)$ degrees, with each case labeled in its title. Each case contains three plots, the magnitude of $|E(\boldsymbol{x})|$, the phase inside the telescope pupil, and the resulting point spread function in the telescope image plane.

A few key observations from this Figure. First, the image formation in the SGL image plane is dominated by the quadrupole moment $I_2$, and the resulting $|E|$ is shaped primarily like the four-pointed hypocycloid, the astroid. The higher order multipole moments have a small effect, but are subdominant for realistic solar values. Second, for larger $\beta_S$ the size of the astroid caustic grows significantly. For the first case with $\beta_S = 11.25^\circ$ the astroid diameter is $\sim 20$ m, but for $\beta_S = 45^\circ$ it is $\sim 200$ m. At the solar equator $\beta_S = 90^\circ$ the astroid diameter reaches its maximum value $\sim 560$ m. This has the implication that any given target planet, which has a unique location on the sky, will suffer from degradation to the average magnification over the pupil for a set telescope diameter, since the positioning and size of the telescope control the fraction of intercepted light. However, this only is true when the telescope diameter is much smaller than the astroid diameter. If the telescope is sufficiently large to capture the entire astroid, then all of the flux can be intercepted, but that would require telescopes much larger than is currently feasible.

\begin{figure}
    \centering
    \includegraphics{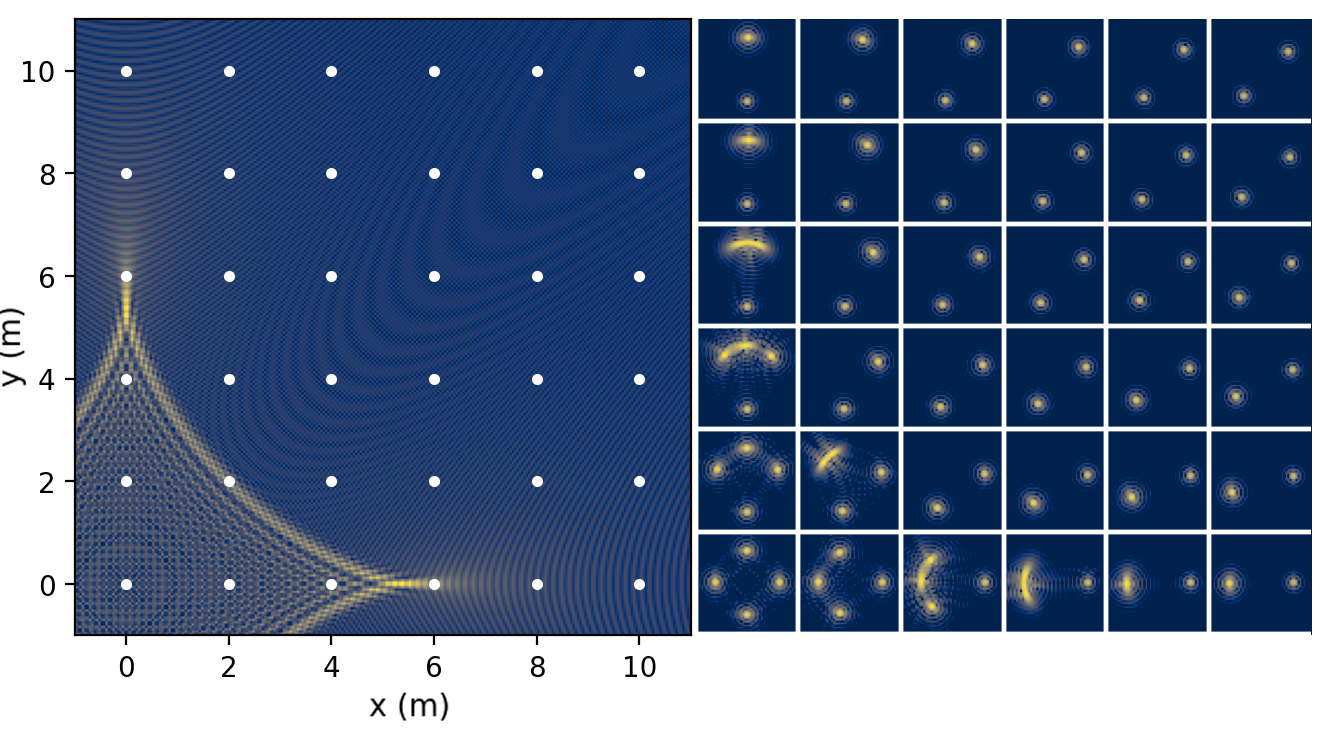}
    \caption{Left: Zoom in on the SGL PSF in the upper right quadrant for a simulation with $\beta_S = 8.13^\circ$  and $\lambda = 1 \mu$m. White points correspond to telescope locations for the right figure. Right: Grid of point-spread functions corresponding to telescope locations on the left Figure. The bottom left is directly aligned with the optical axis and the Einstein Cross is symmetric in the image plane. As the position of the telescope shifts, the four images move around the ring and can merge into arcs. Outside the caustic, the images are formed similar to the monopole only case, with only two points appearing, however, their intensities are asymmetric.}
    \label{fig:psfs}
\end{figure}

Third, the phase of the electromagnetic field appears to have a grid-like pattern which is responsible for the formation of the Einstein Cross. The size scaling of this grid pattern is controlled mostly by the wavelength of the light, with shorter wavelengths having a narrower spacing between the zones, and longer wavelengths having larger zones. It is crucially important that the sampling of the phase inside the pupil is fine enough that the delicate grid pattern is not aliased, and that the pupil area is large enough so that at least one ``grid element" is fully contained inside the pupil. If these conditions are not both met, the resulting point-spread function does not produce the characteristic Einstein Cross, seen on the right. Fourth, the angular size of the four points in the telescope's point-spread function grow linearly with wavelength, as is expected. The positioning of the four point images is just outside the angular extent of the solar disk, shown at a heliocentric distance of $r_\textrm{SGL} = 650$ AU. When the telescope is precisely aligned with the optical axis of the point source and the gravitational lens, the locations of the four point sources is symmetric, however, a lateral misalignment of the telescope intercepts an offset phase pattern, which will cause the locations of the points to be perturbed around the edge of the ring, see Figure (\ref{fig:psfs}). When the telescope is positioned directly on the caustic edge, the images have moved sufficiently around the edge to merge and form into an arc. When the telescope has moved outside the astroid caustic, the result is two point sources similar to the monopole only case. However, the effect of the multipoles is still present and the intensity of each point is asymmetric.

\subsection{The average and asymptotic magnification}

\begin{figure}
    \centering
    \includegraphics[width=\textwidth]{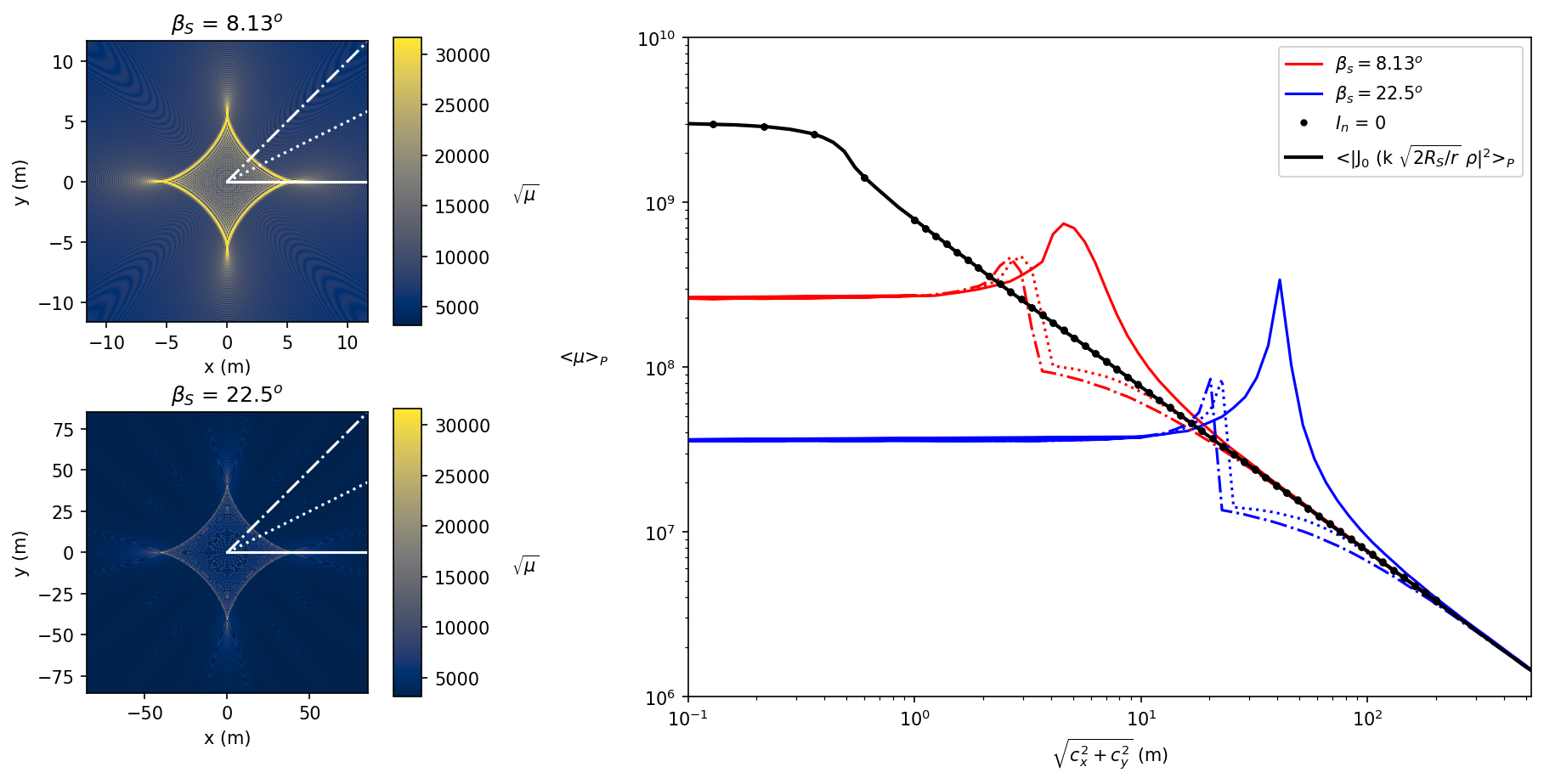}
    \caption{Left: $\mu(\boldsymbol{x})$ in the SGL image plane for two different values of $\beta_S$ visualized on the same color scale, proportional to the electromagnetic field amplitude or $\sqrt{\mu}$. The shape of the SGL PSF around the astroid caustic is apparent, and one can see the effect of larger $\beta_S$, widening the PSF and diminishing the magnification. Additionally, solid/dot/dash-dot radial tracks at different angles correspond to equivalent line styles in the right figure. Right: The average magnification across the pupil versus the radial distance between the optical axis and the center of the telescope. In addition to the multipole-inclusive cases, the average magnification for a pure monopole SGL, with $I_n = 0$ for all n is plotted. The black dots correspond to evaluation of Equation (\ref{eq:approx}) while the solid black line instead evaluates the analytic Bessel function. The equivalence of these two curves provides a sanity check on the numerical integrator. While the monopole case reaches its maximum $\sim 2 \times 10^9$ on the optical axis, the maximum for the multipole cases is somewhat less and located near the caustic. Realistic values for the average magnification across the pupil are found to be in the range of $10^7$ to $10^9$ considering the effect of the multipoles.}
    \label{fig:mu}
\end{figure}

To further investigate the effect of the telescope diameter on the intercepted flux, we compute the average magnification across the telescope pupil $\langle \mu \rangle_P$. Equation (\ref{eq:approx}) is computed numerically for points $(\rho,\phi)$ in the image plane, and the average across the pupil is defined as the weighted sum of $\mu$ in the region inside the telescope pupil $P \in \{(x,y) | \sqrt{(x-c_x)^2 + (y-c_y)^2} \leq D_\textrm{tel}/2\}$.
\begin{equation}
    \langle \mu (\boldsymbol{x}) \rangle_\textrm{P} = \frac{ \iint\limits_\textrm{P} \mu(\boldsymbol{x}) \textrm{d}\boldsymbol{x} }{ \iint\limits_\textrm{P} \textrm{d}\boldsymbol{x} }
\end{equation}
Here $(c_x,c_y)$ is the location of the center of the telescope in Cartesian coordinates, which has diameter $D_\textrm{tel} = 2$ m. This metric is used to compare the multiple scenarios plotted in Figure (\ref{fig:mu}).

Examining the behavior of the magnification in Figure (\ref{fig:mu}), it is apparent that the effect of the higher order multipole moments causes significant alteration to the magnification in the region of the astroid caustic.  This regime is characterized by $\rho \lesssim D_\textrm{astroid}$. Here the maximum of the magnification no longer occurs on the optical axis, but instead in the region of the caustic, and is reduced from the maximum of the monopole by a few orders of magnitude, of which the exact amount depends on the solar co-latitude $\beta_S$ as well as the telescope diameter $D_\textrm{tel}$.

However, far outside the astroid caustic, the average magnification is consistent with the pure monopole case. In this section, we investigate the asymptotic behavior of these functions and demonstrate an analytic comparison between the wave and geometric theory which results in a power law approximation valid in this regime. Using the geometric theory, the magnification of a point mass lens can be described by \citep{Narayan1996}
\begin{equation}\label{eq:7}
    \mu_\textrm{geo} = \frac{1}{1-\Big( \frac{\theta_E}{\theta_\pm} \Big)^4},
\end{equation}
where $\theta_E$ is the angular scale of the Einstein Ring 
\begin{equation}
    \theta_E = \sqrt{2 R_S \frac{D_{LS}}{D_S D_L}},
\end{equation}
and the $\theta_\pm$ are the angular coordinates of the magnified images
\begin{equation}
    \theta_\pm = \frac{1}{2}\Big( \beta \pm \sqrt{\beta^2 + 4 \theta_E} \Big).
\end{equation}
Here $\beta$ is the angular coordinate of the point source and is geometrically related to the radial coordinate $\rho$ in the image plane by
\begin{equation}
    \beta = \arctan\frac{D_S}{\rho} - \arctan\frac{D_L}{\rho} \approx \rho \Big( \frac{1}{D_L} - \frac{1}{D_S}\Big).
\end{equation}
Taking a Taylor expansion of Equation (\ref{eq:7}) in the regime where $\beta \ll \theta_E$, results in
\begin{equation}
    \mu_\textrm{geo} \approx \frac{1}{2} + \frac{\theta_E}{2\beta},
\end{equation}
so evidently the dominant scaling of the geometric magnification is $\mu_\textrm{geo} \propto \frac{1}{\beta} \propto \frac{1}{\rho}$. Taking a Taylor expansion in the regime where $\beta \gg \theta_E$ results in
\begin{equation}
    \mu_\textrm{geo} \approx 1 + (\frac{\theta_E}{\beta}\Big)^4,
\end{equation}
and the dominant term is the constant, which is succinctly summarized in the limiting behavior
$\lim_{\beta \to \infty} \mu_\textrm{geo} = 1$. This limit indicates that far enough away from the lens, the point source simply exists as it would be unperturbed by anything. To compare and contrast this geometric scaling argument with the wave-theoretic treatment described previously, we take an asymptotic expansion of the magnification for a pure monopole SGL. Starting with the expression for the magnification
\begin{equation}
    \mu_\textrm{wave} = 2\pi k R_S \Big|J_0 \Big(k \sqrt{\frac{2 R_S}{r}} \rho\Big)\Big|^2,
\end{equation}
and since the Bessel Functions of the First Kind have asymptotic expansions \citep{abramowitz1968}
\begin{equation}
    J_n(z) = \sqrt{\frac{2\pi}{z}}\Big[\cos \Big(z - \frac{n\pi}{2} - \frac{\pi}{2}\Big) + e^{|\textrm{Im}(z)|}O\Big(\frac{1}{|z|}\Big)\Big],
\end{equation}
we can conclude for $\rho \gtrsim D_\textrm{astroid}$, that
\begin{equation}
    \mu_\textrm{wave} \approx 4 \pi^2 \sqrt{\frac{r R_S}{2}} \rho^{-1} \cos^2\Big( k\sqrt{\frac{2 R_S}{r}}\rho - \frac{\pi}{2}\Big) \propto \frac{1}{\rho},
\end{equation}
which shares the same power law scaling with the geometric theory. Incidentally, the limiting behavior of this function is discrepant as $\lim_{\rho \to \infty} \mu_\textrm{wave} = 0$. This likely is a result of the formalism being used to describe the electromagnetic field in the geometric shadow of the Sun, where the direct path component of the field is obscured by the solar disk. So if $\rho > \rho_\textrm{critical}$, such that $\beta > \theta_E$, then this description is no longer valid. Indeed, in this case it would be simpler to imagine that the lens does not exist at all, and to treat the direct path on its own. To summarize, the monopole approximation is valid and consistent with the geometric power law scaling $\propto \rho^{-1}$ in the regime where $D_\textrm{astroid} \lesssim \rho < \rho_\textrm{critical}$. Inside the astroid one must consider the perturbations of the higher order multipoles and outside the critical distance one may ignore the lensing entirely. This approximation is useful to quickly evaluate the response of a realistic solar gravitational lens at large lateral separations, since numerical evaluation of Equation (\ref{eq:approx}) becomes increasingly challenging at large $\rho.$

\subsection{Imaging of Extended Sources}

With formalism described in Section 3.1 and 3.2, it is possible to calculate the point-spread function of the SGL as well as the point-spread function as it would appear in the image plane of a telescope in the focal plane of the SGL. In this section, we extend our analysis to the imaging of extended sources under the assumptions of incoherent emission and small angular extent of the source. The former assumption allows one to treat the extended source as a superposition of independent point sources, and the latter allows one to ignore minor perturbations to the orientation of the solar rotation axis $(\beta_S,\phi_S)$ that would begin to matter if the angular extent of the extended source was sufficiently large. For justification, an Earth-sized planet at a target distance of $d_S = 100$ pc occupies a solid angle of $\Omega_E = \pi(\frac{R_E}{d_S})^2 \approx 1.4 \times 10^{-23}$ steradians.

If the Cartesian coordinates in the source plane are $(x',y')$, then we can approximate the continuous intensity distribution of the target planet $I(x',y')$ with
\begin{equation}
    I_{n_x,n_y} \approx \langle \Sh_{n_x}(x') \Sh_{n_y}(y') I(x',y')\rangle,
\end{equation}
where $\Sh_{n_x}(x') = \delta(x' - \Delta_x(n_x - N_x/2))$ is a comb function, $\delta$ is the Dirac delta function, $n_x \in [0,1,2,...,N_x]$ is the integer index of the pixel, $\Delta_x$ is the spatial sampling, and the $N_x/2$ is a centering offset, so the origin of $x'=0$ occurs at the halfway index $n_x = N_x/2$. Additionally, $\langle ... \rangle$ represents the average of the intensity distribution in the vicinity of the sampling point, and at the end, $I_{n_x,n_y}$ is a vector (or matrix), indexed by pixel numbers $(n_x,n_y)$ which represents a discrete approximation to the continuous intensity distribution of the source. In our analysis, we use uniform sampling for both $x$ and $y$ and set $\Delta_x = 2 R_\textrm{Earth} / N_x$, $N_x = 64$ pixels. As a consequence,

\begin{eqnarray}
    x'_{n_x} = R_\textrm{Earth} \Big( 2 \frac{n_x}{N_x} - 1 \Big) \\
    y'_{n_y} = R_\textrm{Earth} \Big( 2 \frac{n_y}{N_y} - 1 \Big),
\end{eqnarray}
and both $x',y' \in [-R_\textrm{Earth},R_\textrm{Earth}],$ since $\frac{n_{x,y}}{N_{x,y}} \in [0,1]$.

\begin{figure}
    \centering
    \includegraphics[width=\textwidth]{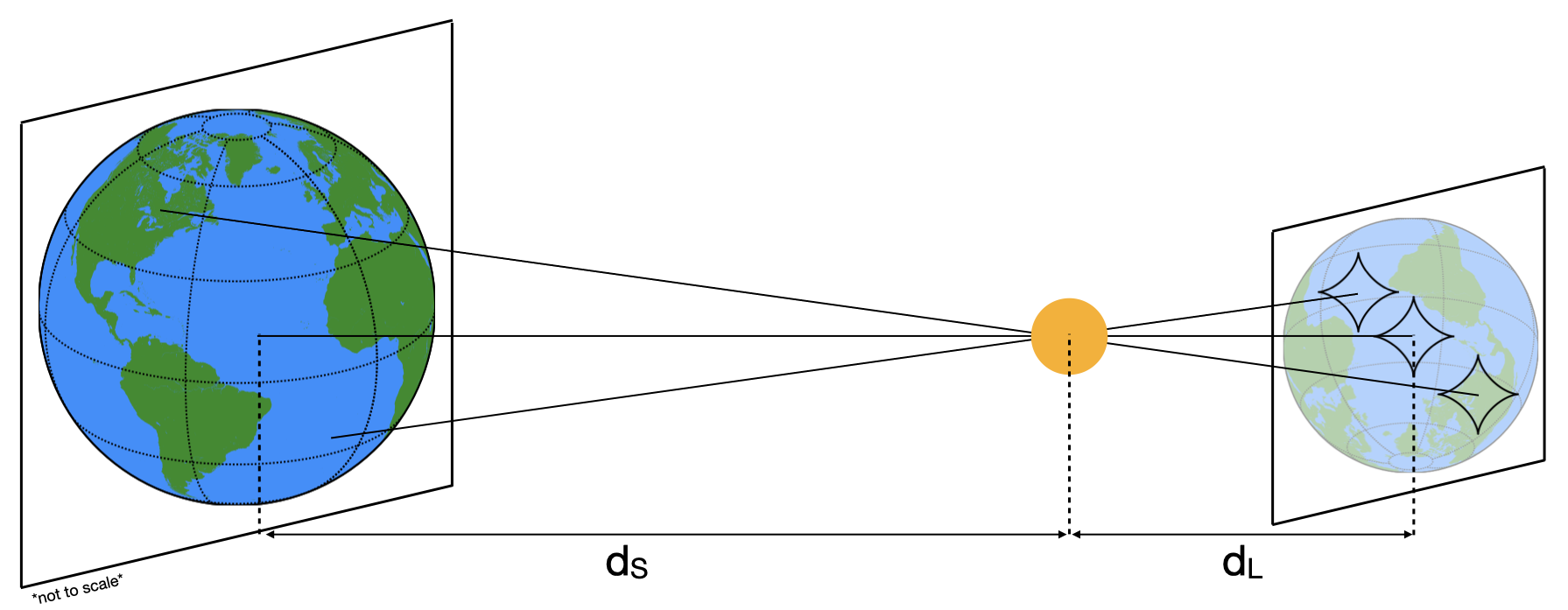}
    \caption{Schematic diagram showing the relationship between a point source's location in the source plane, and its corresponding location in the SGL image plane. Diagram not to scale.}
    \label{fig:extended}
\end{figure}

For our analysis, each pixel in this approximation acts as an independent, incoherent source in the imaging problem. Each point on the source object is reimaged by the SGL and appears as an astroid-shaped PSF in the SGL image plane, with the center of the astroid offset by it location in the source plane multiplied by the SGL plate scale. If the Cartesian coordinates in the SGL image plane are $(x_{n_x},y_{n_y})$, then the location of the astroid center for a point source located at $(x'_{n_x},y'_{n_y})$ is
\begin{eqnarray}
    x_{n_x} = - x'_{n_x} \frac{d_L}{d_S}, \\
    y_{n_y} = - y'_{n_y} \frac{d_L}{d_S},
\end{eqnarray}
where the negative sign appears due to the inversion of the image. Figure (\ref{fig:extended}) depicts the geometry of the problem for convenience. 

In this diagram, which is not to scale, the size of the astroid PSF in the SGL image plane is depicted to be a fraction of the size of the extended object, a diagram of the Earth. One can visualize how a telescope situated directly on the optical axis will see unique point-spread functions for each of the three points highlighted in the diagram. For the central point on the optical axis, the telescope will sit squarely inside the center of the astroid, where the PSF will form an Einstein cross in the telescope image plane. For the two points which are offset from the optical axis, the telescope centered on the optical axis will instead intercept light from outside the edge of the astroid, which will result in simply two point sources in the telescope's image plane. It may be helpful to again reference Figure (\ref{fig:psfs}), which shows the PSFs which form in the telescope image plane for telescope locations offset relative to the astroid.

\begin{figure}
    \centering
    \includegraphics[width=\textwidth]{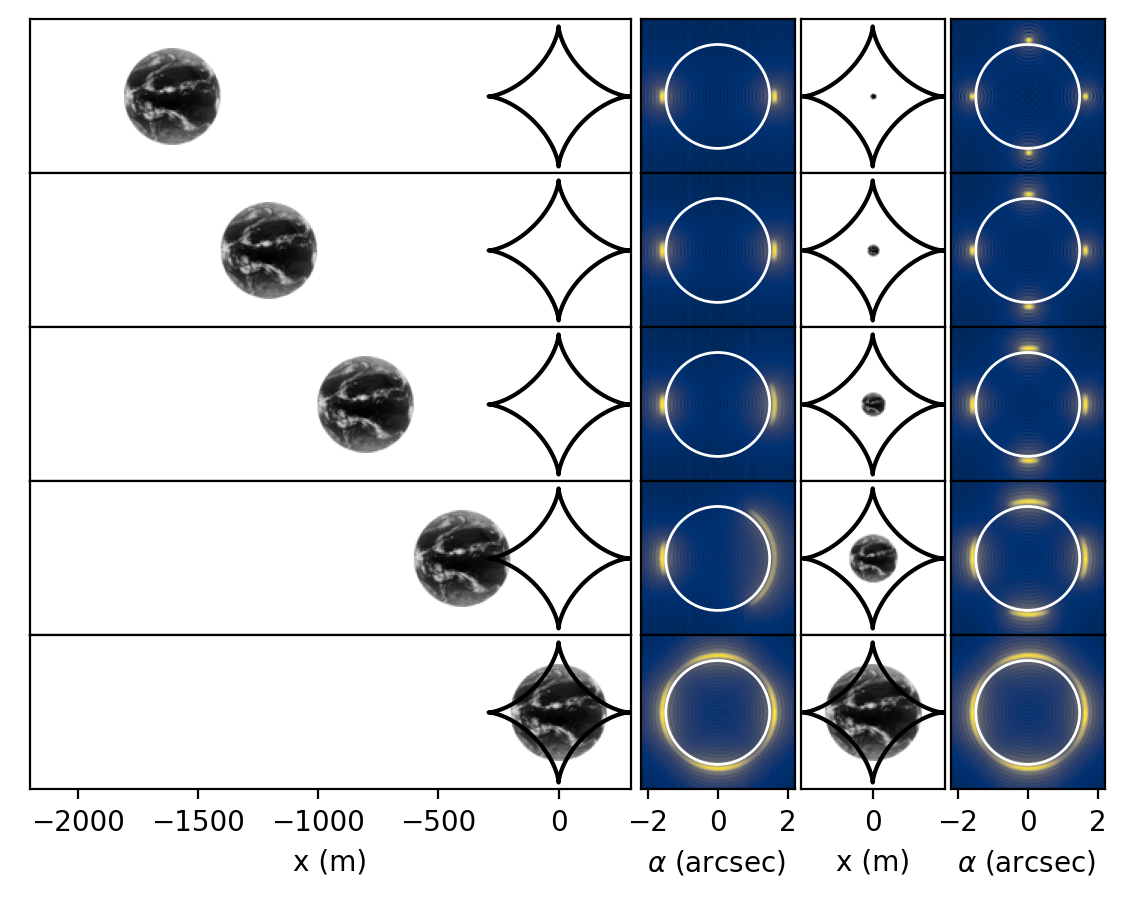}
    \caption{Imaging of an extended source with size comparable to the astroid. In these simulations, $d_S = 100$ pc, $\beta_S = 90^\circ$, and $\lambda = 1$ $\mu$m. The Earth is depicted with a black-hot colormap in reimaged coordinates, and the location of the astroid for the point aligned with the optical axis is plotted. The left half shows the Earth (or equivalently the telescope), moving laterally, and the change in the extended image as the target approaches optical alignment. When far from alignment, the target appears as only two points, although these grow asymmetrically and finally form a ring-like structure, with higher intensity regions in a quadrupolar pattern due to the combined effects of the extended emission and the resulting superposition of Einstein cross-like PSFs in the telescope image plane. The right half depicts an unphysical scenario where the extended source changes in angular size, to compare the limits of small extended sources with point-like objects. One can see that the resulting intensity distribution in the telescope image plane smoothly varies from ring-like in the limit of large sources compared to cross-like in the limit of point like sources.}
    \label{fig:extended_example}
\end{figure}

To obtain the final intensity distribution in the telescope image plane, we simply sum the contributions of each point in the source plane, using the unique point-spread function a telescope would see corresponding to that point. To do this, we compute the point spread function over a grid of telescope positions offset relative to the optical axis. It is equivalent to consider a point source located at $(x',y')$ in the source plane and a telescope centered on the optical axis $(x=0,y=0)$ as it would be to consider a point source on the optical axis $(x'=0,y'=0)$, and have the telescope center offset by negated and plate scale reduced coordinates $(c_x = x'_{n_x}\frac{d_L}{d_S},c_y = y'_{n_y}\frac{d_L}{d_S})$. Thus the relationship between source plane pixel indices and corresponding telescope centering offsets is
\begin{eqnarray}
    c_x = \frac{d_L}{d_S} R_\textrm{Earth} \Big( 2 \frac{n_x}{N_x} - 1 \Big) \\
    c_y = \frac{d_L}{d_S} R_\textrm{Earth} \Big( 2 \frac{n_y}{N_y} - 1 \Big).
\end{eqnarray}
With the vectors of centering offsets forming a grid of telescope locations $(c_x,c_y)$, we compute the PSF in the telescope image plane using Equation (\ref{eq:PSF}), where the pupil transmission function is modified so that $P(\boldsymbol{x}) \in \{(x,y) | \sqrt{(x-c_x)^2 + (y-c_y)^2} \leq D_\textrm{tel}/2\}$ and the telescope is properly centered around $(c_x,c_y)$. As a computational shortcut, $E(\boldsymbol{x})$ in Equation (\ref{eq:approx}) is only computed for points $\boldsymbol{x}$ inside the pupil where $P \neq 0$ to save on computational cost. Additional computational cost is saved by reverting to the analytic monopole expression for $E(\boldsymbol{x}) = J_0(k\sqrt{2R_S/r}\rho)$ when the telescope is sufficiently far from the optical axis. Since Figure (\ref{fig:mu}) demonstrates that the monopole approximation starts to be very accurate for $\rho = \sqrt{c_x^2 + c_y^2} \gtrsim 4 D_\textrm{astroid}$, this is the condition we check before falling back on the fast analytic computation for $E(\boldsymbol{x})$. The final intensity distribution $I_\textrm{extended}$ is computed as a matrix product of the point-spread function grid $\textrm{PSF}_{c_x,c_y}$ and the source image vector $I_{n_x,n_y}$.
\begin{equation}\label{eq:extended}
    I_\textrm{extended}(\alpha,\beta) = \textrm{PSF}_{c_x,c_y}(\alpha,\beta) I_{n_x,n_y}
\end{equation}
Thus with the point-spread function grid previously computed and in memory, the forward model of imaging an extended source can be represented as a linear transformation of the input intensity distribution. However, for this to be achieved with a single matrix operation, the two dimensional indices of source coordinates indexed by pixel numbers $(n_x,n_y)$ and the point source offsets corresponding to telescope centering offsets $(c_x,c_y)$ must be unraveled into a single monotonically increasing index. This choice of ordering is arbitrary and can be done in a combinatorially large number of unique ways, although we find unraveling indices one row at a time to be simplest. Figure (\ref{fig:extended_example}) was created to demonstrate imaging of extended sources.

In general, the size of the astroid is controlled by the quadrupole moment $I_2$ and the solar co-latitude $\beta_S$, see equation (\ref{eq:astroid}), while the size of the image $D_\textrm{image}$ is controlled by the physical dimension of the source, in this case $R_\textrm{Earth}$, as well as the geometry of the problem, including the distances between the source and lens $d_S$, and the telescope and the lens $d_L$. The exact relation is
\begin{equation}
    D_\textrm{image} = 2 R_\textrm{Earth} \frac{d_L}{d_S}.
\end{equation}
For extended sources of different sizes, there are three regimes which are important to consider. One limit is that of very small sources, or $D_\textrm{image} << D_\textrm{astroid}$. In this case, one may think of the extended objects as essentially just a point, and the resulting image formed will be that of the point-spread function. The other limit is that of very large objects, or $D_\textrm{image} >> D_\textrm{astroid}$. In this case, the majority of point sources in the SGL image plane will be farther from the optical axis than the dimension of the astroid, and the resulting point spread function in the telescope image plane for each location on the extended object will simply be two points. One can think of this limit as being equivalent to taking $D_\textrm{astroid} \rightarrow 0$, or $\beta_S \rightarrow 0$, where the telescope is situated directly above or below the solar rotational axis and the projection of the quadrupole moment into the plane of observation becomes negligible. In this limit, the higher order effects from the multipole moments can be neglected, as nearly all points appear to be reimaged into two points in the telescope image plane, which is identical to the monopole-only analysis described in \citep{madurowicz2020}. 

The last regime is that where the dimension of the reimaged target and astroid have similar sizes, or $D_\textrm{image} \sim D_\textrm{astroid}$. In this regime, one cannot ignore the higher order effects of the multipoles, as many of the points on the target will be reimaged multiple times, some of the points may sit on the critical caustic, and some of the points may be outside the astroid entirely. Thus, the final intensity distribution formed in the telescope image plane is a superposition of the multi-faceted PSF distribution and intensity distribution from the source. 

\section{Sources of Contamination}
\subsection{A model of the Earth and Sun}

Actual physical observations using the solar gravitational lens will face an extreme difficulty dealing with contamination of source signal with noise from multiple sources along the line of sight, including the Sun and its corona, zodiacal dust in the solar system, and additional non target sources in the SGL field of view. This section investigates the expected brightness from the Sun itself as primary source of noise, compared to the expected brightness of a potential, Earth-like, target source by collecting data and models built from observations of the Earth and Sun and scaling them to match the geometry of the solar gravitational lens problem.

\begin{figure}
    \centering
    \includegraphics[width=\textwidth]{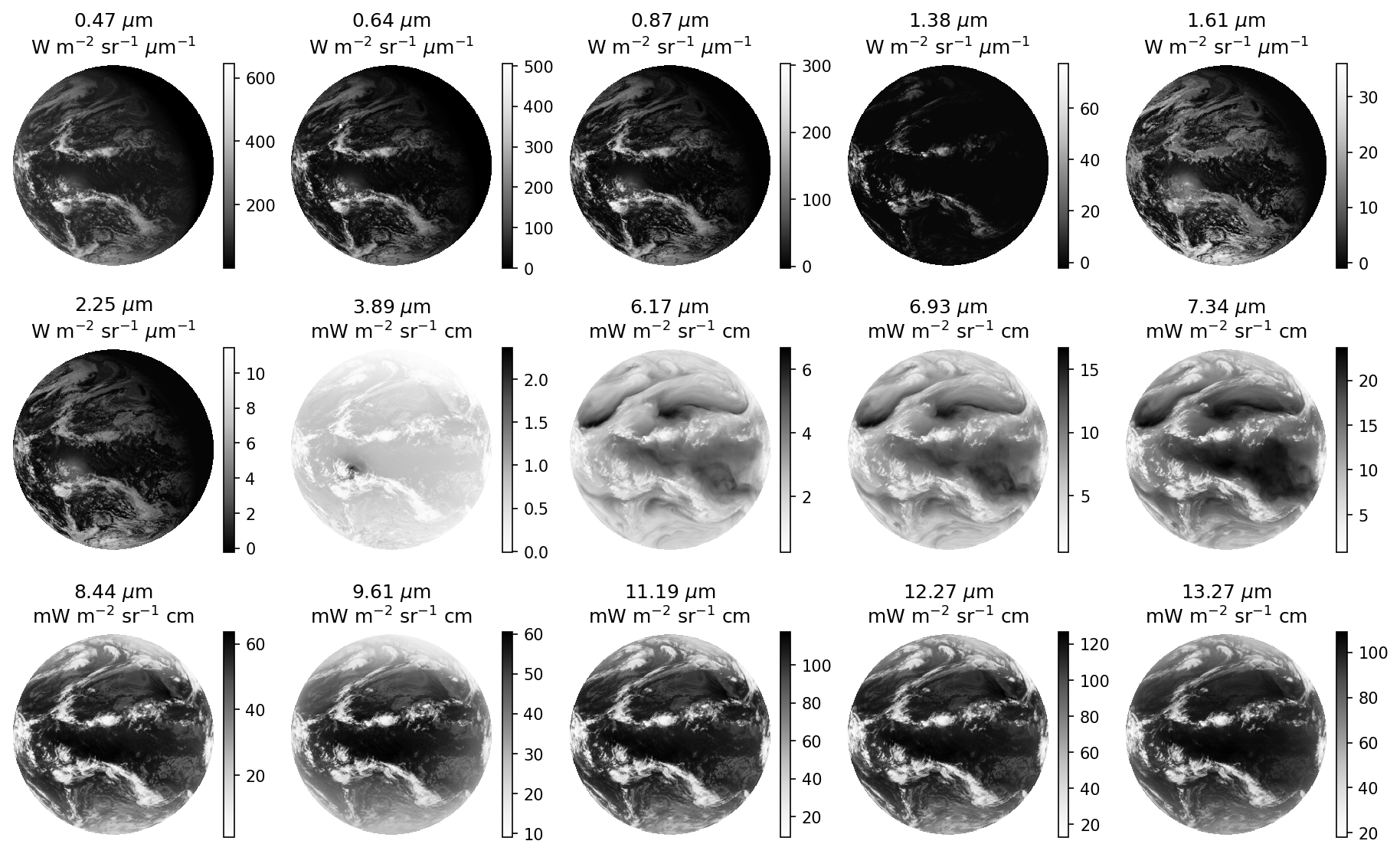}
    \caption{GOES R level-1b radiances across the available wavelengths \citep{goesr}. This specific time of observation corresponds to roughly 85\% phase for illuminations where the reflection of sun light dominates, which can be seen as a slightly dark crescent in the upper right of the images of the reflection dominated channels from $0.47 \mu$m to $2.24 \mu$m. These reflection dominated channels are plotted with a white hot colormap, while the emission dominated thermal channels from $3.89 \mu$m and onward are plotted with a black hot colormap, so that the clouds appear white in each image. The high resolution images have been downsampled here to a resolution of $512$ by $512$ pixels from the original data.}
    \label{fig:earth}
\end{figure}

The first Earth dataset used in this study is a series of images taken from the Advanced Baseline Imager onboard the GOES R weather satellite \citep{goesr}. The data includes extremely high resolution (either $\sim$$5$k by $\sim$$5$k or $\sim$$10$k by $\sim$$10$k, depending on wavelength) images of the Earth taken from geostationary orbit, across a wide range of wavelengths, from $0.47 \mu$m in the visible light spectrum, to $13.27 \mu$m in the mid-infrared. The specific dataset used in this study are the level-1b radiances product, which have been processed and calibrated to represent authentic top-of-the-atmosphere radiances in units of [W m$^{-2}$ sr$^{-1}$ $\mu$m] or [mW m$^{-2}$ sr$^{-1}$ (cm$^{-1}$)$^{-1}$] depending on if the channel is dominated by reflection of sunlight or thermal emission from the Earth itself. The GOES R data is plotted in Figure (\ref{fig:earth}).

\begin{figure}
    \centering
    \includegraphics[width=\textwidth]{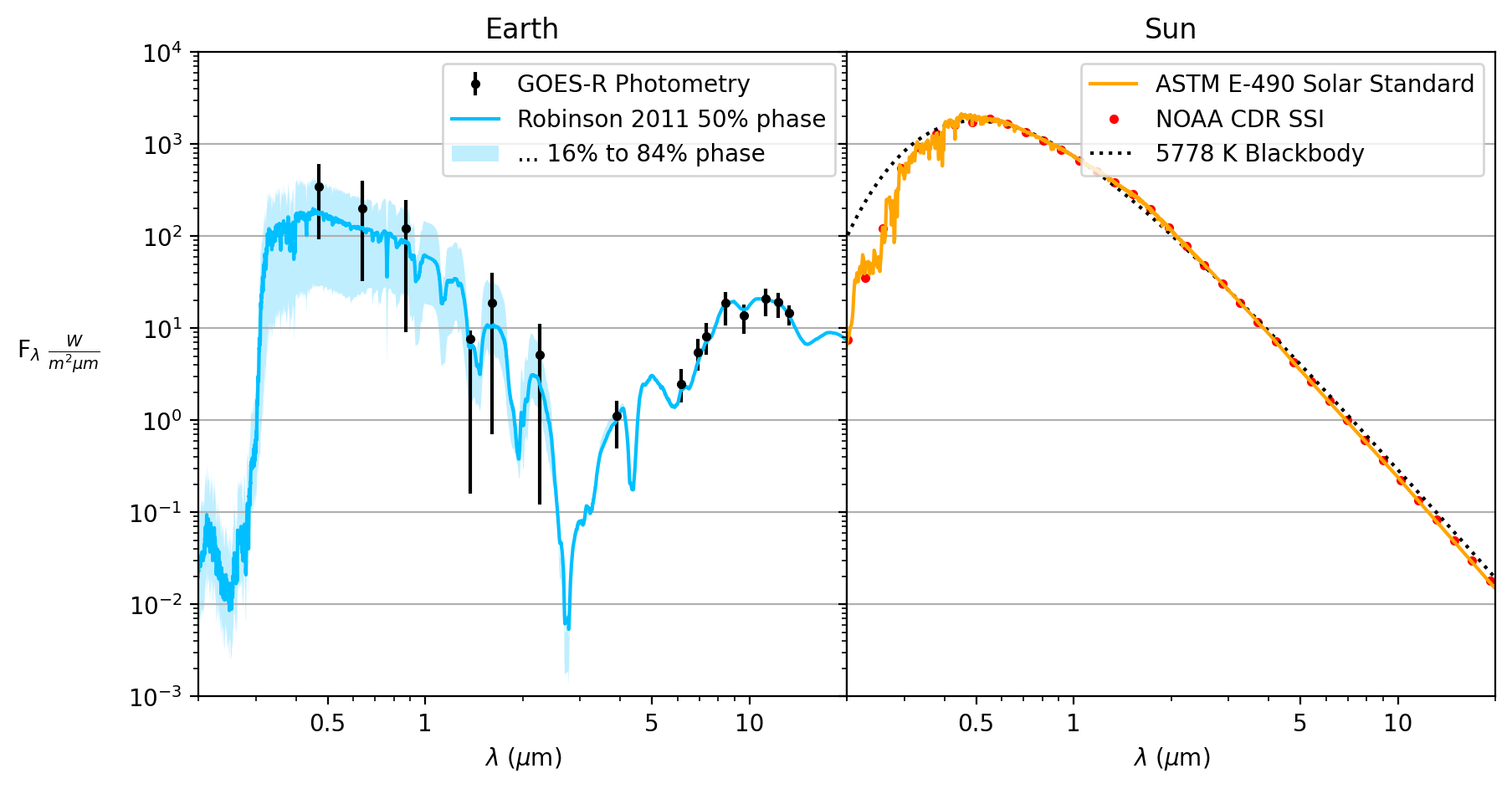}
    \caption{Left: Comparison of Robinson 2011 \citep{robinson2011} Earth spectral model with disk-integrated photometry from the GOES-R ABI \citep{goesr}. The spectral model is plotted in blue at 50\% phase with a shaded region corresponding to the 16\% and 84\% phase to show the variance of possible brightnesses at different observing geometeries for the wavelengths dominated by reflection. The error bars on the GOES data points represent the 16$^{th}$ and 84$^{th}$ percentiles of intensity across the Earth's disk, so the deep error bars correspond to near completely dark regions where the surface of the Earth is in the shadow of the Sun at nighttime. Right: Comparison of the three solar models, the blackbody model, the NOAA CDR \citep{noaaCDR}, and the ASTM-E Solar standard \citep{frohlich1998}.}
    \label{fig:earth_sun}
\end{figure}

While the GOES-R data will be useful to investigate the effect of non-uniform intensity distributions across different wavelengths, the wavelength coverage is not complete. With the goal of investigating higher spectral resolutions, we investigate the Earth model described in \citep{robinson2011}. The Earth spectral model describes disk-integrated specific intensities of the Earth across a massive range in wavelength space, from $0.1\mu$m to $200 \mu$m, with spectral resolution $R = \frac{\lambda}{\Delta \lambda}$ ranging from $\sim$100 at the longest wavelengths to $\sim$10$^{5}$ at the shortest. The Earth spectral model also includes variations in brightness due to the inclination of the orbit and the orbital phase. Both the GOES-R data and the Earth spectral model are plotted in Figure (\ref{fig:earth_sun}) in units of spectral flux density or [W m$^{-2}$ $\mu$m] by multiplying by $\int_0^{2\pi}\int_0^\pi \sin \theta \cos \theta \textrm{d}\theta = \pi$ sr. These units will be convenient for comparison with a model of the spectral flux density of the Sun.

To compare the brightness of the Earth and the Sun across many wavelengths, we use three separate Solar models to act as validation and ensure consistency. The first and simplest model is that of a spherical blackbody with radius $R_\odot$ and temperature $T = 5778$ K, to match the effective temperature of the solar photosphere \citep{Carroll2007}. The blackbody spectrum is governed by the Planck equation
\begin{equation}
    B_\lambda (T) = \frac{2hc^2}{\lambda^5}\big(e^{\frac{hc}{\lambda k T}} - 1\big)^{-1},
\end{equation}
where $h$ is Planck's constant, $c$ is the speed of light, $k$ is the Boltzmann constant, and $T$ is the blackbody temperature. Since the blackbody equation gives specific intensity at the photosphere, to calculate the flux density at the top of Earth's atmosphere, one must rescale by the solid angle of the Sun as seen from Earth
\begin{equation}
    F_\lambda = B_\lambda \pi \Big(\frac{R_\odot}{1 \textrm{ AU}}\Big)^2.
\end{equation}
The blackbody model is quite a good description of the radiation from the Sun, due to its physical elegance and analytic form, but is incomplete due to absorption in the solar atmosphere and additional complications. However, the sun is bright and easily observable, so spectral data at high resolution is plentiful and of high significance. We use two separate Solar spectral models which are validated with real observations. The first model is the National Oceanic and Atmospheric Administration (NOAA) Climate Data Record (CDR) Solar Standard Irradiance (SSI) \citep{noaaCDR}, and the second is the American Society for Testing and Materials (ASTM) E-490 Solar standard, which is compiled from multiple unique observations \citep{Kurucz1993} \citep{frohlich1998} \citep{Woods1996} \citep{Smith1974} to serve as a reference spectrum across a large range of wavelengths. Both of these models are converted into identical units for comparison in Figure (\ref{fig:earth_sun}).

A few observations regarding the spectral models. The Sun is accurately described by a blackbody spectrum, with small perturbations due to absorption and emission lines that are mostly invisible at the plotted spectral resolution. The Earth model has two peaks, one in the visible spectrum corresponding to the reflection of sunlight, and one in the infrared corresponding to the thermal emission from the planet itself. Additionally, the Earth's spectrum is significantly altered by absorption in its atmosphere, with many valleys in different wavelength regions indicative of particular molecules such as H$_2$O and CH$_4$.

\begin{figure}
    \centering
    \includegraphics[width=\textwidth]{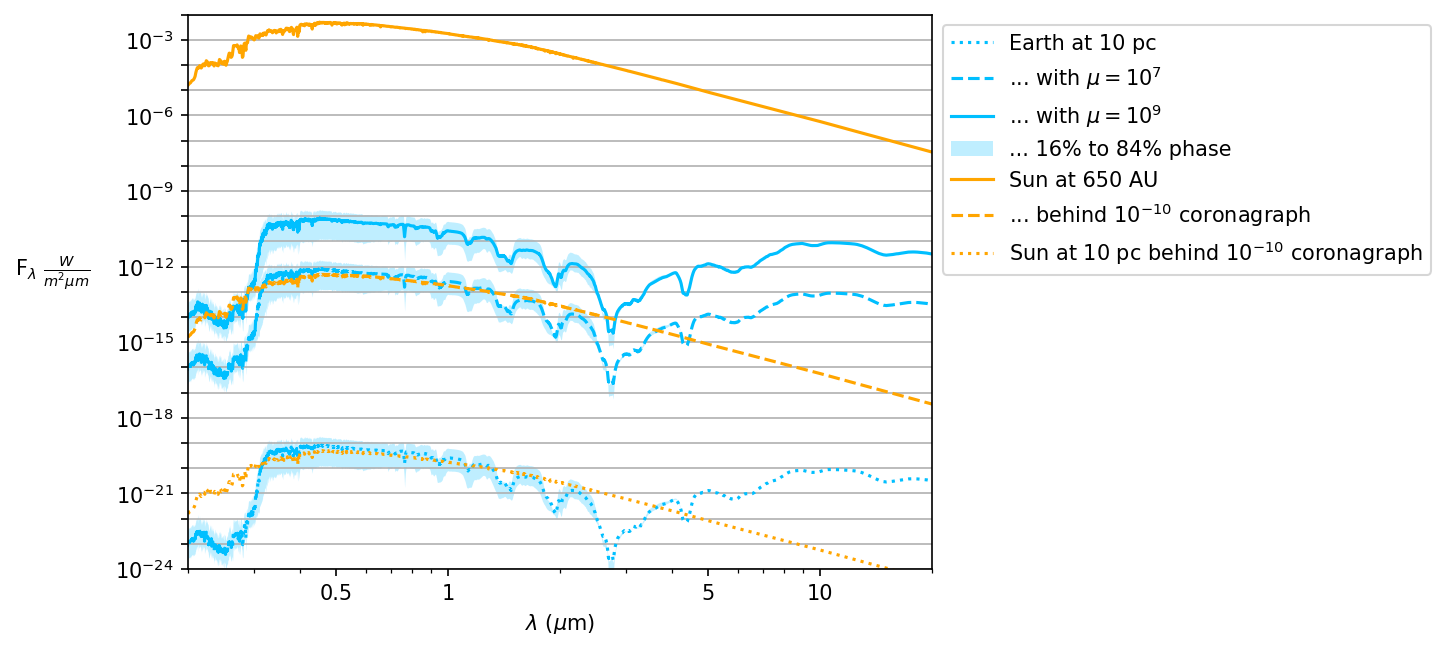}
    \caption{Comparison of spectral flux density for various cases. The pair of dashed lines represents the curves relevant for the solar gravitational lens imaging, while the dotted curves represent the curves relevant to direct imaging of Earth-like planets around Sun like stars without the solar gravitational lens.}
    \label{fig:DI}
\end{figure}

The previous figure depicts the spectral flux density that an observer would see for the Sun and the Earth if that observer was located at the top of Earth's atmosphere. However, this arrangement is not equivalent to the geometry of imaging a target planet through the solar gravitational lens, and so these curves must be rescaled to the appropriate distance using the inverse square law. Additionally, the flux of the planet is magnified by imaging through the SGL, by an amount that depends on its location on sky as well as by the exact alignment between telescope, target, and the Sun. Figure (\ref{fig:mu}) shows that reasonable values of the magnification can range from $\mu = 10^7$ to $10^9$. Additionally, one may obscure the light from the Sun using a starshade \citep{cash2011} or coronagraph \citep{lyot1939}, which may need to be specifically designed to obscure the extended disk of the Sun \citep{ferrari2009}, but could significantly reduced the observed flux from the Sun. 

To make a fair comparison of the brightnesses with all of these factors considered, Figure (\ref{fig:DI}) was created. In this Figure, there are three curves for each the Sun and the Earth. Each starts with the Robinson Earth spectral model or the ASTM-E solar spectral model and rescales the flux amplitude depending on different assumptions. From top to bottom, the Solar curves are rescaled to $d_L = 650$ AU without a coronagraph, $d_L = 650$ AU with a coronagraph with a contrast of $10^{-10}$, and $d_L = 10$ pc behind a similar coronagraph. The rescaling factor is simply a multiplication by $\textrm{contrast} \times (\frac{d_L}{1 \textrm{ AU}})^2$. This level of coronagraphic sensitivity has not yet been reached but is projected to be possible by the HABEX mission concept \citep{habex}. The three curves for the Earth, from top to bottom, all correspond to a target distance $d_S = 10$ pc, but with $\mu = 10^9, 10^7,$ or $0$, corresponding to a possible magnified source behind the solar gravitational lens and an unmagnified source that one would see without the solar gravitational lens.

The principal observation this Figure was designed to showcase is that in terms of flux $\textit{contrast}$ between the target source and the primary noise source, using the solar gravitational lens is comparable to direct imaging. The spectral flux density essentially describes the total number of photons that would be captured by the telescope over some bandpass, but is agnostic to the distribution of that light in the focal plane of the telescope, which is discussed in \ref{sec:44}. The dashed pair of curves correspond to a reasonable expectation of the flux for the standard solar gravitational lens problem, while the dotted pair of curves correspond to a reasonable comparison for direct imaging. The ratio of the flux between these two objects gives the contrast as a function of wavelength, which is quite similar. However, the major advantage that the solar gravitational lens arrangement has over direct imaging is in the absolute brightness of both the target and noise source, being approximately $10^7$ times brighter for both the target and the solar / stellar noise contribution. This advantage may prove useful to obtain extremely high dispersion spectra of an exoplanet atmosphere, since high dispersion implies correspondingly smaller signal to noise per bin, and high dispersion instruments typically struggle from low throughput \citep{vogt1994}.

\subsection{The local astronomical field}

\begin{figure}
    \centering
    \includegraphics[width=\textwidth]{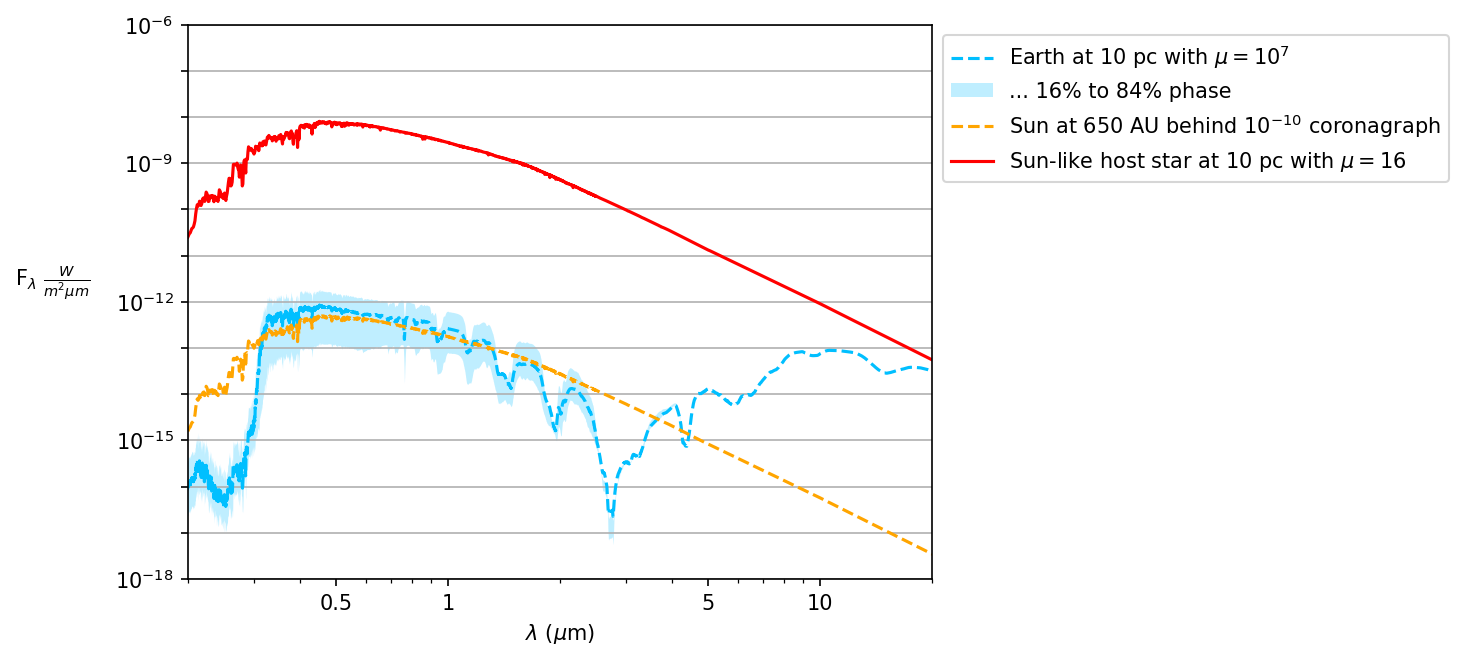}
    \caption{Comparison of spectral flux density for target planet, host star, and the Sun itself in a typical SGL configuration, with $d_L = 650$ AU, $d_S = 10$ pc, $a = 1$ AU. The blue and orange curve correspond to the nominal SGL case depicted in Figure (\ref{fig:DI}). In the optical wavelengths the target planet's host star can contribute four orders of magnitude more power than the target planet, though this is somewhat lessened at thermal IR wavelengths.}
    \label{fig:host_star}
\end{figure}

Aside from the light from the target source and the Sun itself, the Einstein ring contains light from all astrophysical sources in the SGL's field of view. This includes primarily the light from the target planet's host star, but also includes the probability of faint distant sources which are coincidentally close to alignment with the optical axis. In section 3.3 Figure (\ref{fig:mu}) we show the average magnification $\langle \mu \rangle_P$ over the telescope pupil as a function of the telescope's radial offset in the SGL image plane $\rho = \sqrt{c_x^2 + c_y^2}$. For radial separations significantly greater than the dimension of the astroid, the average magnification is a power-law with index $-1$, matching the expectation for a pure monopole SGL.

\begin{equation}\label{eq:mu_powerlaw}
    \log \langle \mu \rangle_P \approx 8.88 - \log \Big( \frac{\rho}{1 \textrm{ m}} \Big), \quad \rho \gg D_\textrm{astroid}
\end{equation}

Since a host star at $d_S = 10$ pc with a target planet at $a = 1$ AU has an angular separation in the source plane $\delta \theta = \frac{a}{d_S} = 0.1$ arcsec, this is equivalent to a telescope radial offset in the SGL image plane of $\rho = d_L \delta \theta = a \frac{d_L}{d_S} \sim 4.7 \times 10^7$ m and thus an average magnification of $\langle \mu \rangle_P \sim 16$. Figure (\ref{fig:host_star}) demonstrates the relative contribution across various wavelengths the host star will have to the observed flux of the target planet and the post-coronagraphic solar contribution.

Examining Figure (\ref{fig:host_star}), it is apparent that the target planet's host star contributes a relatively significant amount of the total flux intercepted. In optical wavelengths, the total contrast between the target planet and its host star is about four orders of magnitude, while at thermal IR wavelengths it is closer to one order of magnitude. This flux will be spatially separated in the telescope image plane, as the host star is far off axis, it will be reimaged into two points on opposite sides of the Einstein ring, similar to the fist panel in Figure (\ref{fig:extended_example}), which could be separated from the light from the planet which forms the entire ring.

The target planet's host star forms the primary source of contamination inside the Einstein Ring, due to its relatively nearby separation from the target planet at $\sim .1$ arcsec. However, astrophysical false positives are known to appear in direct imaging, such as this one case of a white dwarf companion to a K-type star \citep{zurlo2013}. Other astrophysical objects may appear in the field of view despite being gravitationally unbound to the host star, although this can be mitigated by multi-epoch observations demonstrating common proper motion \citep{macintosh2015}. Recent work investigating the probabilities of observing a galactic or extragalactic background source from the Hubble Source catalog, Gaia catalog and Besancon galaxy simulations \citep{cracraft2021} demonstrate that the probability of such a background object is usually low, though it increases at lower galactic latitude.

\begin{figure}
    \centering
    \includegraphics[width=\textwidth]{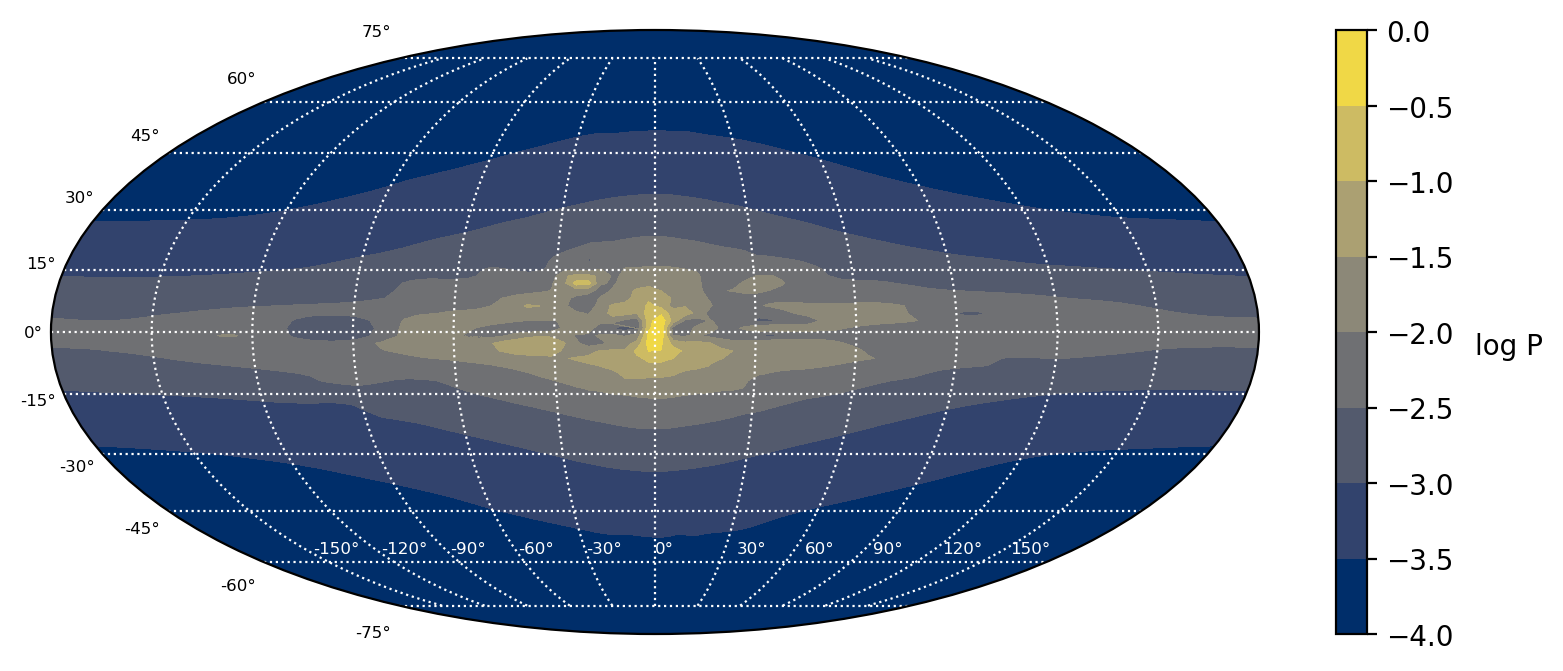}
    \caption{Interpolated map of the Besancon galaxy model star counts across the entire sky. Colormap is proportional to the log probability of finding a star with V magnitude $<$ 21 in a FOV of 1 square arcsec.}
    \label{fig:galaxy}
\end{figure}

A mission to the focus of the solar gravitational lens naturally occurs inside of the Milky Way galaxy and thus must be prepared for the possibility that a background object appears inside the SGL's field of view and contaminates the Einstein Ring. To investigate this probability we rescale star counts from the Besancon galaxy simulation \citep{czekaj2014} across the entire sky. Figure (\ref{fig:galaxy}) shows an interpolated map of star counts of Johnson V-mag $<$ 21 within 1 deg$^2$ converted into probability of finding a such a star within 1 arcsec$^2$. 

Examining the map in the Figure, the probability of some background object ranges by multiple orders of magnitude depending on sky location. In the direction of the galactic center, the probability is nearly certain that at least one such background star will be visible, while away from the galactic center is may be as low as 1 in 10,000. The magnification of such a background object depends strongly on the angular separation between it and the target planet (Equation \ref{eq:mu_powerlaw}), so it is difficult to quantify exactly how bright such a source will appear. If the background object has an angular separation of $\delta \theta = 1$ arcsec, this is equivalent to a telescope radial offset in the SGL image plane of $\rho = d_L \delta \theta \sim 4.7 \times 10^8$ m and thus an average magnification of $\langle \mu \rangle_P \sim 1.6$, which is a factor of 10 less magnification than that of the host star at $0.1$ arcsec. If the background object has an intrinsically low luminosity or is significantly further away in distance than the target, the total flux may or may not be large compared to the light from the planet. There is always a small possibility that the background object coincidentally appears at a small angular separation and becomes further magnified by the SGL. However, the light from the background object will be necessarily restricted to two diametrically opposed points on the Einstein Ring, similarly to the host star, with an azimuthal angle corresponding to the position angle of the background object, and can be spatially mitigated in the telescope's image plane by separating the azimuthal elements of the Einstein ring.

\subsection{The solar corona and zodiacal light}

Astrophysical backgrounds aside, there will be additional contamination of light from astrophysical foregrounds which must be mitigated to obtain precise measurements of the Einstein ring. We consider here primarily the light from the solar corona and zodiacal light from dust populations in the solar system. Since the dawn of civilization, total solar eclipses have provided a natural occultation of the solar disk and allowed people to observe in detail the solar corona \citep{deJong1989} \citep{DREYER1877}. Even in the modern day, with the development of the coronagraphic instruments \citep{lyot1939}, eclipse observations still provide some of the only detailed observations of the solar corona right up to the edge of the solar disk \citep{pasachoff2018} \citep{Judge2019}. These eclipse observations demonstrate the azimuthally asymmetric nature of the coronal emission, which complicates the analysis of the azimuthal variations in the Einstein ring's structure. 

Coronagraphic instruments are typically limited by an inner working angle which prevents observations at small separations, and so cannot observe the corona near the edge of the solar disk. However, unlike eclipse observations, coronagraphic instruments on space-borne telescopes can provide continuous monitoring of the solar coronal environment. These space-borne observations enable investigation of time dependent coronal phenomena, such as the relationship between coronal emission and changing magnetic field patterns related to the solar cycle \citep{Antonucci2020}.

\begin{figure}
    \centering
    \begin{minipage}[b]{0.59\textwidth}
    \includegraphics[width=\textwidth]{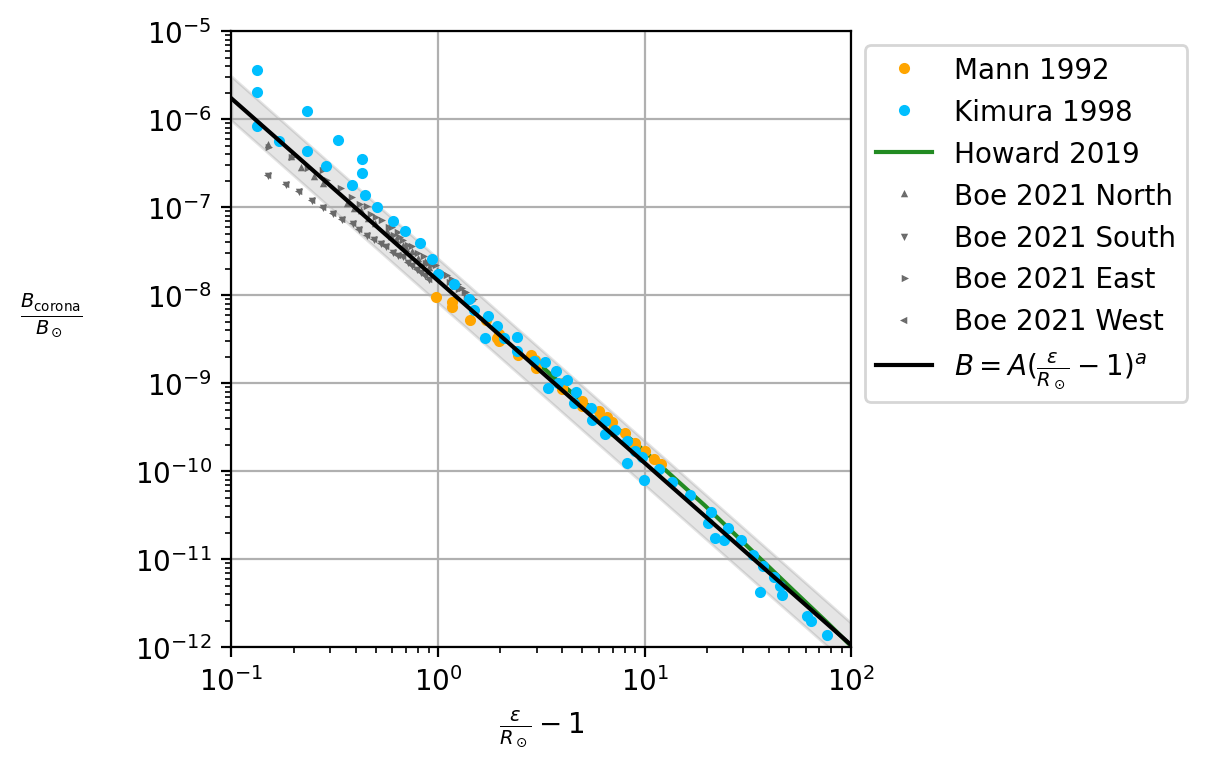}
    \end{minipage}
    \hfill
    \begin{minipage}[b]{0.37\textwidth}
    \includegraphics[width=\textwidth]{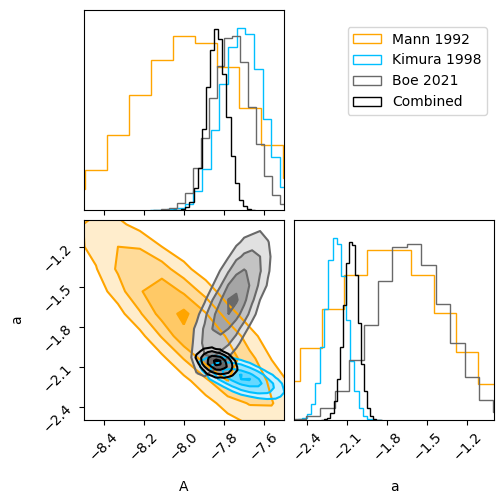}
    \end{minipage}
    \caption{(a): Compilation of coronal brightness measurements from multiple literature sources, and the best fit power law. (b): Triangle plot of power law parameter posterior probability distributions and covariances. The joint analysis in black constrains the model to the region of overlap between the datasets.}
    \label{fig:corona}
\end{figure}

Spectroscopic observations \citep{moore1934} of the solar corona indicates that the light emanating from this region is the result of multiple distinct processes, the scattering of sunlight off of dust particles, scattering off of free electrons, and direct emission from recombination of ionized atoms \citep{ZESSEWITSCH1929} \citep{mierla2008}. These unique components have been termed the F-, K-, and E- coronas respectively. F- for Fraunhofer, who discovered the similarity of absorption lines between the sun's photosphere and corona, indicating the light is reflected by large particles, K- for "continuous" in German, as Doppler broadening of the photosphere absorption lines from the velocity of free electrons creates the appearance of a continuum spectrum, and E- for "emission."  The F- and K- coronal contributions can reliably separated using the difference in spatial dependence between dust and free electrons, \citep{Calbert1972}, while the E- corona may be spectroscopically mitigated due to the narrow emission lines.

The F- corona is largely considered to be a natural extension of the zodiacal light at smaller angular separations or elongations $\epsilon$ with regard to the sun, and contains a wavelength dependence which increases towards long wavelengths due to thermal emission from the grains themselves in addition to the reflection of sunlight \citep{kimura1998}. However, interpretation of the F- coronal brightness and the corresponding dust distribution is complicated by inversion of the line-of-sight integral. Theoretical predictions for the near solar environment have speculated that the F- corona may decrease strength in the vicinity of the sun \citep{lamy1974,mukai1974}, due to orbital differentiation from Poynting-Robertson drag, pressure drag, radiation pressure, and sublimation of the grains. Observations from the WISPR instrument onboard the Parker Solar Probe \citep{Howard2019} have provided the first observation of such an F- coronal decrease, inferred from brightness measurements at various elongations at different heliocentric distance between $0.166$ and $0.336$ AU.

\begin{figure}
    \centering
    \includegraphics{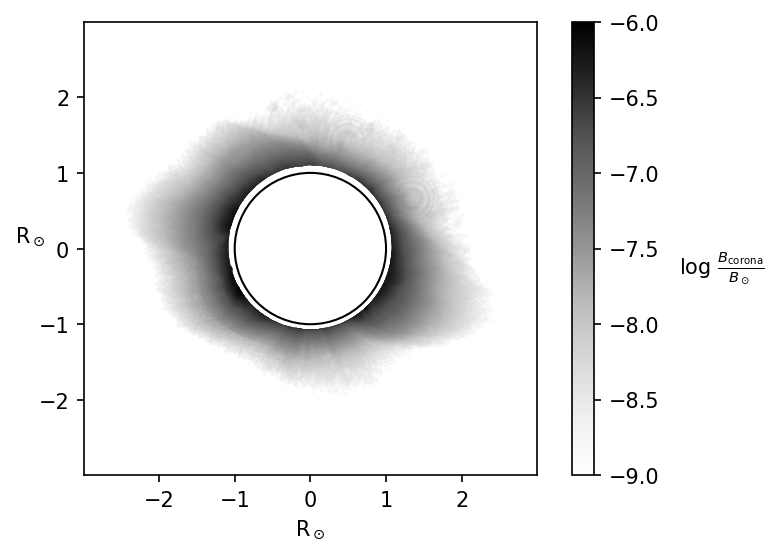}
    \caption{White light coronagraphic observation of the solar corona showing azimuthal variation in intensity. The solar disk is plotted as a black circle in the obstructed region behind the coronagraph.}
    \label{fig:kcor}
\end{figure}

Extensive brightness and color measurements of the solar corona have been taken and compiled previously, and for this work we cross reference multiple sources in the literature. By combining the data from Figure (2) in \citep{mann1992}, Figure (1) in \citep{kimura1998}, Figure (1) in \citep{Howard2019}, and Figure (9) in \citep{boe2021}, we are able to analyze the solar coronal brightness across a wide range of elongations $\epsilon$. The data are reproduced in Figure (\ref{fig:corona} a) for convenience. To build a model of the solar coronal brightness, we perform a joint analysis of the various datasets, by fitting for the maximum likelihood power law model of the data
\begin{equation}
    \log \Big( \frac{B_\textrm{corona}}{B_\odot} \Big)  = A + a \log \Big( \frac{\epsilon}{R_\odot} - 1 \Big),
\end{equation}
which is linear in log coordinates, and by choosing as the base of the exponent as $\frac{\epsilon}{R_\odot} - 1 $, the edge of the solar disk at $\epsilon = R_\odot$ corresponds to the origin of the power law. The resulting posterior distributions on the power law slope $a$ and intercept $A$ are shown in Figure (\ref{fig:corona} b). The covariances of the inferred parameters show constraints given by different individual datasets, while the joint analysis is restricted to the region of overlap between all datasets. The best fit parameters as a result of this analysis are $A = -7.836 \pm 0.05$ and $a = -2.071 \pm 0.055$. The resulting power law curve is shown in black in Figure (\ref{fig:corona} a).

Most of these measurements are taken with visible light wavelengths, although some effort has been made to make measurements in the near-infrared as well, and the results are compiled into Figures (5) in \citep{boe2021} and (2) in \citep{kimura1998}. In Figure (2) of \citep{kimura1998}, the scatter is large, there is a slight increase at longer wavelengths, going from roughly $10^{-9} B_\odot$ at $\lambda = 0.3 \mu$m to $10^{-8} B_\odot$ at $\lambda = 3 \mu$m for elongations of $\epsilon = 4 R_\odot$. Figure (5) of \citep{boe2021} reports relative intensities over $\lambda \in [0.5, 0.8] \mu$m, and measures a spectral power law index of $B_\textrm{corona} \propto \lambda^{0.91 \pm 0.07}$. This reddening effect could be the result of thermal emission from the particles themselves, as well as preferential directional scattering of thermal emission from the sun. 

Figure (7) in \citep{boe2021} also demonstrates the azimuthal variation in intensity, with iso-intensity curves fluctuating from $\epsilon \in [2.0, 2.4] R_\odot$ at $10^{-8} B_\odot$ to $\epsilon \in [1.3,1.5] R_\odot$ at $10^{-7} B_\odot$. To show the azimuthal variation that can reasonably be expected here, we plot data from the HAO COSMO K-coronagraph \citep{wijn2012} \citep{hou2013} in Figure (\ref{fig:kcor}) with a black hot colormap.

The data in Figure (\ref{fig:kcor}) are level-2 PB calibrated data downloaded directly from the \href{https://www2.hao.ucar.edu/mlso/mlso-data-and-movies}{HAO website}. Some of the many calibration steps include camera non-linearity, dark, and gain corrections, polarization demodulation, distortion correction, combining images from multiple cameras, removal of sky polarization, and correction for daily variation in sky transmission. From examining the image, it is clear that the observations are roughly in correspondence with the power law analysis described previously, with the coronal brightness ranging from $10^{-6} B_\odot$ at $\epsilon \sim R_\odot$ to $10^{-8} B_\odot$ at $\epsilon \sim 2 R_\odot$, although it is clear how the coronal brightness varies depending on azimuthal angle. 

\subsection{Implications for Instrumental Design}\label{sec:44}

\subsubsection{Coronagraph Design}

Assessing the feasibility of imaging the Einstein ring at extremely high contrast is complicated by the redistribution of light in angular coordinates. The Einstein Ring has an approximate radius of $\sim 1$ arcsec, while a host star / target planet separation for an orbit of $a = 1$ AU at $d_S = 10$ pc is an order of magnitude smaller at $0.1$ arcsec. Designing a coronagraph to reach high contrast at small inner working angle is challenging \citep{spaan2006} \citep{mawet2012}, and is typically investigated in the context of direct imaging of point sources. This is complicated by the extended solar disk and the minimal separation between the edge of the disk and Einstein ring. The angular separation between the edge of the solar disk and the center of the Einstein ring for a source with $d_S >> d_L$ is \citep{Hippke2020}

\begin{equation}
    \delta \theta \approx \sqrt{\frac{2 R_s}{r}} - \frac{R_\odot}{r},
\end{equation}
which reaches a maximum value of of $\sim$0.4 arcsec at a heliocentric distance of $r \sim 2300$ AU independently of wavelength. Specialized coronagraph designs for this particular case have only recently been investigated. An original suggestion from Turyshev and Toth was the use of an annular coronagraph \citep{TT2020} which would only allow light in from the region of the Einstein ring while blocking light from the solar disk and outer corona. However, as we have seen, the Einstein Ring itself contains contamination from the target planet's host star, as well as potentially other objects in the field and so this may be insufficient. Loutsenko \citep{loutsenko2021} has recently suggested a design which is a combination of an annular mask and a bar mask, with the intention of obstructing light from both the solar disk as well as the magnified off axis contribution from the target planet's host star. Their design claims to reach contrasts of $10^{-8}$ between the solar disk and the region of the Einstein Ring, with similar performance for off axis sources behind the bar mask. However, neither of these coronagraphs on their own can mitigate contamination from the region of the Einstein ring itself, which includes light from the corona and background sources. A spectroscopic mitigation strategy similar to the method described in \citep{ruffio2019} could model all sources of contamination in the final image, using spectral differences between noise sources to obtain high contrast in the absence of or in combination with a coronagraph.

\subsubsection{Background and Foreground Mitigation}\label{sec:441}

In the process of designing a mission to the SGL, one should be prepared to deal with the possibility of a background object contaminating the Einstein ring. The target planet's host star will certainly contaminate two points along the Einstein Ring, and a possible background star could contaminate two additional points in a similar fashion. Even contamination from non-target planets in the target solar system may be relevant if the angular separation between the target and contaminating planet is small enough. Selecting a target planet from a list of multiple potential target planets being considered should choose the safer option which is further from the galactic center. This will lower the probability of a background star, but the probability cannot be eliminated entirely. Early and deep observations of the near target region could provide advanced evidence for a clear background, but a robust design should include a contingency plan for mitigating this contamination if the observation is unlucky, as galactic proper motion could place a background object in the FOV sometime between launch and observation. By spatially separating the light in the Einstein ring, either a hardware or a software mask could be applied to the region around the points of contamination, although scattered light from this unobstructable source will ultimately limit the possible SNR obtainable in uncontaminated pixels.

For proper mitigation of the corona and zodiacal light as astrophysical foregrounds, it will be necessary to model the temporal and azimuthal dependence of the coronal brightness to extract the full information contained in the Einstein ring. One observational strategy to achieve this is a differential measurement. A single telescope equipped with an integral field spectrograph could make spatially- and spectrally- resolved measurements of the Einstein ring and the corona as well as the region of the corona nearby the Einstein ring. By measuring the coronal brightness in the vicinity and extrapolating a model into the region of overlap, a differential measurement may be possible as a mitigation strategy. Another type of instrument design could achieve a similar measurement instead using two separate spacecraft, one the primary observation telescope at the focal plane of the SGL and an additional companion telescope observing the solar corona from a closer heliocentric distance, where the Einstein ring cannot form at $d_L \lesssim 550$ AU. Simultaneous observations could enable a differential measurement of the brightness of the coronal region with and without the Einstein ring present, and would not rely upon fitting a nearby coronal region and extrapolating. However, this concept would suffer from complexities involved with coordinating observations on different telescopes separated by a great distance.

\begin{figure}
    \centering
    \includegraphics[width=0.8\textwidth]{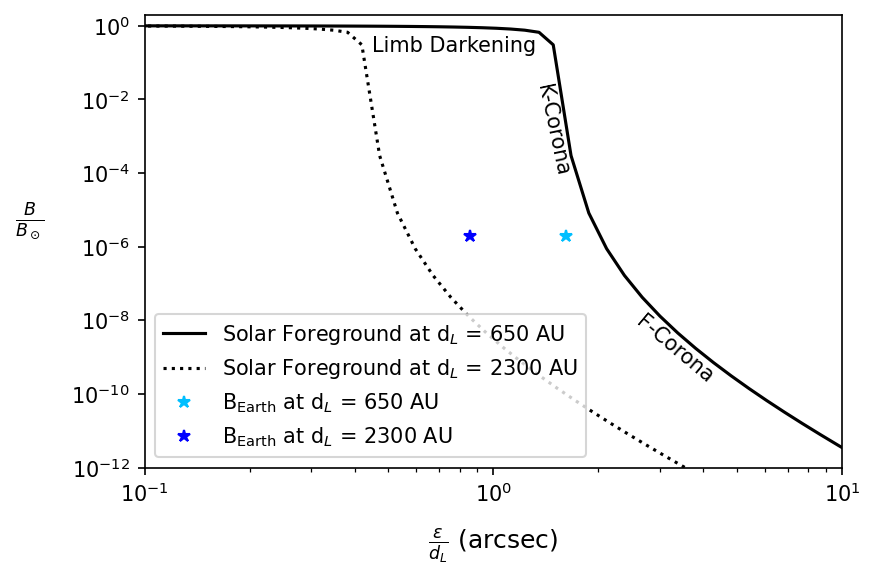}
    \caption{Comparing the specific intensity of the solar foreground and an Earth-like target source at different heliocentric distances $d_L$.}
    \label{fig:brightness}
\end{figure}

Another alternative strategy for mitigation of the corona is by sending the observation craft to a significantly larger heliocentric distance of $d_L = 2300$ AU for observing. Both the angular size of the Einstein ring and the solar disk shrink at larger $d_L$, but the Einstein ring shrinks proportionally to $d_L^{-1/2}$ while the solar disk shrinks proportionally to $d_L^{-1}$. This means at the optimal observation distance $d_L = 2300$ AU, the angular separation between the edge of the solar disk and the center of the Einstein ring is maximal, and the foreground contamination will be minimal as the Einstein ring overlaps with the much fainter F-corona rather than the bright K-corona at small separations. 

The solar foreground curves shown in Figure (\ref{fig:brightness}) is a combination of the coronal model fit in Section (4.3) and a fifth-order polynomial limb darkening model for the brightness of the solar disk described in \citep{Hestroffer1998} and using data from \citep{Pierce1997} and \citep{Neckel1994}. The data correspond to measurements at $\lambda \sim 0.58\mu$m but additional measurements at wavelengths $\lambda \in [0.78,2.44]$ $\mu$m is reported in Figure (6) of \citep{Pierce1954} which demonstrates the increasing effects of limb darkening at longer wavelengths. The specific intensity of the Earth $B_\textrm{Earth}$ is taken from the model in \citep{robinson2011} and the specific intensity of the Sun is taken from the blackbody model described in Section (4.1). The specific intensity ratio $B_\textrm{Earth}/B_\odot \sim 10^{-6}$ is not the typical flux contrast ratio of the Earth and Sun $\sim 10^{-10}$, as it lacks the additional factor of $(R_\textrm{Earth}/R_\odot)^2 \sim 10^{-4}$ to account for the difference in solid angle of the bodies. A version of Liouville's theorem of conservation of phase space volume applied to gravitational lenses \citep{Misner1973} proves that gravitational lenses preserves surface brightness or specific intensity up to a frequency shift due to gravitational redshift, while the magnification acts to modify the apparent solid angle of the object. This justifies our comparison of the specific intensities of the background and foreground along the line of sight, and demonstrates the advantage of observing at a farther heliocentric distance.

\section{Spectral Image Reconstruction}

Investigating different instrument designs such as broadband coronagraph masks and multi spectral IFS imaging spectrometers will require a full forward model of all sources in the field of view, emission from the solar disk and corona, light from the target planet, host star, and other contaminating objects in the field like additional planets or background stars. Such a forward model is within reach by combining all of the models described in this paper, but computationally quite expensive and outside the scope of this paper. An approximation like the quartic polynomial solution described by \citep{TT2021e} could expedite this compute time at the cost of degraded accuracy, but a complete forward model will be necessary as the mission concept matures and specific parameters of the problem can be further narrowed down to construct instrumental design requirements on the performance of the craft. This complete model could be used to simulate instrument design concepts and test post processing algorithms to extract information from the Einstein ring in the presence of realistic astrophysical noise sources in a spatially and spectrally resolved manner.

For now, in the absence of such a complete model, we can still test the reconstruction power of such observations in a toy environment. The forward model of the lensing problem in Equation (\ref{eq:extended}) is a linear transformation of the input intensity distribution $I_{n_x,n_y}$. The linear transformation is given by the matrix PSF$_{c_x,c_y}$ which describes the impulse response of the SGL to points laterally offset from the optical axis and captures all of the complexity of lensing in the presence of the solar gravitational potential's zonal spherical harmonics. To test the reconstruction power of an ideal measurement of the extended source's Einstein ring $I_\textrm{extended}$, we can pseduo-invert the forward model using the singular value decomposition. The SVD can be represented as a matrix factorization of the form \citep{HENDLER19941}
\begin{equation}
    \textrm{PSF}_{c_x,c_y} = U \Sigma V^*.
\end{equation}
Where $U$ and $V$ are square, real, orthonormal, and unitary matrices, and $\Sigma$ is a diagonal matrix containing the singular values. This is a useful method to decompose the linear transformation $\textrm{PSF}_{c_x,c_y}$, because $U$ and $V$ are unitary operators, they can be thought of as acting to rotate the basis elements of the space, while $\Sigma$ acts to stretch the rotated vector along the intermediary axis. This combination of rotate, stretch, derotate naturally allows one to find the pseudo-inverse, by the means of de-rotating, un-stretching, and re-rotating, written mathematically as \citep{Gregorcic2001}
\begin{equation}
    \textrm{PSF}^+ = V\Sigma^{-1}U^*.
\end{equation}
The inverse of $\Sigma$ is simply the inverse of each of the diagonal elements, which is why numerically small singular values can cause the pseudo-inverse to be discontinuous, and many routines to find the pseudo-inverse will introduce a cutoff, below which the singular values are set to zero. This cutoff parameter is known as $\texttt{rcond}$, and is typically set to values near the machine epsilon, where numerical noise can become problematic.

\begin{figure}[H]
    \centering
    \includegraphics[width=\textwidth]{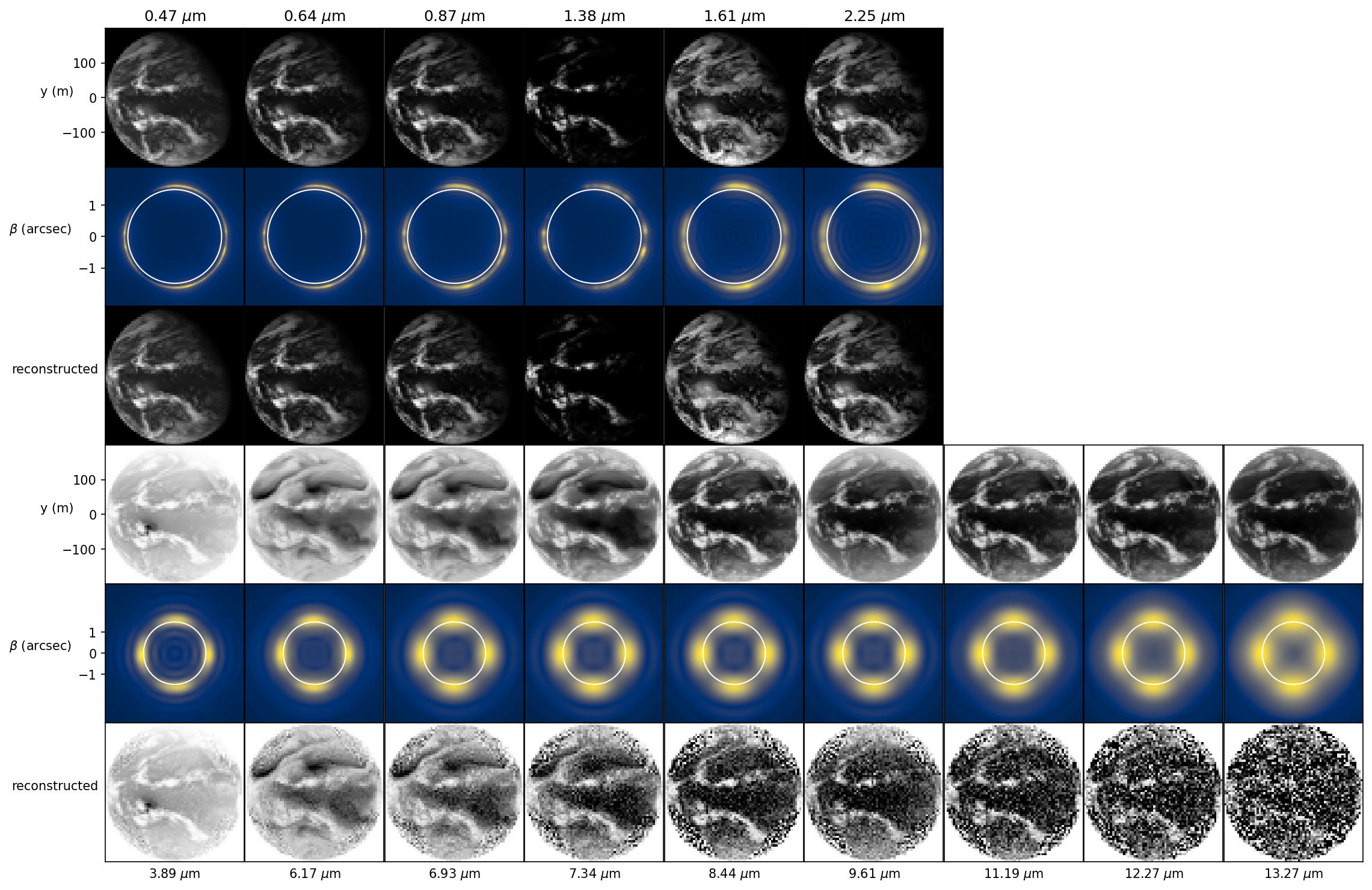}
    \caption{Image reconstruction across multiple wavelengths $\lambda$, with $d_S = 100$ pc, $d_L = 650$ AU, $\beta_S = 90^\circ$, and SNR = 2$\times10^5$ Upper half: Reflection dominated channels in a white-hot colormap. Lower half: Thermal emission dominated channels shown with a black-hot colormap. First row: Input intensity distribution taken from GOES-R data in reduced coordinates corresponding to the image projection. Second row: Forward model of the Einstein ring. The extent of the solar disk is shown as a white circle. Third row: Reconstructed intensity distribution using SNR = 2$\times10^5$. This SNR is unrealistically large considering all of the possible sources of contamination, but demonstrates that the Einstein ring theoretically contains most of the information of the input intensity distribution for regions inside the projection of the caustic. The reconstruction is increasingly challenging at longer wavelengths as the PSF grows larger and washes out the delicate intensity fluctuations in the ring needed to infer the original brightness distribution.}
    \label{fig:reconstruction}
\end{figure}

To evaluate the reconstruction performance of these observations, we start by computing the forward model of the intensity distribution, add Gaussian noise to simulate imperfect measurement, and apply the pseudo-inverse. The measurement noise takes the form of a sample generated from a Gaussian distribution with zero mean and variance proportional to the intensity in the brightest pixels $\textrm{max}(I_\textrm{extended})$ and  inversely proportional to the SNR, which acts as a tunable parameter.
\begin{equation}
    \hat{I}_\textrm{reconstructed} = \textrm{PSF}^+(I_\textrm{extended} + \mathcal{N}(0,\textrm{max}(I_\textrm{extended})/\textrm{SNR})).
\end{equation}
To investigate the properties of this reconstruction schema across multiple different wavelengths, we compute the PSF forward model matrix at a spatial sampling of 256 $\times$ 256 pixels across a finite wavelength grid, with $\lambda \in [0.5, 1, 2, 4, 8, 16]$ $\mu$m, and interpolate the PSF grids to match the wavelengths of the GOES-R radiances described in Section (4.1), which act as the input intensity distributions and have been downsampled to a resolution of 64 $\times$ 64 pixels. The GOES-R data ranges in wavelength from $\lambda \in [0.47, 13.27]$ $\mu$m and we interpolate using the following process. For the $0.47$ $\mu$m channel, we directly use the PSFs computed at $0.5$ $\mu$m for simplicity and the increased computational cost of integrating Equation (\ref{eq:approx}) at short wavelengths. For the rest of the channels, since each lies between two wavelength grid points, we linearly interpolate the resulting PSF grids with a weighting factor corresponding to the linear distance between the wavelength $\lambda$, and the two nearest grid points $\lambda_0,\lambda_1$. The resulting model is
\begin{equation}
    \textrm{PSF}_\lambda = \textrm{PSF}_{\lambda_0}\frac{\lambda - \lambda_1}{|\lambda_1 - \lambda_0|} +  \textrm{PSF}_{\lambda_1}\frac{\lambda - \lambda_0}{|\lambda_1 - \lambda_0|}.
\end{equation}
The input intensities, forward model of the Einstein ring, and corresponding reconstruction for all of the channels in the GOES-R data are shown in Figure (\ref{fig:reconstruction}).

Examining the results of Figure (\ref{fig:reconstruction}), it is clear that in the limit of large SNR, that measurements of the Einstein ring can be used to reconstruct the input intensity distribution. It is important to note that this example relies on super-resolution since the number of reconstructed pixels is larger than the number of independent resolution elements available in the focal plane, which is only possible with very high SNR. For the reflection-dominated channels in the upper half of the figure, the input intensity distribution is a complex function of the distribution of clouds in the planetary atmosphere, due to their high albedo. In the thermal emission dominated channels, the surface emits more than the cold clouds at high altitudes, and the intensity distribution is mostly a full disk, with cold pockets due to absorption from the clouds. This difference in input intensity manifests as a unique distribution of light in the Einstein ring, visible most apparently in the reflection dominated channels, where the bright clouds on the planet's surface result in a non-uniform azimuthal distribution of intensity in the ring. These azimuthal variations encode the information of the original image. The effect is still relevant for the long wavelength channels but less noticeable by eye. Since the distribution of thermal emission is nearly a uniform disk, the resulting Einstein ring profile is mostly a quadrupolar pattern.

\begin{figure}[H]
    \centering
    \includegraphics[width=\textwidth]{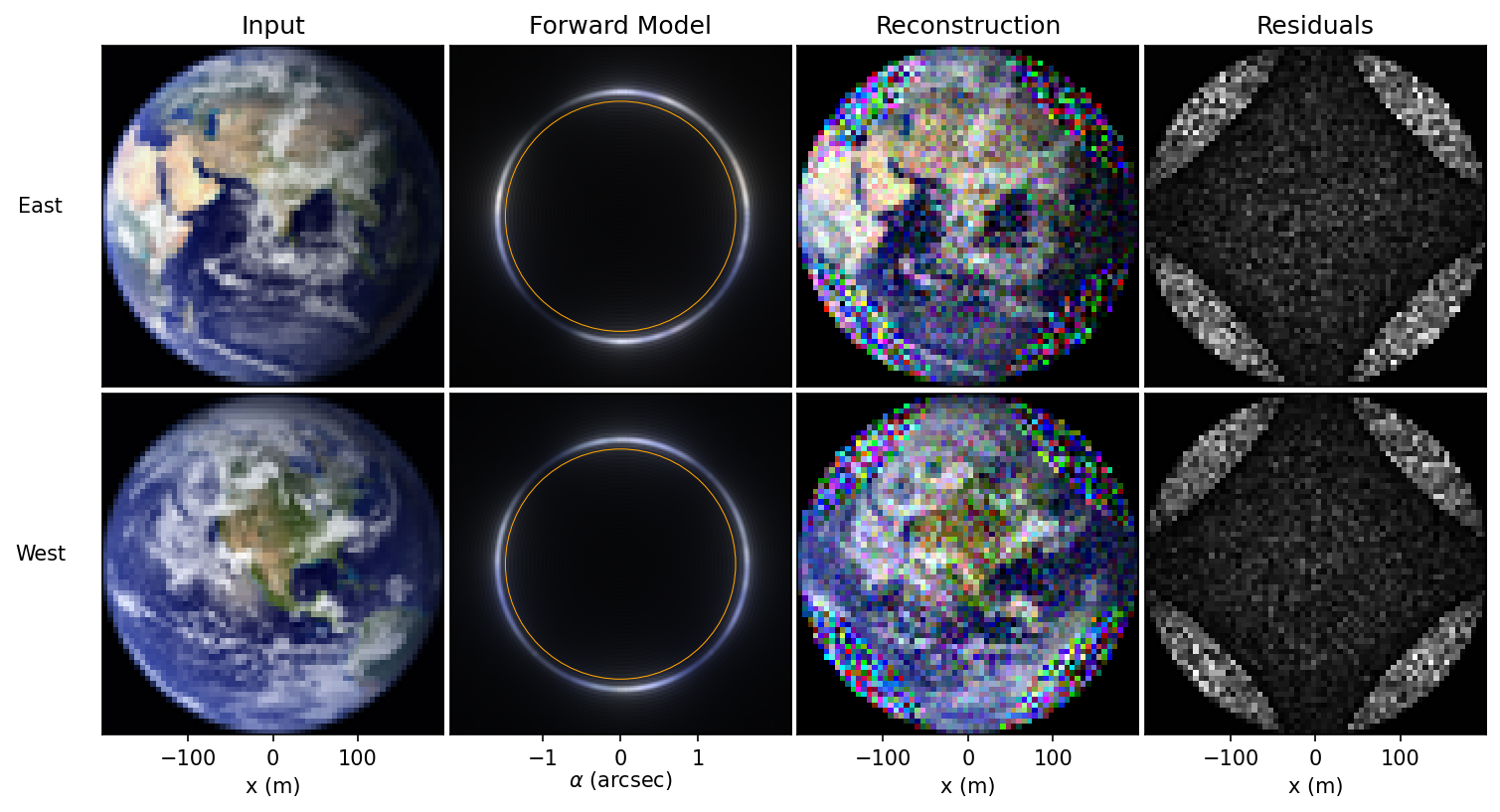}
    \caption{Visible color reconstruction with $d_S = 100$ pc, $d_L = 650$ AU, $\lambda = 0.5$ $\mu$m, $\beta_S = 90^\circ$, and SNR $= 4000$. The first three plots are the same as in Figure (\ref{fig:reconstruction}), with the exception that the solar disk is plotted with an orange circle. The last plot shows the average residuals across all three channels in a white-hot colormap, making apparent the importance of the caustic in imaging reconstruction.}
    \label{fig:color}
\end{figure}

This reconstruction scheme allows one to easily adjust the SNR of the observations and test how the reconstructor fails in the presence of noise. The extremely high SNR = 2$\times 10^5$ shown in Figure (\ref{fig:reconstruction}) is unrealistic considering all of the sources of contamination, but represents the limit of essentially infinite signal to noise ratio. In this limiting case, only the longest wavelength channels have significant degradation across the entire image, as the PSF grows with wavelength and washes out fine structure hidden within the ring. However, degradation can be seen in shorter wavelength channels as well, specifically around the edge of the caustic. To highlight this effect, another reconstruction was done at SNR = $4000$ for $\lambda = 0.5$ $\mu$m. This reconstruction uses visible color imagery of the Earth as the input intensity distribution, and neglects small discrepancies in wavelengths between the different color channels for simplicity. This reconstruction is shown in Figure (\ref{fig:color}).

For points on the surface of the planet which are inside the projection of the caustic, the lens reimages those points into four distinct locations around the edge of the Einstein ring in an asymmetric fashion demonstrated in Figure (\ref{fig:psfs}), while points outside the caustic are reimaged symmetrically to two points just similarly to the lensing of a point mass or the monopole-term only. The implications of this were understood in \citep{madurowicz2020}, where it was demonstrated that the monopole only forward model of the lens has a certain anti-symmetric null space where it is nearly impossible to tell the difference between an input image and the same image rotated by 180 degrees from just the azimuthal profile of the ring. This anti-symmetric null space then requires a certain degree of \textit{navigational symmetry breaking}, where the craft moves laterally to scan over the image in the SGL focal plane to make inferences about the original intensity distribution. This is problematic for rapidly evolving images such as the distribution of clouds across a planet which is rotating and experience changing weather patterns, and especially difficult if long integration times are required to reach sufficient SNR in the presence of foreground contamination.

However, in the presence of the quadrupole moment, this is not the case. The asymmetric reimaging of the points inside the critical caustic allow for direct disentangling of the original intensity distribution inside the projection of the caustic. This demonstrates a fundamental trade off for imaging and spectroscopy with the SGL which must be considered when selecting targets. For targets with small solar co-latitude $\beta_S$, with alignment near the axis of the sun's rotation, the magnification of the SGL is largest, and the total number of photons received from the source is maximal, as the diameter of the astroid caustic $D_\textrm{astroid}$ becomes small, going to exactly zero at $\beta_S = 0$ degrees. For these targets, direct reconstruction of the original intensity distribution from single measurements of the Einstein ring profile is \textit{not possible}, as nearly all of the points are outside the projection of the critical caustic and \textit{reimaged symmetrically}. For target planets in alignment with the solar equator where $\beta_S = 90$ degrees, the magnification of the SGL is small compared to polar alignments, but has the advantage that the astroid caustic is comparatively quite large. With a majority of points reimaged \textit{antisymmetrically} into the Einstein ring, recovering the original intensity distribution from a \textit{single}, high-SNR measurement of the intensity in the Einstein ring is possible. This arrangement could enable \textit{continuous monitoring} of exoplanetary systems for detailed study of the atmospheric dynamics and composition of other worlds.

\section{Conclusion}

The prospect of leveraging the solar gravitational lens to spatially and spectrally resolve the surfaces and atmospheres of extrasolar planets is exciting and fascinating. The combination of extremely high angular resolution and magnification provides unprecedented observational information inaccessible to any other instrument concept. However, the concept is constrained by a number of engineering challenges which are difficult but not insurmountable. Operating a telescope at a minimum heliocentric distance of 650 AU would require extreme patience with conventional and existing rocket technology, with travel times of order $\sim 100$ years, or advancements in propulsion to achieve greater departure velocity, such as a solar sail \citep{TT2020b}. Furthermore, navigational acceleration of order $\textrm{d}v \lesssim 80 \frac{\textrm{m}}{\textrm{s}} + 6.7 \frac{\textrm{m}}{\textrm{s}} \frac{t}{\textrm{year}}$ is needed to obtain and maintain alignment of the craft, sun, and a fixed point on the sky in the worst-case scenario. The fixed cost term is due to the relative lateral motion between the craft and solar system barycenter associated with the escaping hyperbolic trajectory's conservation of specific angular momentum, and the linear cost is associated with perturbations of the solar center of mass with respect to the solar system barycenter due to the gravitational interactions with the other planets in the solar system. Additional acceleration may be needed due to the target planet's system's proper acceleration and relative orbital motion, but this is complicated and could potentially reduce the needed acceleration depending on the alignment of the multiple bodies and orbits. Furthermore, no existing exoplanet characterization methods have obtained sufficient precision to locate the position of the image of the target exoplanet to a comparable degree as the size of the projected image itself, even with GRAVITY-interferometer levels of astrometric precision. Complete astrometric coverage of the target planet will assist in constraining the image location posterior and predicting its future location, which is necessary for navigation. 

The optical properties of the solar gravitational lens depend crucially on the target planet's position on the sky, so ultimately the target selection and mission design processes should become interconnected. The presence of a significant quadrupole moment due to oblateness from the rotation of the sun introduces a critical caustic in the lensing problem which scales proportionally to $\sin^2\beta_S$, the angle between the solar rotational axis and the line of sight. This effect reduces the magnifying power of the lens and modifies the distribution of light in the Einstein ring. For planets in alignment with the solar rotation axis the effect disappears and the lensing problem reduces to a monopole-only gravitational potential, but at the solar equator the effect is maximal. The antisymmetric imaging of points inside the critical caustic can be exploited to break the rotational symmetry of the monopole-only lensing forward model, allowing for direct disentanglement of the original intensity distribution from single high SNR measurements of the Einstein ring. However, for this to be possible the critical caustic must be as large as the projection of the planet, which is only possible for Earth sized planets at a minimum distance of 100 pc. Furthermore, the large SNR required for super-resolution limits the practicality of high resolution image reconstruction using the quadrupole based symmetry breaking on its own, especially at long wavelengths. Low resolution image reconstruction would not require super-resolution and would be possible at smaller SNR. However, other strategies exist which can break this degeneracy, which are especially relevant for nearby or polar-aligned targets. Scanning a single telescope across the projected image must account for temporal changes in the target illumination, which could be solved with a multi-telescope swarm architecture. For single telescopes, it may just be simpler to observe a target during crescent or quarter-illumination phases where the degeneracy is irrelevant, but this is only possible for reflection dominated wavelengths during a fraction of the orbit depending on the target's orbital inclination. Additionally, the rotation of the planet and atmospheric variability introduce additional complications in reconstructing images from observations taken over time. The combination of all of these effects should be investigated in the design of an optimal reconstruction architecture.

Obtaining high SNR measurements of the Einstein ring will be difficult, as contamination from other sources along the line of sight obstruct the faint signal. For visible light, the disk of the solar photosphere has a specific intensity $B_\odot$, while the solar corona is 10$^{-6}$ to 10$^{-9}$ $B_\odot$ depending on elongation and azimuthal angle. The target planet's host star is reimaged into two points in the Einstein ring and can be magnified by a factor $\sim$10 depending on angular separation, and has an intrinsic brightness $\sim$ $B_\odot$ depending on spectral type. An Earth-like target planet has specific intensity $\sim 10^{-6} B_\odot$ and can be magnified by a factor of $\sim$10$^7$ to $\sim$10$^9$ depending on sky location. Background stars or other astronomical objects could be reimaged into the Einstein ring, and the probability of observing a background star depends on the target's galactic coordinates, with the direction of the galactic center being least optimal. A complete strategy to characterize, model, and mitigate all sources of noise has yet to be demonstrated. A pure coronagraphic approach would be limited to a narrow bandpass, and a major advantage of the SGL over direct imaging is the enormous quantity of available photons for high dispersion spectroscopy. A starshade could be an achromatic solution to obtain high contrast across a wide bandpass. The strategy of spectrally and spatially resolving the light in the vicinity of the Einstein ring potentially has the capability to disentangle all sources of noise, but our results are not conclusive. A complete instrument model is outside the scope of this work. As the desire for knowledge locked behind immense observational barriers increases, further understanding of instrumentation from a universal perspective will be necessary to overcome human limitations in the construction of physical and optical instruments.

\hfill

\textit{Acknowledgements:} A.M. would like to thank B.M. for continued financial support and academic freedom in the pursuit of grand ideas. This work benefited from NASA’s Nexus for Exoplanet System Science (NExSS) research coordination network sponsored by NASA’s Science Mission Directorate. This research was sponsored by a grant from NASA NNX15AD95G. A.M. would like to thank Jean-Baptiste Ruffio, Rob de Rosa, Kieran Gilmore, Eric Nielsen, Jason Wang, Sarah Blunt, Lea Hirsch, Michael Hippke, Jason Wright, and Simon Birrer for their helpful discussions, as well as the anonymous referee whose comments greatly helped improve the quality of the manuscript.

\software{Python \citep{python}
        Numpy \citep{numpy}
        Matplotlib \citep{matplotlib}
        Scipy \citep{scipy}
        REBOUND \citep{rebound}
        orbitize! \citep{Blunt_2020}
        Astropy \citep{astropy2018}}
          
\bibliography{main}{}

\begin{thebibliography}{}
\expandafter\ifx\csname natexlab\endcsname\relax\def\natexlab#1{#1}\fi
\providecommand{\url}[1]{\href{#1}{#1}}
\providecommand{\dodoi}[1]{doi:~\href{http://doi.org/#1}{\nolinkurl{#1}}}
\providecommand{\doeprint}[1]{\href{http://ascl.net/#1}{\nolinkurl{http://ascl.net/#1}}}
\providecommand{\doarXiv}[1]{\href{https://arxiv.org/abs/#1}{\nolinkurl{https://arxiv.org/abs/#1}}}

\bibitem[{Abramowitz \& Stegun(1968)}]{abramowitz1968}
Abramowitz, M., \& Stegun, I. 1968, Handbook of Mathematical Functions with
  Formulas, Graphs, and Mathematical Tables, Applied mathematics series No. v.
  55, no. 1972 (U.S. Government Printing Office).
\newblock \url{https://books.google.com/books?id=ZboM5tOFWtsC}

\bibitem[{Akeson {et~al.}(2013)Akeson, Chen, Ciardi, Crane, Good, Harbut,
  Jackson, Kane, Laity, Leifer, Lynn, McElroy, Papin, Plavchan, Ram{\'{\i}}rez,
  Rey, von Braun, Wittman, Abajian, Ali, Beichman, Beekley, Berriman, Berukoff,
  Bryden, Chan, Groom, Lau, Payne, Regelson, Saucedo, Schmitz, Stauffer, Wyatt,
  \& Zhang}]{Akeson2013}
Akeson, R.~L., Chen, X., Ciardi, D., {et~al.} 2013, Publications of the
  Astronomical Society of the Pacific, 125, 989, \dodoi{10.1086/672273}

\bibitem[{Antonucci {et~al.}(2020)Antonucci, Harra, Susino, \&
  Telloni}]{Antonucci2020}
Antonucci, E., Harra, L., Susino, R., \& Telloni, D. 2020, Space Science
  Reviews, 216, 117, \dodoi{10.1007/s11214-020-00743-1}

\bibitem[{{Astropy Collaboration} {et~al.}(2018){Astropy Collaboration},
  {Price-Whelan}, {Sip{\H{o}}cz}, {G{\"u}nther}, {Lim}, {Crawford}, {Conseil},
  {Shupe}, {Craig}, {Dencheva}, {Ginsburg}, {Vand erPlas}, {Bradley},
  {P{\'e}rez-Su{\'a}rez}, {de Val-Borro}, {Aldcroft}, {Cruz}, {Robitaille},
  {Tollerud}, {Ardelean}, {Babej}, {Bach}, {Bachetti}, {Bakanov}, {Bamford},
  {Barentsen}, {Barmby}, {Baumbach}, {Berry}, {Biscani}, {Boquien}, {Bostroem},
  {Bouma}, {Brammer}, {Bray}, {Breytenbach}, {Buddelmeijer}, {Burke},
  {Calderone}, {Cano Rodr{\'\i}guez}, {Cara}, {Cardoso}, {Cheedella}, {Copin},
  {Corrales}, {Crichton}, {D'Avella}, {Deil}, {Depagne}, {Dietrich}, {Donath},
  {Droettboom}, {Earl}, {Erben}, {Fabbro}, {Ferreira}, {Finethy}, {Fox},
  {Garrison}, {Gibbons}, {Goldstein}, {Gommers}, {Greco}, {Greenfield},
  {Groener}, {Grollier}, {Hagen}, {Hirst}, {Homeier}, {Horton}, {Hosseinzadeh},
  {Hu}, {Hunkeler}, {Ivezi{\'c}}, {Jain}, {Jenness}, {Kanarek}, {Kendrew},
  {Kern}, {Kerzendorf}, {Khvalko}, {King}, {Kirkby}, {Kulkarni}, {Kumar},
  {Lee}, {Lenz}, {Littlefair}, {Ma}, {Macleod}, {Mastropietro}, {McCully},
  {Montagnac}, {Morris}, {Mueller}, {Mumford}, {Muna}, {Murphy}, {Nelson},
  {Nguyen}, {Ninan}, {N{\"o}the}, {Ogaz}, {Oh}, {Parejko}, {Parley}, {Pascual},
  {Patil}, {Patil}, {Plunkett}, {Prochaska}, {Rastogi}, {Reddy Janga},
  {Sabater}, {Sakurikar}, {Seifert}, {Sherbert}, {Sherwood-Taylor}, {Shih},
  {Sick}, {Silbiger}, {Singanamalla}, {Singer}, {Sladen}, {Sooley},
  {Sornarajah}, {Streicher}, {Teuben}, {Thomas}, {Tremblay}, {Turner},
  {Terr{\'o}n}, {van Kerkwijk}, {de la Vega}, {Watkins}, {Weaver}, {Whitmore},
  {Woillez}, {Zabalza}, \& {Astropy Contributors}}]{astropy2018}
{Astropy Collaboration}, {Price-Whelan}, A.~M., {Sip{\H{o}}cz}, B.~M., {et~al.}
  2018, \aj, 156, 123, \dodoi{10.3847/1538-3881/aabc4f}

\bibitem[{Blunt {et~al.}(2020)Blunt, Wang, Angelo, Ngo, Cody, Rosa, Graham,
  Hirsch, Nagpal, Nielsen, Pearce, Rice, \& Tejada}]{Blunt_2020}
Blunt, S., Wang, J.~J., Angelo, I., {et~al.} 2020, The Astronomical Journal,
  159, 89, \dodoi{10.3847/1538-3881/ab6663}

\bibitem[{Boe {et~al.}(2021)Boe, Habbal, Downs, \& Druckmüller}]{boe2021}
Boe, B., Habbal, S., Downs, C., \& Druckmüller, M. 2021, The Astrophysical
  Journal, 912, 44, \dodoi{10.3847/1538-4357/abea79}

\bibitem[{Brown(2021)}]{gaia2021}
Brown, A.~G. 2021, Annual Review of Astronomy and Astrophysics, 59, 59,
  \dodoi{10.1146/annurev-astro-112320-035628}

\bibitem[{Bryan {et~al.}(2020)Bryan, Chiang, Bowler, Morley, Millholland,
  Blunt, Ashok, Nielsen, Ngo, Mawet, \& Knutson}]{Bryan_2020}
Bryan, M.~L., Chiang, E., Bowler, B.~P., {et~al.} 2020, The Astronomical
  Journal, 159, 181, \dodoi{10.3847/1538-3881/ab76c6}

\bibitem[{{Calbert} \& {Beard}(1972)}]{Calbert1972}
{Calbert}, R., \& {Beard}, D.~B. 1972, The Astrophysical Journal, 176, 497,
  \dodoi{10.1086/151652}

\bibitem[{Carroll \& Ostlie(2007)}]{Carroll2007}
Carroll, B.~W., \& Ostlie, D.~A. 2007, {A}n {I}ntroduction to {M}odern
  {A}strophysics, 2nd edn. (Addison-Wesley, San Francisco: Pearson).
\newblock \url{https://books.google.com/books?id=PY0wDwAAQBAJ}

\bibitem[{Cash(2011)}]{cash2011}
Cash, W. 2011, The Astrophysical Journal, 738, 76,
  \dodoi{10.1088/0004-637x/738/1/76}

\bibitem[{{Cionco} \& {Pavlov}(2018)}]{barycenter2018}
{Cionco}, R.~G., \& {Pavlov}, D.~A. 2018, Astronomy and Astrophysics, 615,
  A153, \dodoi{10.1051/0004-6361/201732349}

\bibitem[{Coddington {et~al.}(2015)Coddington, Lean, Lindholm, Pilewskie, Snow,
  \& Program}]{noaaCDR}
Coddington, O., Lean, J.~L., Lindholm, D., {et~al.} 2015, NOAA National Centers
  for Environmental Information, \dodoi{10.7289/V51J97P6}

\bibitem[{Cracraft {et~al.}(2021)Cracraft, Rosa, Sparks, Bailey, \&
  Turnbull}]{cracraft2021}
Cracraft, M., Rosa, R.~D., Sparks, W., Bailey, V., \& Turnbull, M. 2021,
  arXiv:2110.08097.
\newblock \doarXiv{2110.08097}

\bibitem[{Crossfield {et~al.}(2014)Crossfield, Biller, Schlieder, Deacon,
  Bonnefoy, Homeier, Allard, Buenzli, Henning, Brandner, Goldman, \&
  Kopytova}]{Crossfield2014}
Crossfield, I. J.~M., Biller, B., Schlieder, J.~E., {et~al.} 2014, Nature, 505,
  654, \dodoi{10.1038/nature12955}

\bibitem[{{Czekaj, M. A.} {et~al.}(2014){Czekaj, M. A.}, {Robin, A. C.},
  {Figueras, F.}, {Luri, X.}, \& {Haywood, M.}}]{czekaj2014}
{Czekaj, M. A.}, {Robin, A. C.}, {Figueras, F.}, {Luri, X.}, \& {Haywood, M.}
  2014, Astronomy and Astrophysics, 564, A102,
  \dodoi{10.1051/0004-6361/201322139}

\bibitem[{de~Jong \& van Soldt(1989)}]{deJong1989}
de~Jong, T., \& van Soldt, W.~H. 1989, Nature, 338, 238,
  \dodoi{10.1038/338238a0}

\bibitem[{de~Wijn {et~al.}(2012)de~Wijn, Burkepile, Tomczyk, Nelson, Huang, \&
  Gallagher}]{wijn2012}
de~Wijn, A.~G., Burkepile, J.~T., Tomczyk, S., {et~al.} 2012, Ground-based and
  Airborne Telescopes IV, 8444, 1201 , \dodoi{10.1117/12.926511}

\bibitem[{Deeg \& Alonso(2018)}]{Deeg2018}
Deeg, H.~J., \& Alonso, R. 2018, Transit Photometry as an Exoplanet Discovery
  Method (Cham: Springer International Publishing), 633--657,
  \dodoi{10.1007/978-3-319-55333-7_117}

\bibitem[{Dreyer(1877)}]{DREYER1877}
Dreyer, J. L.~E. 1877, Nature, 16, 549, \dodoi{10.1038/016549b0}

\bibitem[{{Eggenberger, A.} \& {Udry, S.}(2010)}]{Eggenberger2010}
{Eggenberger, A.}, \& {Udry, S.} 2010, EAS Publications Series, 41, 27,
  \dodoi{10.1051/eas/1041002}

\bibitem[{{Eshleman}(1979)}]{ehsleman1979}
{Eshleman}, V.~R. 1979, Science, 205, 1133,
  \dodoi{10.1126/science.205.4411.1133}

\bibitem[{Ferrari {et~al.}(2009)Ferrari, Aime, \& Soummer}]{ferrari2009}
Ferrari, A., Aime, C., \& Soummer, R. 2009, The Astrophysical Journal, 708,
  218, \dodoi{10.1088/0004-637X/708/1/218}

\bibitem[{{Foreman-Mackey} {et~al.}(2013){Foreman-Mackey}, {Hogg}, {Lang}, \&
  {Goodman}}]{ForemanMackey2013}
{Foreman-Mackey}, D., {Hogg}, D.~W., {Lang}, D., \& {Goodman}, J. 2013, \pasp,
  125, 306, \dodoi{10.1086/670067}

\bibitem[{Fröhlich \& Lean(1998)}]{frohlich1998}
Fröhlich, C., \& Lean, J. 1998, Symposium - International Astronomical Union,
  185, 89–102, \dodoi{10.1017/S0074180900238357}

\bibitem[{Gaudi {et~al.}(2018)Gaudi, Seager, Mennesson, Kiessling, Warfield,
  Kuan, Cahoy, Clarke, Domagal-Goldman, Feinberg, Guyon, Kasdin, Mawet,
  Robinson, Rogers, Scowen, Somerville, Stapelfeldt, Stark, \& Turner}]{habex}
Gaudi, B., Seager, S., Mennesson, B., {et~al.} 2018, arXiv: Instrumentation and
  Methods for Astrophysics.
\newblock \doarXiv{2001.06683}

\bibitem[{Gillon {et~al.}(2017)Gillon, Triaud, Demory, Jehin, Agol, Deck,
  Lederer, de~Wit, Burdanov, Ingalls, Bolmont, Leconte, Raymond, Selsis,
  Turbet, Barkaoui, Burgasser, Burleigh, Carey, Chaushev, Copperwheat, Delrez,
  Fernandes, Holdsworth, Kotze, Van~Grootel, Almleaky, Benkhaldoun, Magain, \&
  Queloz}]{Gillon2017}
Gillon, M., Triaud, A. H. M.~J., Demory, B.-O., {et~al.} 2017, Nature, 542,
  456, \dodoi{10.1038/nature21360}

\bibitem[{Giorgini {et~al.}(1996)Giorgini, Yeomans, Chamberlin, Chodas,
  Jacobson, Keesey, Lieske, Ostro, Standish, \& Wimberly}]{jplhorizons}
Giorgini, J., Yeomans, D., Chamberlin, A., {et~al.} 1996, Bulletin of the
  American Astronomical Society, 28, 1158.
\newblock \url{https://ssd.jpl.nasa.gov/horizons/app.html}

\bibitem[{Goebel \& Katz(2008)}]{goebel2008}
Goebel, D., \& Katz, I. 2008, Fundamentals of Electric Propulsion: Ion and Hall
  Thrusters, JPL Space Science and Technology Series (Wiley).
\newblock \url{https://books.google.com/books?id=P5OGFXcBKcwC}

\bibitem[{{GOES-R Calibration Working Group} \& {GOES-R Series
  Program}(2017)}]{goesr}
{GOES-R Calibration Working Group}, \& {GOES-R Series Program}. 2017, NOAA
  National Centers for Environmental Information, \dodoi{10.7289/V5BV7DSR}

\bibitem[{Goodman(1996)}]{goodman1996}
Goodman, J. 1996, Introduction to Fourier Optics, McGraw-Hill Series in
  Electrical and Computer Engineering: Communications and Signal Processing
  (McGraw-Hill).
\newblock \url{https://books.google.com/books?id=QllRAAAAMAAJ}

\bibitem[{{GRAVITY Collaboration} {et~al.}(2019){GRAVITY Collaboration},
  {Lacour, S.}, {Nowak, M.}, {Wang, J.}, {Pfuhl, O.}, {Eisenhauer, F.},
  {Abuter, R.}, {Amorim, A.}, {Anugu, N.}, {Benisty, M.}, {Berger, J. P.},
  {Beust, H.}, {Blind, N.}, {Bonnefoy, M.}, {Bonnet, H.}, {Bourget, P.},
  {Brandner, W.}, {Buron, A.}, {Collin, C.}, {Charnay, B.}, {Chapron, F.},
  {Cl\'enet, Y.}, {Coud\'e du Foresto, V.}, {de Zeeuw, P. T.}, {Deen, C.},
  {Dembet, R.}, {Dexter, J.}, {Duvert, G.}, {Eckart, A.}, {F\"orster Schreiber,
  N. M.}, {F\'edou, P.}, {Garcia, P.}, {Garcia Lopez, R.}, {Gao, F.}, {Gendron,
  E.}, {Genzel, R.}, {Gillessen, S.}, {Gordo, P.}, {Greenbaum, A.}, {Habibi,
  M.}, {Haubois, X.}, {Hau\ss{}mann, F.}, {Henning, Th.}, {Hippler, S.},
  {Horrobin, M.}, {Hubert, Z.}, {Jimenez Rosales, A.}, {Jocou, L.}, {Kendrew,
  S.}, {Kervella, P.}, {Kolb, J.}, {Lagrange, A.-M.}, {Lapeyr\`ere, V.}, {Le
  Bouquin, J.-B.}, {L\'ena, P.}, {Lippa, M.}, {Lenzen, R.}, {Maire, A.-L.},
  {Molli\`ere, P.}, {Ott, T.}, {Paumard, T.}, {Perraut, K.}, {Perrin, G.},
  {Pueyo, L.}, {Rabien, S.}, {Ram\'{\i}rez, A.}, {Rau, C.},
  {Rodr\'{\i}guez-Coira, G.}, {Rousset, G.}, {Sanchez-Bermudez, J.},
  {Scheithauer, S.}, {Schuhler, N.}, {Straub, O.}, {Straubmeier, C.}, {Sturm,
  E.}, {Tacconi, L. J.}, {Vincent, F.}, {van Dishoeck, E. F.}, {von Fellenberg,
  S.}, {Wank, I.}, {Waisberg, I.}, {Widmann, F.}, {Wieprecht, E.}, {Wiest, M.},
  {Wiezorrek, E.}, {Woillez, J.}, {Yazici, S.}, {Ziegler, D.}, \& {Zins,
  G.}}]{gravity2019}
{GRAVITY Collaboration}, {Lacour, S.}, {Nowak, M.}, {et~al.} 2019, Astronomy
  and Astrophysics, 623, L11, \dodoi{10.1051/0004-6361/201935253}

\bibitem[{Gregorcic(2001)}]{Gregorcic2001}
Gregorcic, G. 2001.
\newblock
  \url{https://www.cs.bgu.ac.il/~na131/wiki.files/SVD_application_paper.pdf}

\bibitem[{Guo \& Farquhar(2008)}]{Guo2008}
Guo, Y., \& Farquhar, R.~W. 2008, Space Science Reviews, 140, 49,
  \dodoi{10.1007/s11214-007-9242-y}

\bibitem[{Harris {et~al.}(2020)Harris, Millman, van~der Walt, Gommers,
  Virtanen, Cournapeau, Wieser, Taylor, Berg, Smith, Kern, Picus, Hoyer, van
  Kerkwijk, Brett, Haldane, del R{\'{i}}o, Wiebe, Peterson,
  G{\'{e}}rard-Marchant, Sheppard, Reddy, Weckesser, Abbasi, Gohlke, \&
  Oliphant}]{numpy}
Harris, C.~R., Millman, K.~J., van~der Walt, S.~J., {et~al.} 2020, Nature, 585,
  357, \dodoi{10.1038/s41586-020-2649-2}

\bibitem[{Hecht(2002)}]{Hecht2002}
Hecht, E. 2002, Optics (Addison-Wesley).
\newblock \url{https://books.google.com/books?id=7aG6QgAACAAJ}

\bibitem[{Hendler \& Shrager(1994)}]{HENDLER19941}
Hendler, R.~W., \& Shrager, R.~I. 1994, Journal of Biochemical and Biophysical
  Methods, 28, 1, \dodoi{https://doi.org/10.1016/0165-022X(94)90061-2}

\bibitem[{{Hestroffer} \& {Magnan}(1998)}]{Hestroffer1998}
{Hestroffer}, D., \& {Magnan}, C. 1998, Astronomy and Astrophysics, 333, 338.
\newblock \url{https://ui.adsabs.harvard.edu/abs/1998A&A...333..338H}

\bibitem[{Hippke(2020)}]{Hippke2020}
Hippke, M. 2020, arXiv: Instrumentation and Methods for Astrophysics.
\newblock \doarXiv{2009.01866}

\bibitem[{Hou {et~al.}(2013)Hou, de~Wijn, \& Tomczyk}]{hou2013}
Hou, J., de~Wijn, A.~G., \& Tomczyk, S. 2013, The Astrophysical Journal, 774,
  85, \dodoi{10.1088/0004-637x/774/1/85}

\bibitem[{Howard {et~al.}(2019)Howard, Vourlidas, Bothmer, Colaninno, DeForest,
  Gallagher, Hall, Hess, Higginson, Korendyke, Kouloumvakos, Lamy, Liewer,
  Linker, Linton, Penteado, Plunkett, Poirier, Raouafi, Rich, Rochus,
  Rouillard, Socker, Stenborg, Thernisien, \& Viall}]{Howard2019}
Howard, R.~A., Vourlidas, A., Bothmer, V., {et~al.} 2019, Nature, 576, 232,
  \dodoi{10.1038/s41586-019-1807-x}

\bibitem[{Hunter(2007)}]{matplotlib}
Hunter, J.~D. 2007, Computing in Science \& Engineering, 9, 90,
  \dodoi{10.1109/MCSE.2007.55}

\bibitem[{Judge {et~al.}(2019)Judge, Berkey, Boll, Bryans, Burkepile, Cheimets,
  DeLuca, de~Toma, Gibson, Golub, Hannigan, Madsen, Marquez, Richards, Samra,
  Sewell, Tomczyk, \& Vera}]{Judge2019}
Judge, P., Berkey, B., Boll, A., {et~al.} 2019, Solar Physics, 294, 166,
  \dodoi{10.1007/s11207-019-1550-3}

\bibitem[{{Kimura} \& {Mann}(1998)}]{kimura1998}
{Kimura}, H., \& {Mann}, I. 1998, Earth, Planets and Space, 50, 493,
  \dodoi{10.1186/BF03352140}

\bibitem[{Konopacky {et~al.}(2016)Konopacky, Marois, Macintosh, Galicher,
  Barman, Metchev, \& Zuckerman}]{Konopacky_2016}
Konopacky, Q.~M., Marois, C., Macintosh, B.~A., {et~al.} 2016, The Astronomical
  Journal, 152, 28, \dodoi{10.3847/0004-6256/152/2/28}

\bibitem[{Kraus {et~al.}(2020)Kraus, Bouquin, Kreplin, Davies, Hone, Monnier,
  Gardner, Kennedy, \& Hinkley}]{Kraus_2020}
Kraus, S., Bouquin, J.-B.~L., Kreplin, A., {et~al.} 2020, The Astrophysical
  Journal, 897, L8, \dodoi{10.3847/2041-8213/ab9d27}

\bibitem[{{Kurucz}(1993)}]{Kurucz1993}
{Kurucz}, R. 1993, ATLAS9 Stellar Atmosphere Programs and 2 km/s grid. CD-ROM
  No. 13. Cambridge, 13.
\newblock \url{https://ui.adsabs.harvard.edu/abs/1993KurCD..13.....K}

\bibitem[{{Lamy}(1974)}]{lamy1974}
{Lamy}, P.~L. 1974, Astronomy and Astrophysics, 33, 191.
\newblock \url{https://ui.adsabs.harvard.edu/abs/1974A&A....33..191L}

\bibitem[{{Lesage, A.-L.} \& {Wiedemann, G.}(2014)}]{Lesage2014}
{Lesage, A.-L.}, \& {Wiedemann, G.} 2014, Astronomy and Astrophysics, 563, A86,
  \dodoi{10.1051/0004-6361/201322964}

\bibitem[{{Lindegren, L.} {et~al.}(2018){Lindegren, L.}, {Hern\'andez, J.},
  {Bombrun, A.}, {Klioner, S.}, {Bastian, U.}, {Ramos-Lerate, M.}, {de Torres,
  A.}, {Steidelm\"uller, H.}, {Stephenson, C.}, {Hobbs, D.}, {Lammers, U.},
  {Biermann, M.}, {Geyer, R.}, {Hilger, T.}, {Michalik, D.}, {Stampa, U.},
  {McMillan, P.J.}, {Casta\~neda, J.}, {Clotet, M.}, {Comoretto, G.},
  {Davidson, M.}, {Fabricius, C.}, {Gracia, G.}, {Hambly, N.C.}, {Hutton, A.},
  {Mora, A.}, {Portell, J.}, {van Leeuwen, F.}, {Abbas, U.}, {Abreu, A.},
  {Altmann, M.}, {Andrei, A.}, {Anglada, E.}, {Balaguer-N\'u\~nez, L.},
  {Barache, C.}, {Becciani, U.}, {Bertone, S.}, {Bianchi, L.}, {Bouquillon,
  S.}, {Bourda, G.}, {Br\"usemeister, T.}, {Bucciarelli, B.}, {Busonero, D.},
  {Buzzi, R.}, {Cancelliere, R.}, {Carlucci, T.}, {Charlot, P.}, {Cheek, N.},
  {Crosta, M.}, {Crowley, C.}, {de Bruijne, J.}, {de Felice, F.}, {Drimmel,
  R.}, {Esquej, P.}, {Fienga, A.}, {Fraile, E.}, {Gai, M.}, {Garralda, N.},
  {Gonz\'alez-Vidal, J.J.}, {Guerra, R.}, {Hauser, M.}, {Hofmann, W.}, {Holl,
  B.}, {Jordan, S.}, {Lattanzi, M.G.}, {Lenhardt, H.}, {Liao, S.}, {Licata,
  E.}, {Lister, T.}, {L\"offler, W.}, {Marchant, J.}, {Martin-Fleitas, J.-M.},
  {Messineo, R.}, {Mignard, F.}, {Morbidelli, R.}, {Poggio, E.}, {Riva, A.},
  {Rowell, N.}, {Salguero, E.}, {Sarasso, M.}, {Sciacca, E.}, {Siddiqui, H.},
  {Smart, R.L.}, {Spagna, A.}, {Steele, I.}, {Taris, F.}, {Torra, J.}, {van
  Elteren, A.}, {van Reeven, W.}, \& {Vecchiato, A.}}]{gaia2018}
{Lindegren, L.}, {Hern\'andez, J.}, {Bombrun, A.}, {et~al.} 2018, A\&A, 616,
  A2, \dodoi{10.1051/0004-6361/201832727}

\bibitem[{Loutsenko(2018)}]{Loutsenko2018}
Loutsenko, I. 2018, Progress of Theoretical and Experimental Physics, 2018,
  \dodoi{10.1093/ptep/pty119}

\bibitem[{Loutsenko \& Yermolayeva(2021)}]{loutsenko2021}
Loutsenko, I., \& Yermolayeva, O. 2021, Journal of Astronomical
  Instrumentation, 10, 2150002, \dodoi{10.1142/S2251171721500021}

\bibitem[{{LUVOIR Collaboration}(2019)}]{luvoir}
{LUVOIR Collaboration}. 2019, arXiv: Instrumentation and Methods for
  Astrophysics.
\newblock \doarXiv{1912.06219}

\bibitem[{{Lyot}(1939)}]{lyot1939}
{Lyot}, B. 1939, Monthly Notices of the Royal Astronomical Society, 99, 580,
  \dodoi{10.1093/mnras/99.8.580}

\bibitem[{Macintosh {et~al.}(2015)Macintosh, Graham, Barman, Rosa, Konopacky,
  Marley, Marois, Nielsen, Pueyo, Rajan, Rameau, Saumon, Wang, Patience,
  Ammons, Arriaga, Artigau, Beckwith, Brewster, Bruzzone, Bulger, Burningham,
  Burrows, Chen, Chiang, Chilcote, Dawson, Dong, Doyon, Draper, Duchêne,
  Esposito, Fabrycky, Fitzgerald, Follette, Fortney, Gerard, Goodsell,
  Greenbaum, Hibon, Hinkley, Cotten, Hung, Ingraham, Johnson-Groh, Kalas,
  Lafreniere, Larkin, Lee, Line, Long, Maire, Marchis, Matthews, Max, Metchev,
  Millar-Blanchaer, Mittal, Morley, Morzinski, Murray-Clay, Oppenheimer,
  Palmer, Patel, Perrin, Poyneer, Rafikov, Rantakyrö, Rice, Rojo, Rudy,
  Ruffio, Ruiz, Sadakuni, Saddlemyer, Salama, Savransky, Schneider,
  Sivaramakrishnan, Song, Soummer, Thomas, Vasisht, Wallace, Ward-Duong,
  Wiktorowicz, Wolff, \& Zuckerman}]{macintosh2015}
Macintosh, B., Graham, J.~R., Barman, T., {et~al.} 2015, Science, 350, 64,
  \dodoi{10.1126/science.aac5891}

\bibitem[{Madurowicz(2020)}]{madurowicz2020}
Madurowicz, A. 2020, arXiv:2003.13918.
\newblock \doarXiv{2003.13918}

\bibitem[{Madurowicz {et~al.}(2019)Madurowicz, Macintosh, Chilcote, Perrin,
  Poyneer, Pueyo, Ruffio, Bailey, Barman, Bulger, Cotten, Rosa, Doyon,
  Duchêne, Esposito, Fitzgerald, Follette, Gerard, Goodsell, Graham,
  Greenbaum, Hibon, Hung, Ingraham, Kalas, Konopacky, Maire, Marchis, Marley,
  Marois, Metchev, Millar-Blanchaer, Nielsen, Oppenheimer, Palmer, Patience,
  Rajan, Rameau, Rantakyrö, Savransky, Sivaramakrishnan, Song, Soummer,
  Tallis, Thomas, Wang, Ward-Duong, \& Wolff}]{madurowicz2019}
Madurowicz, A., Macintosh, B., Chilcote, J., {et~al.} 2019, Journal of
  Astronomical Telescopes, Instruments, and Systems, 5, 1 ,
  \dodoi{10.1117/1.JATIS.5.4.049003}

\bibitem[{{Mann}(1992)}]{mann1992}
{Mann}, I. 1992, Astronomy and Astrophysics, 261, 329.
\newblock \url{https://ui.adsabs.harvard.edu/abs/1992A&A...261..329M}

\bibitem[{Masuda(2018)}]{Masuda2018}
Masuda, K. 2018, Measurements of Stellar Obliquities (Singapore: Springer
  Singapore), 21--34, \dodoi{10.1007/978-981-10-8453-9_2}

\bibitem[{Mawet {et~al.}(2012)Mawet, Pueyo, Lawson, Mugnier, Traub, Boccaletti,
  Trauger, Gladysz, Serabyn, Milli, Belikov, Kasper, Baudoz, Macintosh, Marois,
  Oppenheimer, Barrett, Beuzit, Devaney, Girard, Guyon, Krist, Mennesson,
  Mouillet, Murakami, Poyneer, Savransky, Vérinaud, \& Wallace}]{mawet2012}
Mawet, D., Pueyo, L., Lawson, P., {et~al.} 2012, Proceedings of SPIE, 8442, 62
  , \dodoi{10.1117/12.927245}

\bibitem[{Mayor \& Queloz(1995)}]{Mayor1995}
Mayor, M., \& Queloz, D. 1995, Nature, 378, 355, \dodoi{10.1038/378355a0}

\bibitem[{McClain \& Vallado(2001)}]{vallado2001}
McClain, W., \& Vallado, D. 2001, Fundamentals of Astrodynamics and
  Applications, Space Technology Library (Springer Netherlands).
\newblock \url{https://books.google.com/books?id=PJLlWzMBKjkC}

\bibitem[{Mecheri {et~al.}(2004)Mecheri, Abdelatif, Irbah, Provost, \&
  Berthomieu}]{Mecheri2009}
Mecheri, R., Abdelatif, T., Irbah, A., Provost, J., \& Berthomieu, G. 2004,
  Solar Physics, 222, \dodoi{10.1023/B:SOLA.0000043563.96766.21}

\bibitem[{{Mierla} {et~al.}(2008){Mierla}, {Schwenn}, {Teriaca}, {Stenborg}, \&
  {Podlipnik}}]{mierla2008}
{Mierla}, M., {Schwenn}, R., {Teriaca}, L., {Stenborg}, G., \& {Podlipnik}, B.
  2008, Astronomy and Astrophysics, 480, 509,
  \dodoi{10.1051/0004-6361:20078329}

\bibitem[{{Misner} {et~al.}(1973){Misner}, {Thorne}, \& {Wheeler}}]{Misner1973}
{Misner}, C.~W., {Thorne}, K.~S., \& {Wheeler}, J.~A. 1973, {Gravitation} (W.
  H. Freeman Princeton University Press).
\newblock \url{http://adsabs.harvard.edu/abs/1973grav.book.....M}

\bibitem[{Moore(1934)}]{moore1934}
Moore, J.~H. 1934, Publications of the Astronomical Society of the Pacific, 46,
  298, \dodoi{10.1086/124502}

\bibitem[{{Mukai} {et~al.}(1974){Mukai}, {Yamamoto}, {Hasegawa}, {Fujiwara}, \&
  {Koike}}]{mukai1974}
{Mukai}, T., {Yamamoto}, T., {Hasegawa}, H., {Fujiwara}, A., \& {Koike}, C.
  1974, Publications of the Astronomical Society of Japan, 26, 445.
\newblock \url{https://ui.adsabs.harvard.edu/abs/1974PASJ...26..445M}

\bibitem[{Musielak \& Quarles(2014)}]{Musielak_2014}
Musielak, Z.~E., \& Quarles, B. 2014, Reports on Progress in Physics, 77,
  065901, \dodoi{10.1088/0034-4885/77/6/065901}

\bibitem[{{Narayan} \& {Bartelmann}(1996)}]{Narayan1996}
{Narayan}, R., \& {Bartelmann}, M. 1996, arXiv e-prints.
\newblock \doarXiv{astro-ph/9606001}

\bibitem[{{Neckel} \& {Labs}(1994)}]{Neckel1994}
{Neckel}, H., \& {Labs}, D. 1994, Solar Physics, 153, 91,
  \dodoi{10.1007/BF00712494}

\bibitem[{Ohta {et~al.}(2005)Ohta, Taruya, \& Suto}]{Ohta_2005}
Ohta, Y., Taruya, A., \& Suto, Y. 2005, The Astrophysical Journal, 622, 1118,
  \dodoi{10.1086/428344}

\bibitem[{Pasachoff {et~al.}(2018)Pasachoff, Lockwood, Meadors, Yu, Perez,
  Peñaloza-Murillo, Seaton, Voulgaris, Dantowitz, Rušin, \&
  Economou}]{pasachoff2018}
Pasachoff, J.~M., Lockwood, C., Meadors, E., {et~al.} 2018, Frontiers in
  Astronomy and Space Sciences, 5, 37, \dodoi{10.3389/fspas.2018.00037}

\bibitem[{{Pierce}(1954)}]{Pierce1954}
{Pierce}, A.~K. 1954, The Astrophysical Journal, 120, 221,
  \dodoi{10.1086/145905}

\bibitem[{{Pierce} \& {Slaughter}(1977)}]{Pierce1997}
{Pierce}, A.~K., \& {Slaughter}, C.~D. 1977, Solar Physics, 51, 25,
  \dodoi{10.1007/BF00240442}

\bibitem[{{Pueyo}(2018)}]{pueyo2018}
{Pueyo}, L. 2018, in Handbook of Exoplanets, ed. H.~J. {Deeg} \& J.~A.
  {Belmonte} (Springer, Cham), 10, \dodoi{10.1007/978-3-319-55333-7\_10}

\bibitem[{{Rein} \& {Liu}(2012)}]{rebound}
{Rein}, H., \& {Liu}, S.~F. 2012, Astronomy and Astrophysics, 537, A128,
  \dodoi{10.1051/0004-6361/201118085}

\bibitem[{{Rein} \& {Spiegel}(2015)}]{reboundias15}
{Rein}, H., \& {Spiegel}, D.~S. 2015, Monthly Notices of the Royal Astronomical
  Society, 446, 1424, \dodoi{10.1093/mnras/stu2164}

\bibitem[{Robinson {et~al.}(2011)Robinson, Meadows, Crisp, Deming, A'Hearn,
  Charbonneau, Livengood, Seager, Barry, Hearty, Hewagama, Lisse, McFadden, \&
  Wellnitz}]{robinson2011}
Robinson, T.~D., Meadows, V.~S., Crisp, D., {et~al.} 2011, Astrobiology, 11,
  393, \dodoi{10.1089/ast.2011.0642}

\bibitem[{{Roxburgh, I. W.}(2001)}]{Roxbrugh2001}
{Roxburgh, I. W.} 2001, A\&A, 377, 688, \dodoi{10.1051/0004-6361:20011104}

\bibitem[{Ruffio {et~al.}(2019)Ruffio, Macintosh, Konopacky, Barman, Rosa,
  Wang, Wilcomb, Czekala, \& Marois}]{ruffio2019}
Ruffio, J.-B., Macintosh, B., Konopacky, Q.~M., {et~al.} 2019, The Astronomical
  Journal, 158, 200, \dodoi{10.3847/1538-3881/ab4594}

\bibitem[{Schaub \& Junkins(2003)}]{schaub2003}
Schaub, H., \& Junkins, J. 2003, Analytical Mechanics of Space Systems, AIAA
  education series (American Institute of Aeronautics and Astronautics).
\newblock \url{https://books.google.com/books?id=qXvESNWrfpUC}

\bibitem[{Seager(2013)}]{seager2013}
Seager, S. 2013, Science, 340, 577, \dodoi{10.1126/science.1232226}

\bibitem[{Silverwood \& Easther(2019)}]{silverwood_easther_2019}
Silverwood, H., \& Easther, R. 2019, Publications of the Astronomical Society
  of Australia, 36, e038, \dodoi{10.1017/pasa.2019.25}

\bibitem[{Smith \& Gottlieb(1974)}]{Smith1974}
Smith, E. V.~P., \& Gottlieb, D.~M. 1974, Space Science Reviews, 16, 771,
  \dodoi{10.1007/BF00182600}

\bibitem[{{Souami, D.} \& {Souchay, J.}(2012)}]{invariable_plane}
{Souami, D.}, \& {Souchay, J.} 2012, Astronomy and Astrophysics, 543, A133,
  \dodoi{10.1051/0004-6361/201219011}

\bibitem[{Spaan \& Greenaway(2006)}]{spaan2006}
Spaan, F. H.~P., \& Greenaway, A.~H. 2006, Proceedings of SPIE, 6265, 419 ,
  \dodoi{10.1117/12.670248}

\bibitem[{Stern {et~al.}(2015)Stern, Bagenal, Ennico, Gladstone, Grundy,
  McKinnon, Moore, Olkin, Spencer, Weaver, Young, Andert, Andrews, Banks,
  Bauer, Bauman, Barnouin, Bedini, Beisser, Beyer, Bhaskaran, Binzel, Birath,
  Bird, Bogan, Bowman, Bray, Brozovic, Bryan, Buckley, Buie, Buratti, Bushman,
  Calloway, Carcich, Cheng, Conard, Conrad, Cook, Cruikshank, Custodio, Ore,
  Deboy, Dischner, Dumont, Earle, Elliott, Ercol, Ernst, Finley, Flanigan,
  Fountain, Freeze, Greathouse, Green, Guo, Hahn, Hamilton, Hamilton, Hanley,
  Harch, Hart, Hersman, Hill, Hill, Hinson, Holdridge, Horanyi, Howard, Howett,
  Jackman, Jacobson, Jennings, Kammer, Kang, Kaufmann, Kollmann, Krimigis,
  Kusnierkiewicz, Lauer, Lee, Lindstrom, Linscott, Lisse, Lunsford, Mallder,
  Martin, McComas, McNutt, Mehoke, Mehoke, Melin, Mutchler, Nelson, Nimmo,
  Nunez, Ocampo, Owen, Paetzold, Page, Parker, Parker, Pelletier, Peterson,
  Pinkine, Piquette, Porter, Protopapa, Redfern, Reitsema, Reuter, Roberts,
  Robbins, Rogers, Rose, Runyon, Retherford, Ryschkewitsch, Schenk, Schindhelm,
  Sepan, Showalter, Singer, Soluri, Stanbridge, Steffl, Strobel, Stryk,
  Summers, Szalay, Tapley, Taylor, Taylor, Throop, Tsang, Tyler, Umurhan,
  Verbiscer, Versteeg, Vincent, Webbert, Weidner, Weigle, White, Whittenburg,
  Williams, Williams, Williams, Woods, Zangari, \& Zirnstein}]{pluto2015}
Stern, S.~A., Bagenal, F., Ennico, K., {et~al.} 2015, Science, 350, aad1815,
  \dodoi{10.1126/science.aad1815}

\bibitem[{Stone \& Leigh(2019)}]{Stone2019}
Stone, N.~C., \& Leigh, N. W.~C. 2019, Nature, 576, 406,
  \dodoi{10.1038/s41586-019-1833-8}

\bibitem[{{Turyshev} {et~al.}(2020){Turyshev}, {Helvajian}, {Friedman},
  {Heinsheimer}, {Garber}, {Davoyan}, \& {Toth}}]{TT2020b}
{Turyshev}, S.~G., {Helvajian}, H., {Friedman}, L.~D., {et~al.} 2020, arXiv
  e-prints, arXiv:2007.05623.
\newblock \doarXiv{2007.05623}

\bibitem[{Turyshev \& Toth(2020)}]{TT2020}
Turyshev, S.~G., \& Toth, V.~T. 2020, Phys. Rev. D, 102, 024038,
  \dodoi{10.1103/PhysRevD.102.024038}

\bibitem[{{Turyshev} \& {Toth}(2021{\natexlab{a}})}]{TT2021a}
{Turyshev}, S.~G., \& {Toth}, V.~T. 2021{\natexlab{a}}, Phys. Rev. D, 103,
  064076, \dodoi{10.1103/PhysRevD.103.064076}

\bibitem[{{Turyshev} \& {Toth}(2021{\natexlab{b}})}]{TT2021c}
---. 2021{\natexlab{b}}, Phys. Rev. D, 104, 024019,
  \dodoi{10.1103/PhysRevD.104.024019}

\bibitem[{{Turyshev} \& {Toth}(2021{\natexlab{c}})}]{TT2021d}
---. 2021{\natexlab{c}}, Phys. Rev. D, 104, 044032,
  \dodoi{10.1103/PhysRevD.104.044032}

\bibitem[{{Turyshev} \& {Toth}(2021{\natexlab{d}})}]{TT2021e}
---. 2021{\natexlab{d}}, Phys. Rev. D., 104, 124033,
  \dodoi{10.1103/PhysRevD.104.124033}

\bibitem[{{Turyshev} \& {Toth}(2022)}]{TT2021b}
---. 2022, Phys. Rev. D, 105, 044012, \dodoi{10.1103/PhysRevD.105.044012}

\bibitem[{Turyshev {et~al.}(2020)Turyshev, Shao, Toth, Friedman, Alkalai,
  Mawet, Shen, Swain, Zhou, Helvajian, {et~al.}}]{turyshev2020}
Turyshev, S.~G., Shao, M., Toth, V.~T., {et~al.} 2020, arXiv preprint
  arXiv:2002.11871.
\newblock \doarXiv{2002.11871}

\bibitem[{Van~Rossum \& Drake(2009)}]{python}
Van~Rossum, G., \& Drake, F.~L. 2009, Python 3 Reference Manual (Scotts Valley,
  CA: CreateSpace)

\bibitem[{Virtanen {et~al.}(2020)Virtanen, Gommers, Oliphant, Haberland, Reddy,
  Cournapeau, Burovski, Peterson, Weckesser, Bright, {van der Walt}, Brett,
  Wilson, Millman, Mayorov, Nelson, Jones, Kern, Larson, Carey, Polat, Feng,
  Moore, {VanderPlas}, Laxalde, Perktold, Cimrman, Henriksen, Quintero, Harris,
  Archibald, Ribeiro, Pedregosa, {van Mulbregt}, \& {SciPy 1.0
  Contributors}}]{scipy}
Virtanen, P., Gommers, R., Oliphant, T.~E., {et~al.} 2020, Nature Methods, 17,
  261, \dodoi{10.1038/s41592-019-0686-2}

\bibitem[{Vogt {et~al.}(1994)Vogt, Allen, Bigelow, Bresee, Brown, Cantrall,
  Conrad, Couture, Delaney, Epps, Hilyard, Hilyard, Horn, Jern, Kanto, Keane,
  Kibrick, Lewis, Osborne, Pardeilhan, Pfister, Ricketts, Robinson, Stover,
  Tucker, Ward, \& Wei}]{vogt1994}
Vogt, S.~S., Allen, S.~L., Bigelow, B.~C., {et~al.} 1994, in Instrumentation in
  Astronomy VIII, ed. D.~L. Crawford \& E.~R. Craine, Vol. 2198, International
  Society for Optics and Photonics (SPIE), 362 -- 375,
  \dodoi{10.1117/12.176725}

\bibitem[{Woods {et~al.}(1996)Woods, Prinz, Rottman, London, Crane, Cebula,
  Hilsenrath, Brueckner, Andrews, White, VanHoosier, Floyd, Herring, Knapp,
  Pankratz, \& Reiser}]{Woods1996}
Woods, T.~N., Prinz, D.~K., Rottman, G.~J., {et~al.} 1996, Journal of
  Geophysical Research: Atmospheres, 101, 9541,
  \dodoi{https://doi.org/10.1029/96JD00225}

\bibitem[{Zessewitsch \& Nikonow(1929)}]{ZESSEWITSCH1929}
Zessewitsch, W., \& Nikonow, W. 1929, Nature, 123, 909,
  \dodoi{10.1038/123909b0}

\bibitem[{{Zurlo, A.} {et~al.}(2013){Zurlo, A.}, {Vigan, A.}, {Hagelberg, J.},
  {Desidera, S.}, {Chauvin, G.}, {Almenara, J. M.}, {Biazzo, K.}, {Bonnefoy,
  M.}, {Carson, J. C.}, {Covino, E.}, {Delorme, P.}, {D\'{}Orazi, V.},
  {Gratton, R.}, {Mesa, D.}, {Messina, S.}, {Moutou, C.}, {Segransan, D.},
  {Turatto, M.}, {Udry, S.}, \& {Wildi, F.}}]{zurlo2013}
{Zurlo, A.}, {Vigan, A.}, {Hagelberg, J.}, {et~al.} 2013, Astronomy and
  Astrophysics, 554, A21, \dodoi{10.1051/0004-6361/201321179}

\end{thebibliography}
\bibliographystyle{aasjournal}

\end{document}